\documentclass[sort&compress,super,comma,journal=apchd5,manuscript=article,layout=twocolumn]{achemso}

\usepackage[version=3]{mhchem} % Formula subscripts using \ce{}
\setkeys{acs}{usetitle = true, doi = true}
\usepackage[nospread]{cuted}
\usepackage{amsmath}
\usepackage{amssymb} 
\usepackage[colorlinks=true,allcolors=blue]{hyperref}

\newcommand{\dbar}[1]{\overline{\overline{#1}}}
\newcommand{\vect}[1]{\boldsymbol{#1}}
\author{Zoya Eremenko}
\email{eremenko@pks.mpg.de}
\email{zoya.eremenko@gmail.com}
\alsoaffiliation{Leibniz Institute for Solid State and Materials Research, 01069 Dresden, Germany}
\alsoaffiliation{Max Planck Institute for the Physics of Complex Systems, 01187 Dresden, Germany}
\affiliation{O.~Ya.~Usikov Institute for Radiophysics and Electronics of the National Academy of Sciences of Ukraine, 61085 Kharkiv, Ukraine}
\author{Igor Volovichev}
\affiliation{O.~Ya.~Usikov Institute for Radiophysics and Electronics of the National Academy of Sciences of Ukraine, 61085 Kharkiv, Ukraine}
\title[The SPP propagation length]
{Enhancing the Propagation Length of Graphene Surface Plasmon Polaritons using a Metamaterial Substrate with a Near-Zero Refractive Index }

\begin{document}
%%%%%%%%%%%%%%%%%%%%%%%%%%%%%%%%%%%%%%%%%%%%%%%%%%%%%%%%%%%%%%%%%%%%%
%% The "tocentry" environment can be used to create an entry for the
%% graphical table of contents. It is given here as some journals
%% require that it is printed as part of the abstract page. It will
%% be automatically moved as appropriate.
%%%%%%%%%%%%%%%%%%%%%%%%%%%%%%%%%%%%%%%%%%%%%%%%%%%%%%%%%%%%%%%%%%%%%
\begin{tocentry}

\begin{center}
\includegraphics[width=1.0\linewidth, height=4.5cm]{"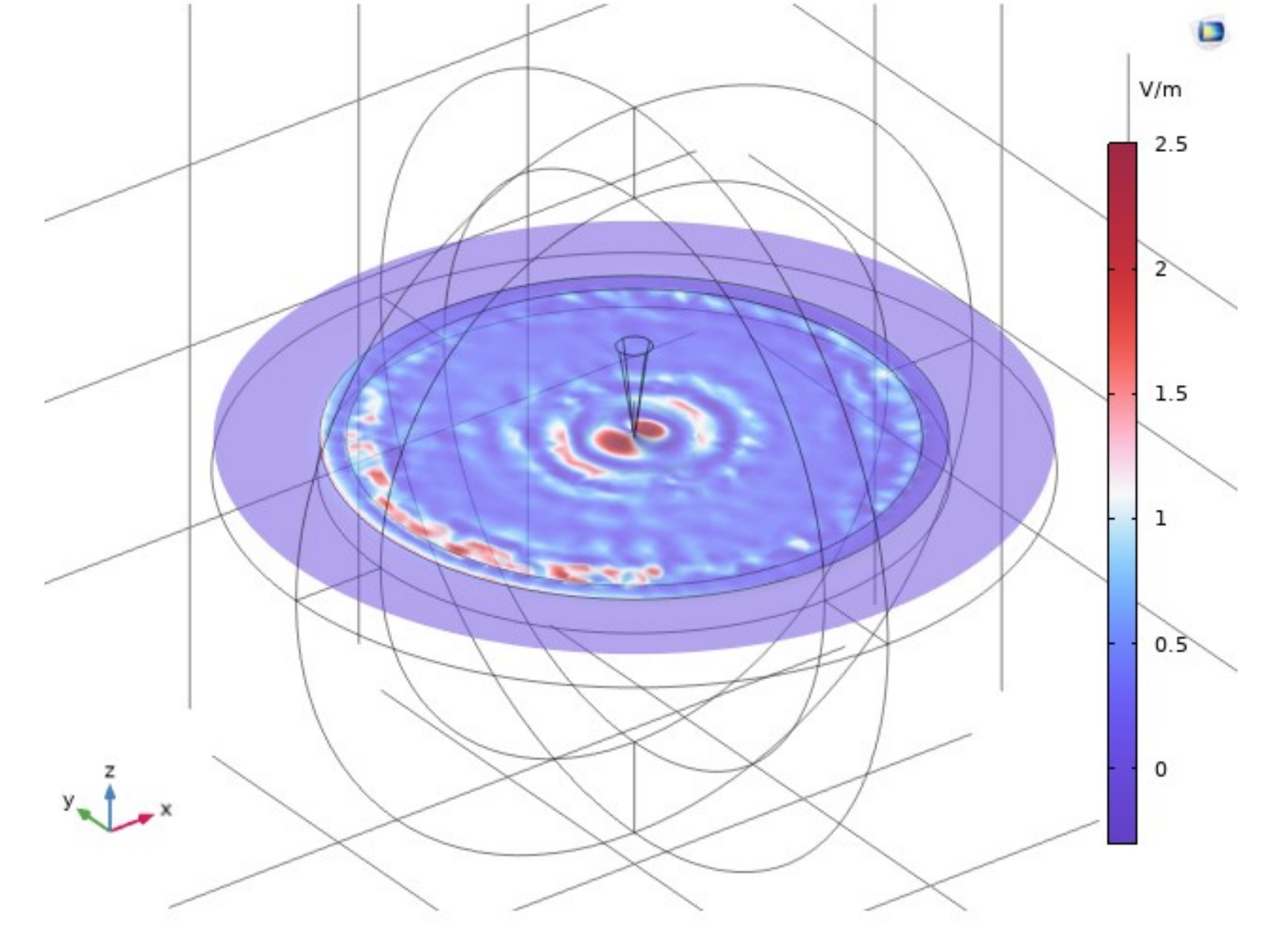"} 
% Some text to explain the graphic.
\end{center}
\end{tocentry}

%\onecolumn
%\begin{multicols}{2}
	
%%%%%%%%%%%%%%%%%%%%%%%%%%%%%%%%%%%%%%%%%%%%%%%%%%%%%%%%%%%%%%%%%%%%%
%% The abstract environment will automatically gobble the contents
%% if an abstract is not used by the target journal.
%%%%%%%%%%%%%%%%%%%%%%%%%%%%%%%%%%%%%%%%%%%%%%%%%%%%%%%%%%%%%%%%%%%%%
\begin{abstract}
This paper aims to investigate the conditions necessary to control, enhance, and modify the propagation length of graphene surface plasmon polaritons (SPPs) at room temperature, using an all-dielectric metamaterial substrate in comparison to suspended graphene. The analysis is conducted within a photonic crystal framework using COMSOL Multiphysics 6.2 to study the resonant modes of the all-dielectric metamaterial. Our results confirm the existence of an near-zero effective refractive index (NZERI) regime at the $\Gamma\text{-point}$ point in the photonic crystal approach. At this NZERI regime a consequence of triply degenerate eigenmodes in a certain frequency range occurs when the effective refractive index of the metasurface approaches zero. Our central idea is that the NZERI regime in the all-dielectric metasurface of a graphene substrate can be used to control, enhance, and modify the propagation of SPPs. We applied several independent theoretical methods to obtain the effective refractive index of the metasurface at NZERI frequency range for two-dimensional and three-dimensional metasurface structures with a graphene layer. Simulation results demonstrate that the effective permittivity and permeability simultaneously attain near-zero values at closely spaced yet distinct frequencies, thereby establishing spectral regions with an effectively vanishing refractive index. 
Key contributions include the first demonstration of NZERI metasurfaces as graphene-supporting platforms for enhancing SPPs, a quantitative approach to balancing propagation distance and field confinement,
and practical design guidelines that align with current nanofabrication capabilities.
Our simulations further demonstrate that when graphene is placed on (or between two) all-dielectric metasurfaces operating in the NZERI regime, the SPP propagation length can be significantly increased. 
\end{abstract}

%%%%%%%%%%%%%%%%%%%%%%%%%%%%%%%%%%%%%%%%%%%%%%%%%%%%%%%%%%%%%%%%%%%%%
%% Start the main part of the manuscript here.
%%%%%%%%%%%%%%%%%%%%%%%%%%%%%%%%%%%%%%%%%%%%%%%%%%%%%%%%%%%%%%%%%%%%%
\section{Introduction}

Research in subwavelength light-matter interactions \cite{Pourmand2023}  has advanced toward the precise engineering of nanostructures with tailored optical properties. The flexibility associated with the application of such nanostructures as advanced artificial two-dimensional (2D) materials (metasurfaces) opens new areas in optics and photonics \cite{Rivera,Ji2023,Hu2021,Fu2022,Liu2023,Chen2024}. In particular, it allows one to design materials with previously unattainable characteristics. 

The results of light-matter interactions can be surface plasmon polaritons (SPPs). This phenomenon refers to an electromagnetic surface wave propagating along a metal–dielectric interface, typically observed at infrared or visible frequencies. The term "surface plasmon polariton" explains that the wave involves both charge motion in the metal ("surface plasmon") and electromagnetic waves in the air or dielectric ("polariton") \cite{X.Luo}.
 
Dynamic control of SPPs in graphene has emerged as a cornerstone for next-generation nanophotonic devices \cite{Li2020}, offering unprecedented capabilities for tunable light-matter interactions at deep subwavelength scales \cite{Jablan2009}.

Compared with SPPs in noble metals (silver or gold), SPPs in graphene have the following advantages \cite{wen2015helicity}: 1) SPPs in graphene exhibit much stronger mode confinement and relatively longer propagation length; 2) it is convenient to control the properties of graphene SPPs because the Fermi energy of graphene can be tuned via the gate in real time; 3) ability to have broadband operation frequencies from GHz to THz.
 
The fundamental limits to SPP dissipation in graphene were determined in Ref.~\cite{G.X.Ni}. The authors used nanometer-scale infrared imaging to study propagating SPPs in high-mobility encapsulating graphene at cryogenic temperatures. In this regime, the propagation of SPPs is limited by the dielectric losses in the encapsulated layers, with a negligible contribution from electron-phonon interactions. At liquid nitrogen temperatures, the intrinsic plasmonic propagation length can exceed 10 $\mu m$.

The maximum SPP propagation length values observed in suspended graphene at room temperature is approximately $ 3~\mu\text{m} $ \cite{Hu2022}. The propagation length of SPPs decreases when graphene is placed on a dielectric substrate due to inherent dielectric losses. This attenuation poses a significant challenge for practical plasmonic applications. Therefore, a critical issue in SPP propagation is the development of strategies to minimize dielectric losses while maintaining efficient waveguiding.

To increase SPP propagation length on graphene layer, for example, in Ref.~\cite{Landa2024} the authors 
examine the frequency and Fermi-level domain for quasi-phase-matched second-harmonic generation of graphene SPP across mid-infrared and terahertz ranges, showing significant SPP amplification with effective loss mitigation. These method achieves propagation lengths exceeding twice those reported in current cryogenic-temperature implementations. 
In a waveguide structure featuring parabolic ridges, the interaction between light and graphene, as well as
the subwavelength confinement of the fundamental SPP mode, can be greatly enhanced. When the graphene chemical potential is set to 1.0 eV, the designed hybrid waveguide demonstrates excellent SPP propagation characteristics, achieving extended propagation lengths ranging from approximately 12 to 17~$\mu$m \cite{Ye2018GraphHyb}.
The SPP propagation length  can be improve by designing a symmetric hybrid waveguide structure with coupled graphene nanoribbons and silica layers around a silicon core, enabling long-range surface plasmon polariton  modes with ultra-long propagation lengths  of order of 10 $\mu$m\cite{Liu2016GrapheneSPP}.

Our goal is to employ substrate engineering to enhance the SPP propagation length in overlaid graphene well in excess of that achievable in suspended graphene sheets. Our proposition is to use an effective near-zero refractive index (NZERI) metamaterial described in Refs.\cite{Moitra2013,Vertchenko2023,Huang2011} as a graphene substrate. 
In these papers the authors theoretically and  experimentally demonstrated an impedance-matched zero-index metamaterial operating at optical frequencies, composed entirely of dielectric materials.
However, its realization remains challenging, as it demands highly specific geometries to satisfy the conditions imposed by effective medium theory. In this paper, we demonstrate computationally the realization of the NZERI regime under specific physical and geometrical parameters of an all-dielectric metasurface (2D arrangements of subwavelength scatterers) unit cell. Using the all-dielectric metasurface in NZERI regime as a graphene layer substrate we present the enhancement of the SPP propagation length in the graphene under specific geometric and physical structure parameters.

This idea based on the following consideration. The phase velocity of the wave is given by $v = {c}/{n} = {\lambda}/{T}$, where $ v $ is the phase velocity, $ c $ is the speed of light in vacuum, $n$ is the refractive index of a material, $ \lambda $ is the wavelength of light inside the material, and $ T $ is the period of the wave. 
As the refractive index $n$ approaches zero, the wavelength tends toward infinity, resulting in a vanishing phase shift. 
This regime, characteristic of NZERI materials, fundamentally alters light-matter interactions, enabling exotic wavefront manipulation and enhanced field confinement. However, the behavior of SPPs on graphene, when coupled to NZERI metasurfaces, remains poorly understood, particularly in terms of SPP propagation length.

This work investigates the propagation dynamics of graphene SPPs coupled to NZERI metasurfaces, combining analytical modeling and full-wave simulations. Understanding these dynamics is critical for applications in ultra-compact plasmonic interconnects, NZERI photonics, and tunable THz devices. Our results will provide a framework for engineering low-loss, high-confinement SPPs in extreme-parameter regimes.

At first, we demonstrate that such an enhancement of the SPP propagation length in graphene on such a substrate indeed occurs. To this end we use Comsol Multiphysics commercial software (see section "Computational methods") to simulate the propagation of SPP in a graphene layer encapsulated between two homogeneous dielectric media (air and NZERI material) and excited by the electromagnetic plane wave. The results of these calculations confirm the increase of SPP propagation length in graphene on a homogeneous substrate with $n < 1$.  The numerical experiment conclusively verifies that, in scenarios where the effective refractive index of the graphene substrate approaches zero, both the SPP wavelength and SPP propagation length are enhanced. In this work we demonstrate that employing an NZERI refractive-index material as a substrate for graphene presents a viable strategy to enhance or modify the SPP propagation length as well as its wavelength.

%The main part of the paper is organized as follows.
%"Problem Statement" presents the structure under study.%"Results and Discussion" section focused subsections such as
%"2D Metasurface Structure in NZERI Regime" examines frequency-dependent NZERI characteristics and provides experimental verification of NZERI conditions in two-dimensional all-dielectric metasurfaces.
%"3D Metasurface Structure in NZERI Regime" extends the analysis to three-dimensional configurations, with comprehensive characterization of NZERI properties.
%"Graphene Layer with Substrates in NZERI Regime" investigates surface plasmon polariton (SPP) propagation enhancements under various model configurations and substrate parameters.
%"Computational Methods" details the numerical simulation framework, including implementation specifics using commercial COMSOL Multiphysics.
%Appendices provide supplementary theoretical foundations.
%Our computational investigation combines first-principles eigensolver analysis (Appendix 1) with two complementary effective medium theories (Appendix 2) to establish design principles for NZERI-enhanced graphene plasmonics.
%It presents two complementary effective medium approaches:
%the magnetodielectric composites theory and
%the surface-wave dispersion retrieval and synthesis technique for metasurfaces
%These methodologies collectively enable systematic determination of NZERI properties for graphene-based metasurface substrates.

\section{Problem statement}	

\begin{figure}[tbh]
	\centering
	\includegraphics[width=0.9\columnwidth]{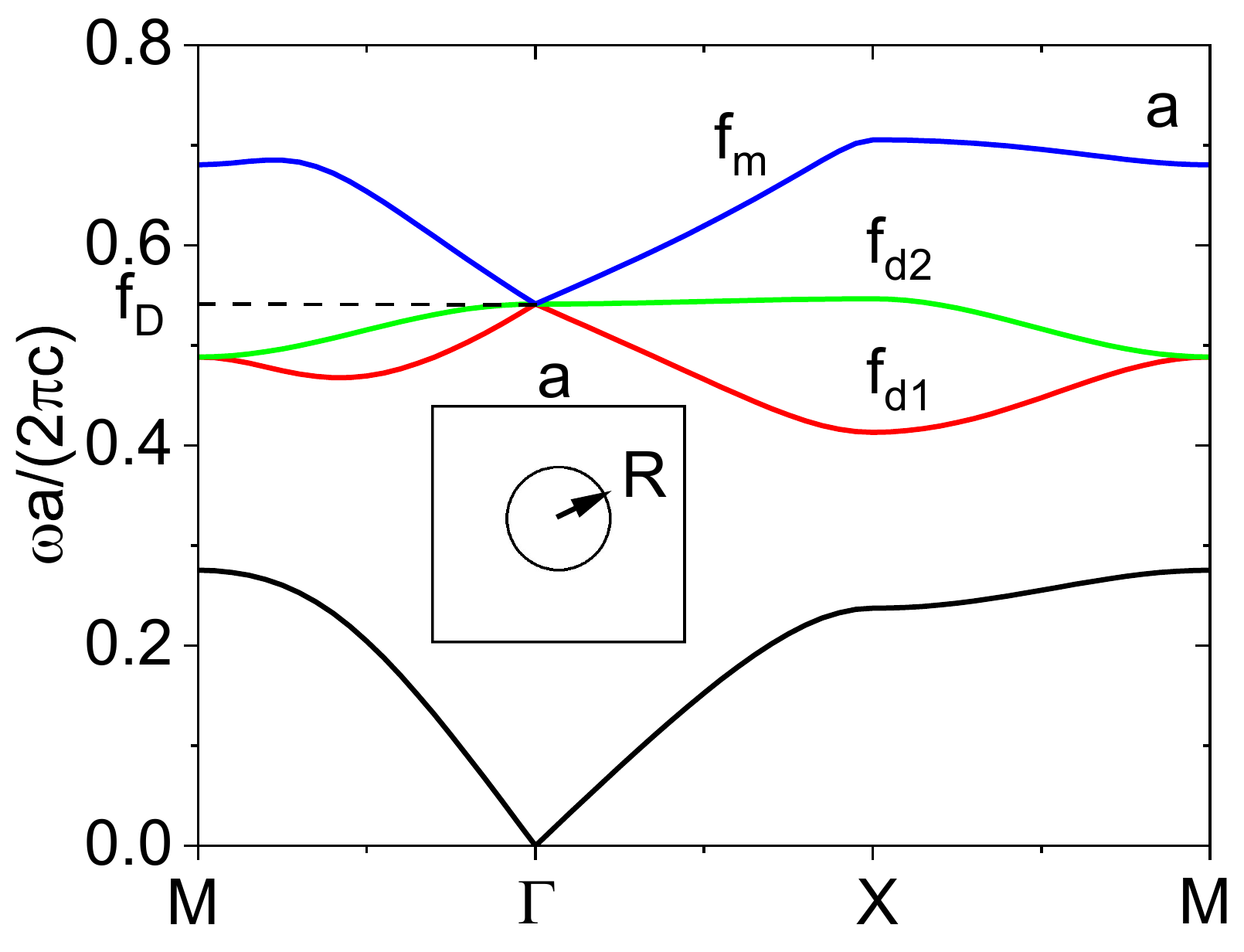}
	\vspace{0.2cm} % Small vertical space between images
	
	\begin{tabular}{ccc}
		\includegraphics[width=0.29\columnwidth]{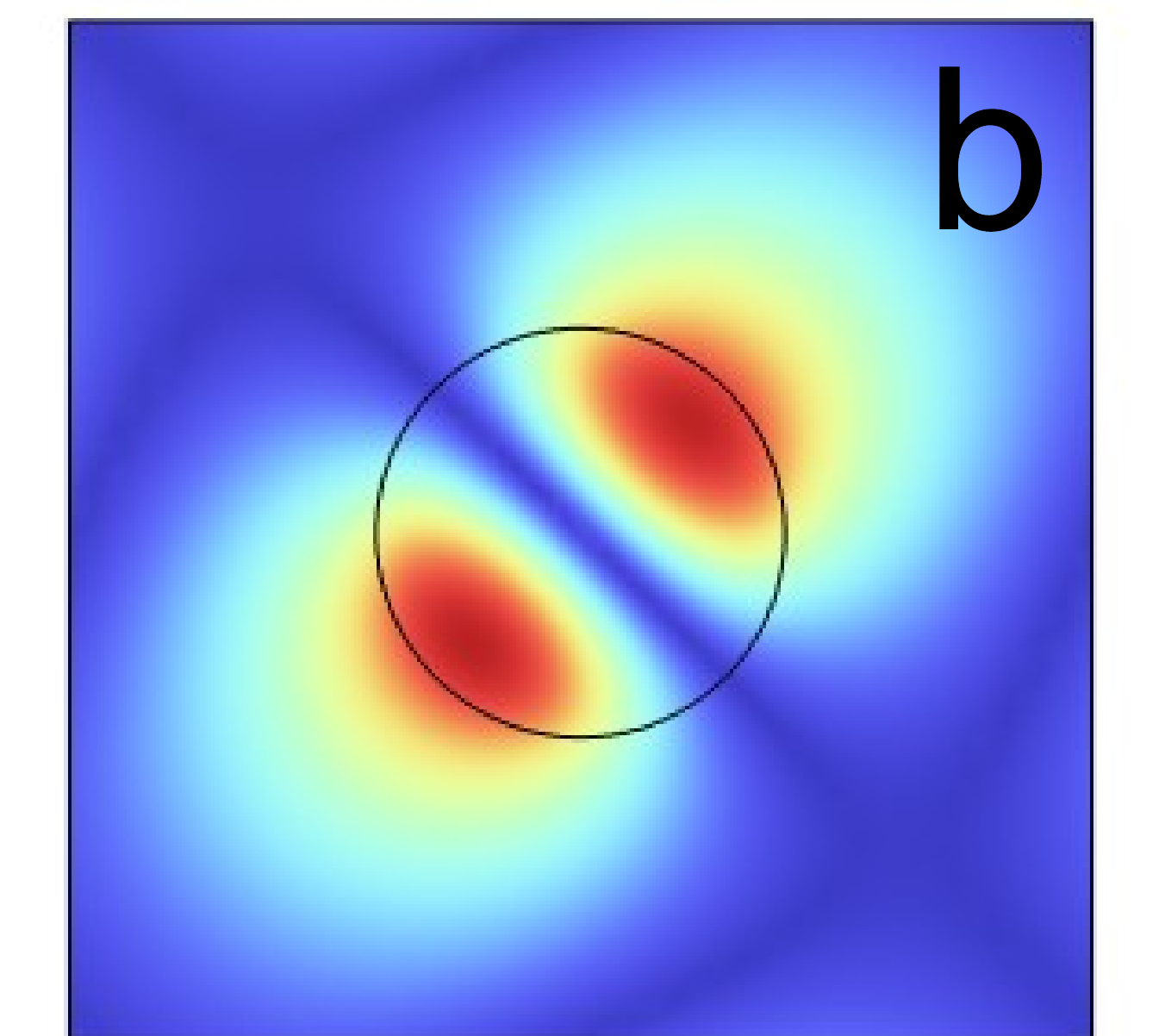} &
		\includegraphics[width=0.29\columnwidth]{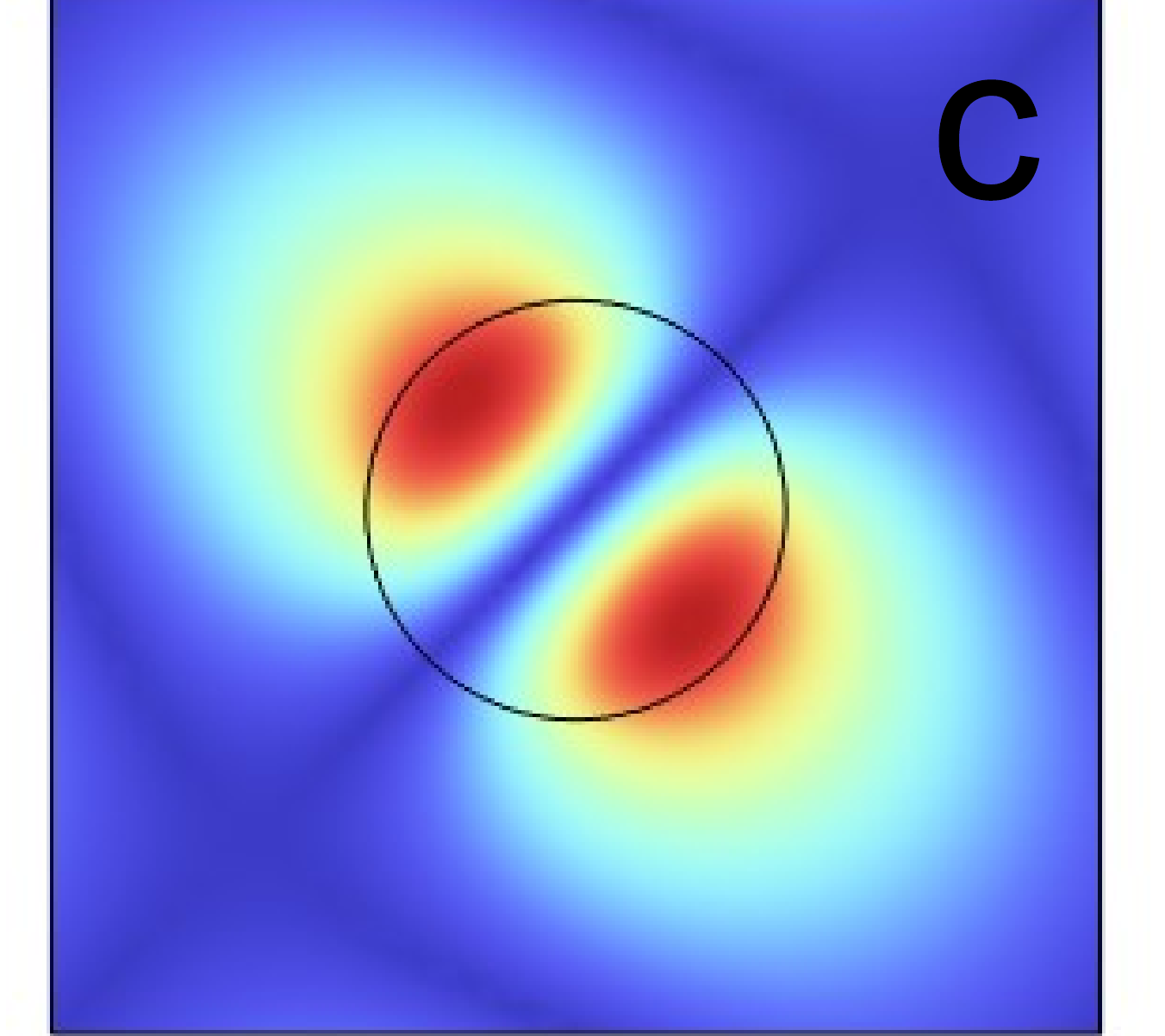} & 
		\includegraphics[width=0.29\columnwidth]{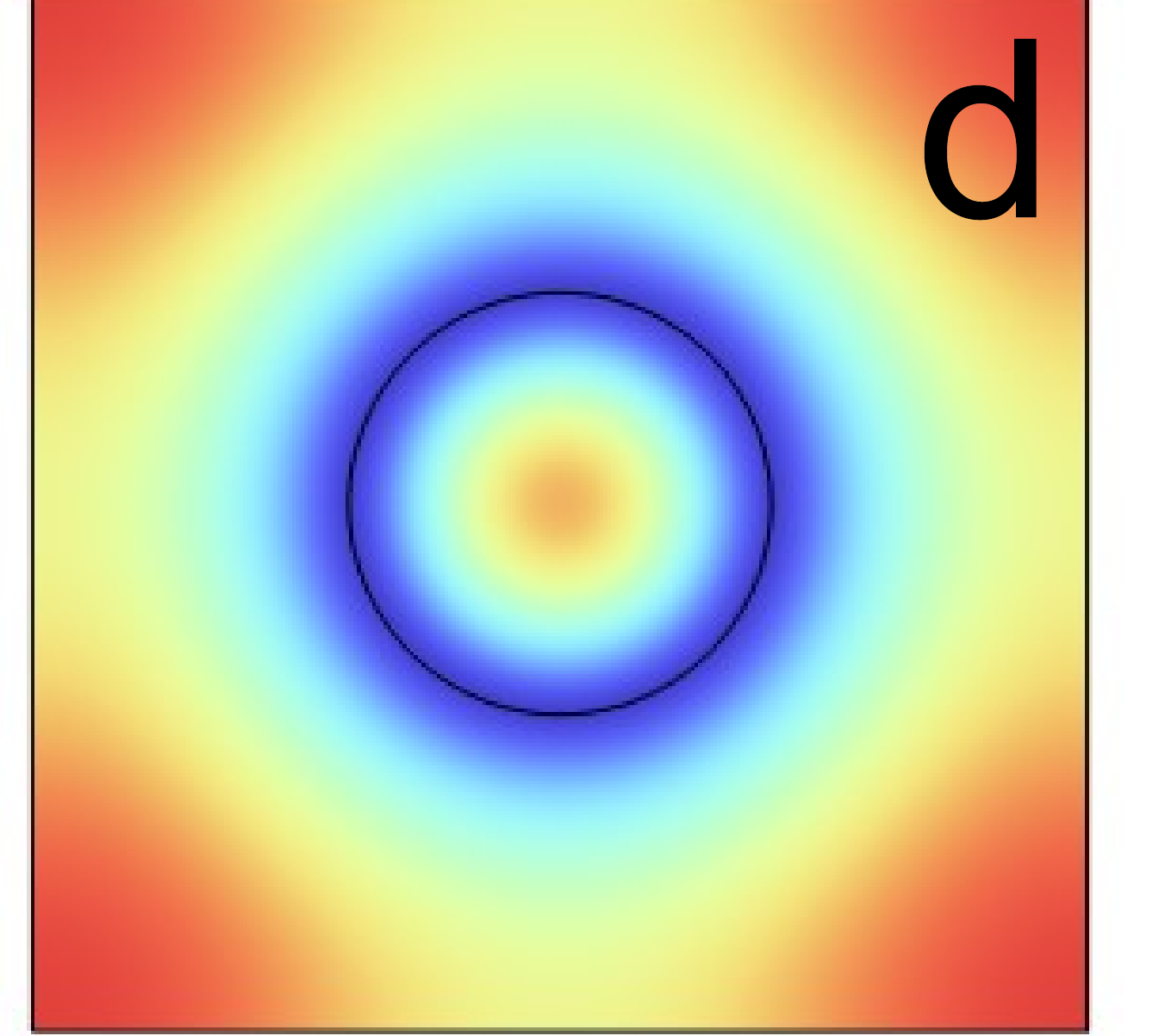} \\
	\end{tabular}
	\caption{The photonic crystal band diagram (a) with triple degenerated dispersion bands intersect at the Dirac point frequency $ f_D~=~0.541c/a$ with the insert of the 2D metasurface square unit cell view  at vertices at $\Gamma$, $X$, and $M$. The plots b,c,d are the electric field distribution for two TM dipoles (b, c) for eigenfrequencies $f_{d1}$, $f_{d2}$ and one TE (d) monopole $f_{m}$ at the frequency $ f_D$ intersection area (NZERI regime), where the red color correspond to the electric field strength $2 V/m $, the blue one is $0~V/m $, $ f_D=$ 32.456 THz  for lattice parameter as a side of the unit cell square $a~=~5 \mu m$ and the central rod radius $R~=~0.2a$.}
	\label{fig:1}
\end{figure}

We consider a 2D/3D all-dielectric metasurface operating in the THz frequency range, serving as a substrate in the NZERI regime for a graphene layer. Although the structure is modeled in 2D and 3D, it can be treated as a metasurface since its thickness (along the $z$-axis) is negligible compared to its in-plane dimensions ($x-y$ plane), and the system is assumed to be periodic (effectively infinite in the transverse directions). 
The metasurface consists of a unit cell containing a single dielectric disk (sphere) made of silicon with permittivity $ \varepsilon^{\prime}+j\varepsilon^{\prime\prime}=12.5+j0.0001$ and permeability 
$\mu~=~1.0$  inside of the square unit cell (see the insert in Figure~\ref{fig:1}a for 2D metasurface structure). In order to realize the regime with NZERI in the graphene substrate material we propose a resonant all-dielectric metasurface study with photonic crystal (PC) approach  presented in Ref.~\cite{Vertchenko2023}. 

%F:\MSCA4UA\Publications Eremenko\2024\paper1
%2D photonic crystal bamd diagram__.opju
%F:\MSCA4UA\Publications Eremenko\2024\paper1\ACS_Photonics_TEX_paper===\FiguresForPaper1\
%D:\COMSOL PROJECTS\2D metasurface as photonic crystal
% Photonic Crystal Band Diagram  in-of-plane solition mikm eps rod = 12.5 a=5 mkm k=0 main point !!!!!!.ini.mph

\begin{figure}[!tbh]
	\centering
	{\includegraphics[width=0.9\linewidth]{"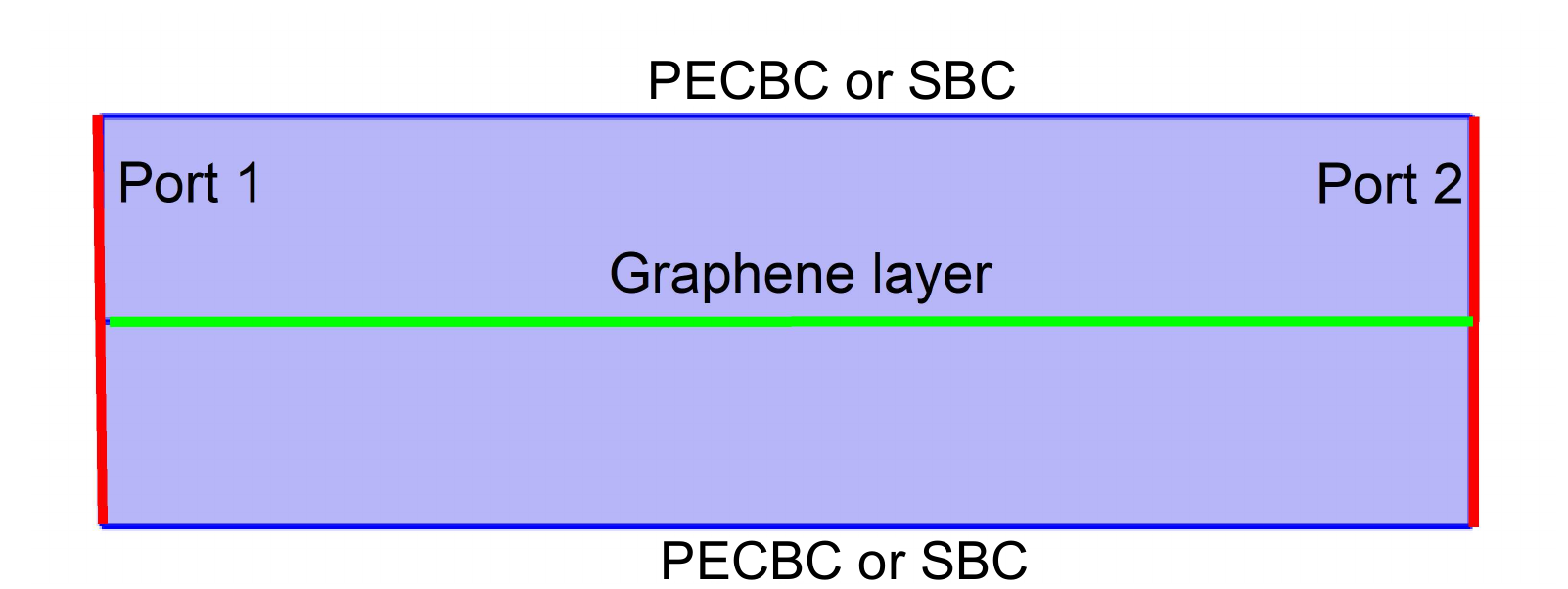"}}
	\caption{The 2D structure with the graphene layer (in green) and  superstrate and substrate areas (in blue) with height of $0.3\mu$m and the width of $2\mu$m. Port 1 and 2 are input and output ports (in red) of the electromagnetic wave propagation along the graphene layer with PECBC or SBC (dark blue borders) on the top and bottom of the structure. }
	\label{fig:fig2Dstructure}
\end{figure}

Let us consider the 2D COMSOL model structure featuring a graphene layer sandwiched between dielectric substrates, as shown in Figure~\ref{fig:fig2Dstructure}. The  port 1, with excitation enabled, launches the SPP mode, while the port 2, with excitation disabled, is designed to absorb the SPPs without reflection. The structure is enclosed by perfect electric conductor boundary conditions (PECBC) or scattering boundary conditions (SBC) on the top and bottom of the structure.

%disper graphene Ef=0.8eV air-air.opju
%F:\MSCA4UA\Publications Eremenko\2024\paper1

The electromagnetic response of a graphene layer can be characterized by its surface optical conductivity. Both intraband $\sigma_{\text{intra}}(\omega$) and interband  $\sigma_{\text{inter}}(\omega$) electronic transitions contribute to the total optical conductivity, using the Kubo formulation~\cite{Andryieuski2013}, it is given by  

%\begin{strip} 
\begin{equation} \label{eq:grcond1}
	\sigma(\omega) = \sigma_{\text{intra}}(\omega) + \sigma_{\text{inter}}(\omega),
\end{equation}
\begin{multline} \label{eq:grcond2}
	\sigma_{\text{intra}}(\omega) = \left( \frac{2 k_B T e^2}{\pi \hbar^2} \right) \times  \\
	\left( \ln \left( 2 \cosh\left[\frac {E_f} {2 k_B T}\right]\right) \frac{-j}{\omega - j/\tau} \right),
\end{multline}
\begin{multline} \label{eq:grcond3}
	\sigma_{\text{inter}}(\omega) = \frac{e^2}{4\hbar} \times\\ 
	\left[ H\left(\frac{\omega}{2}\right) -
	j\frac{4\omega}{\pi} \int_0^\infty \frac{H(\Omega) - H\left(\frac{\omega}{2}\right)}{\omega^2 - 4\Omega^2} d\Omega \right],
\end{multline}
\begin{equation} \label{eq:grcond4}
	H(\Omega) = \frac{\sinh\left(\frac{\hbar \Omega}{k_B T}\right)}{\cosh\left(\frac{\hbar \Omega}{k_B T}\right) + \cosh\left(\frac{E_F}{k_B T}\right)}, \nonumber
\end{equation}
%\end{strip}
where $ k_B $ is the Boltzmann constant, $ \hbar $ is the reduced Planck constant, $ T $ is the temperature, $ e $ is the elementary charge, $ E_F $ is the Fermi energy, $ \omega $ is the angular frequency, and $ \tau $ is the carrier relaxation time.   

In the  Eqs.(\ref{eq:grcond2}-\ref{eq:grcond3}) for COMSOL Multiphysics calculation purpose we changed $+i$ into $-j$ in comparison to the graphene optical conductivity relations presented in Ref.~\cite{Andryieuski2013}. Using 2D model with graphene layer between two dielectric substrates, for example, air, we calculated the dispersion dependence for complex wavevector along the graphene layer for $TM$ mode. 
To check our simulations, using implicit complex SPP dispersion expression (Eqs.~(\ref{eq:grdy1})) in general case, when $\varepsilon_1 $ does not equal to $\varepsilon_2 $, we compared the modeling results with an analytical dispersion expression  (Eqs.~(\ref{eq:grdy2}))  for the graphene layer sandwiched between two identical dielectrics \cite{Bludov2013} and they coincide for air substate expression. Note that there is no closed-form solution to the dispersion relation in the general case (i.e. when the dielectrics are different). However, in the case of strong localization of the SPP, an approximate closed-form solution is reported in Ref.~\cite{Nosich}. But in the case of identical dielectrics, the exact closed-form solution to the dispersion relation can be derived from~\cite{Hanson2008,Bludov2013}
\begin{equation} \label{eq:grdy1}
	\varepsilon_1 k_2 + \varepsilon_2 k_1 + \frac{j k_1 k_2 \sigma}{\varepsilon_0 \omega} = 0, 
\end{equation} 
\begin{equation} \label{eq:grdy1a}
	{k_i}^2 = q^2 - k_0^2\varepsilon_i, 
\end{equation} 
where $	{k_i}$ is cross-sectional propagation number, $\varepsilon_i $ is the permittivity of two substrates, respectively , $q=q^{\prime}+jq^{\prime\prime}$ is complex propagation number along the graphene layer, $k_0=\omega/c$, $\omega$ is the circular frequency, c is the light velocity.
If $\varepsilon_1=~\varepsilon_2=~\varepsilon$ the dispersion equation is as follows 

\begin{equation} \label{eq:grdy2} 
	q^2 = k_0^2 \varepsilon\left( 1 - 4\varepsilon \frac{\varepsilon_0^2 c^2}{\sigma^2} \right). 
\end{equation} 

\begin{figure}[!tbh]
	\centering
	\includegraphics[width=0.9\columnwidth]{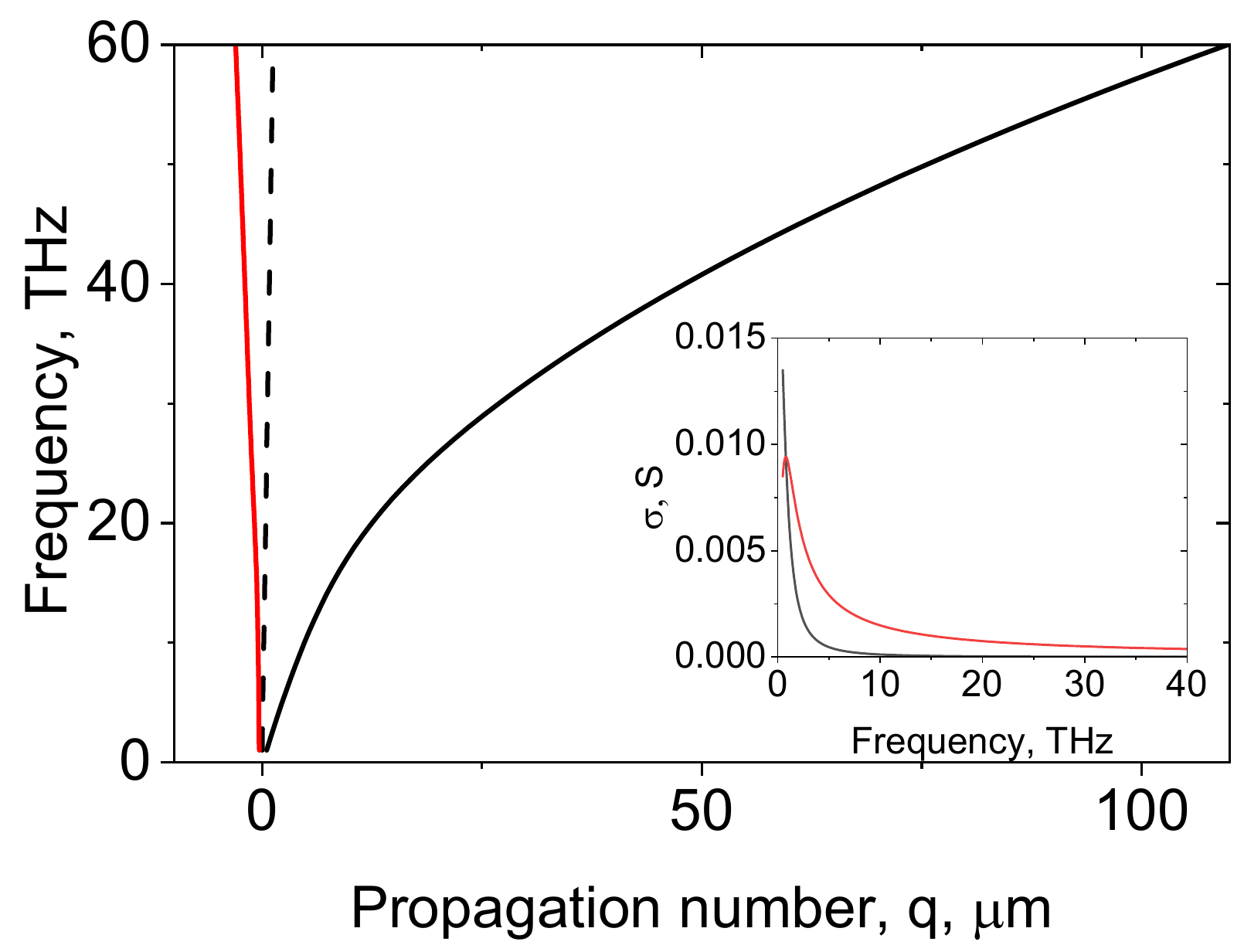}
	\caption{The SPP frequency dispersion dependence on the complex propagation number $q$ along the graphene layer at $\varepsilon_1=~\varepsilon_2=~\varepsilon =~1$.  The real part of $q$ is in black, the imaginary part is in red. The insert is the optical conductivity of the graphene layer at Fermi energy $E_F=0.8$~eV, relaxation time $\tau=2 \cdot 10^{-13}$~s, graphene layer thickness $d_{eff}=0.345$ nm, temperature $T  = 300^0$~K.}
	\label{fig:fig7}
\end{figure}

Figure~\ref{fig:fig7} presents the computed dispersion law with the equation Eqs.~(\ref{eq:grdy1}) for real and imaginary parts of the complex propagation number $q$ along the graphene layer coincide with Comsol dispersion solutions at $\varepsilon_1=~\varepsilon_2=~\varepsilon = 1$.

% graphene spp air-air Ef=0.8 eV gr layer -0.345nm 27-42 THz tau=2E-13.mph
%D:\COMSOL PROJECTS\2D graphene substrate\Ef=0.8eV with substrate - neff\Comparison trans bound conditions
%Sigma graphene Ef= 0.8 eV.opju
% !!!! disper graphene Ef=0.8eV air-air.opju
%F:\MSCA4UA\Publications Eremenko\2024\paper1

\section{Results and Discussion}
\subsection{2D metasurface structure in NZERI regime}
We carry out the eigenfrequency analysis of 2D PC with the single dielectric disk (sphere) unit cell 
to a build photon band diagram  using the known technique presented in Appendix~1.
The metasurface unit cell parameters are the disk radius of $R=0.2a$, where 
$ a=0.5\lambda $ is the lattice parameter of the unit cell (the metasurface unit cell square size), $ \lambda $ is a free-space wavelength of the incident electromagnetic wave taken to be about $ 10~\mu\text{m} $. Such geometric metasurface unit cell parameters were taken to obtain  NZERI regime~\cite{Vertchenko2023,Huang2011}.
The frequency band diagram structure of such a PC is shown in Figure~\ref{fig:1}a presented in 
Ref.~\cite{Huang2011}.
The band structures were computed for transverse magnetic ($ TM $) polarization, where the electric field component is aligned parallel to the disk axis. 
The PC structure was confirmed to exhibit triple-degenerate eigenmodes at the Dirac point, consisting of two $TM$ dipole modes and one transverse electric $TE$ monopole mode (Figure~\ref{fig:1}b-d), similar plots have been illustrated in Ref.~\cite{Huang2011}.

To demonstrate explicitly that the studied metasurface features in NZERI regime near $ f_D $, we use the approach proposed in Ref.~\cite{Vertchenko2023}  that is as follows. We designed a 2D PC model with finite number $N \times N$ of the unit cells and the lattice parameter $a$ and with a bit different disk radius $R^\prime$ in comparison to mentioned above ($R$): $R^\prime~=~0.2085a$ (to excite only the monopole resonance based on the  NZERI regime conditions description in Ref.~\cite{Vertchenko2023}) surrounded by a perfect magnetic conductor (PMC). 
Using PMC allows us to obtain results on a bounded structure that corresponds to an infinite metasurface.
It is worth noting that it is the mandatory condition to excite NZERI regime resonance in the studied metasurface.

\begin{figure}[!tbh]
	\centering
	\includegraphics[width=0.9\columnwidth]{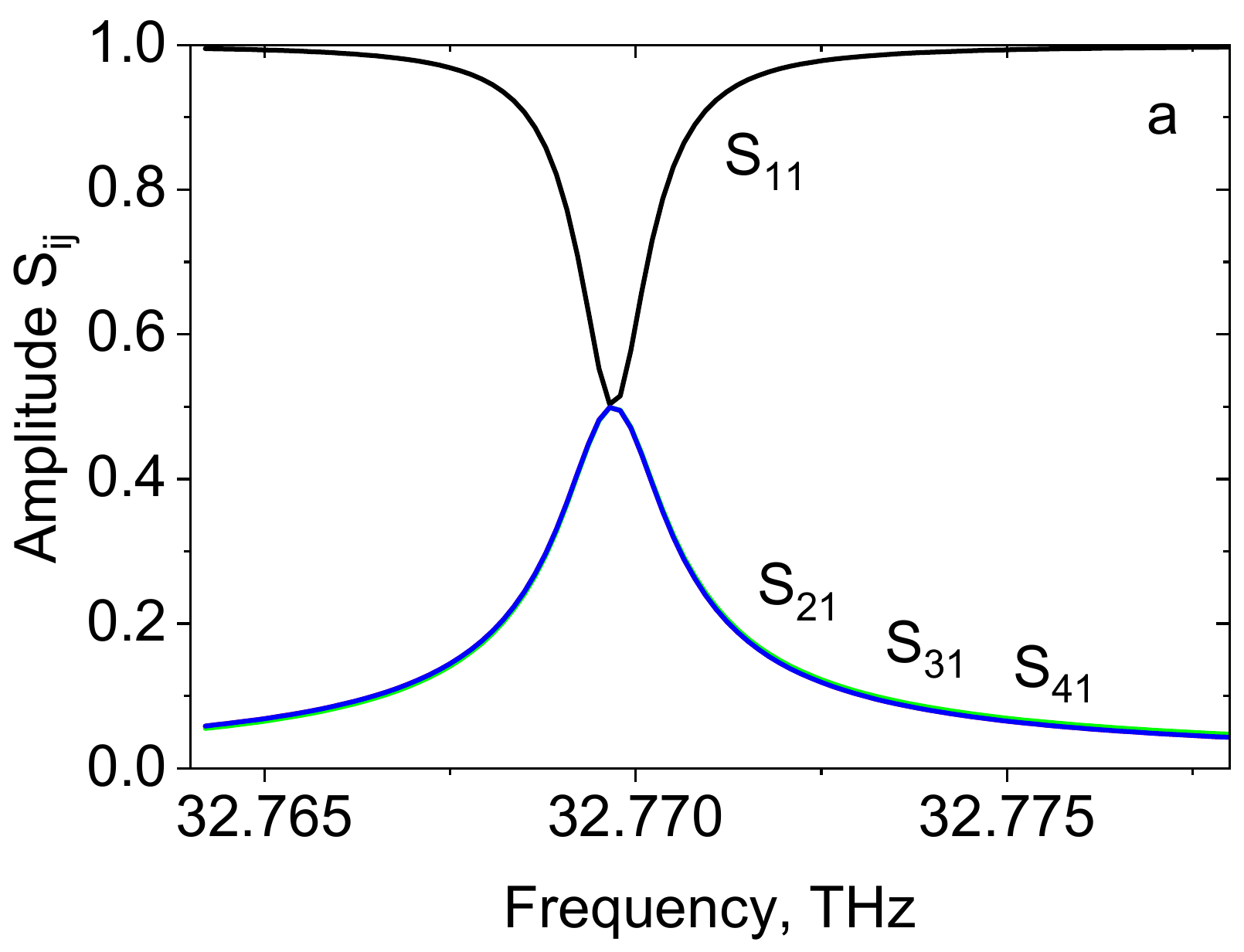} 
	\includegraphics[width=0.9\columnwidth]{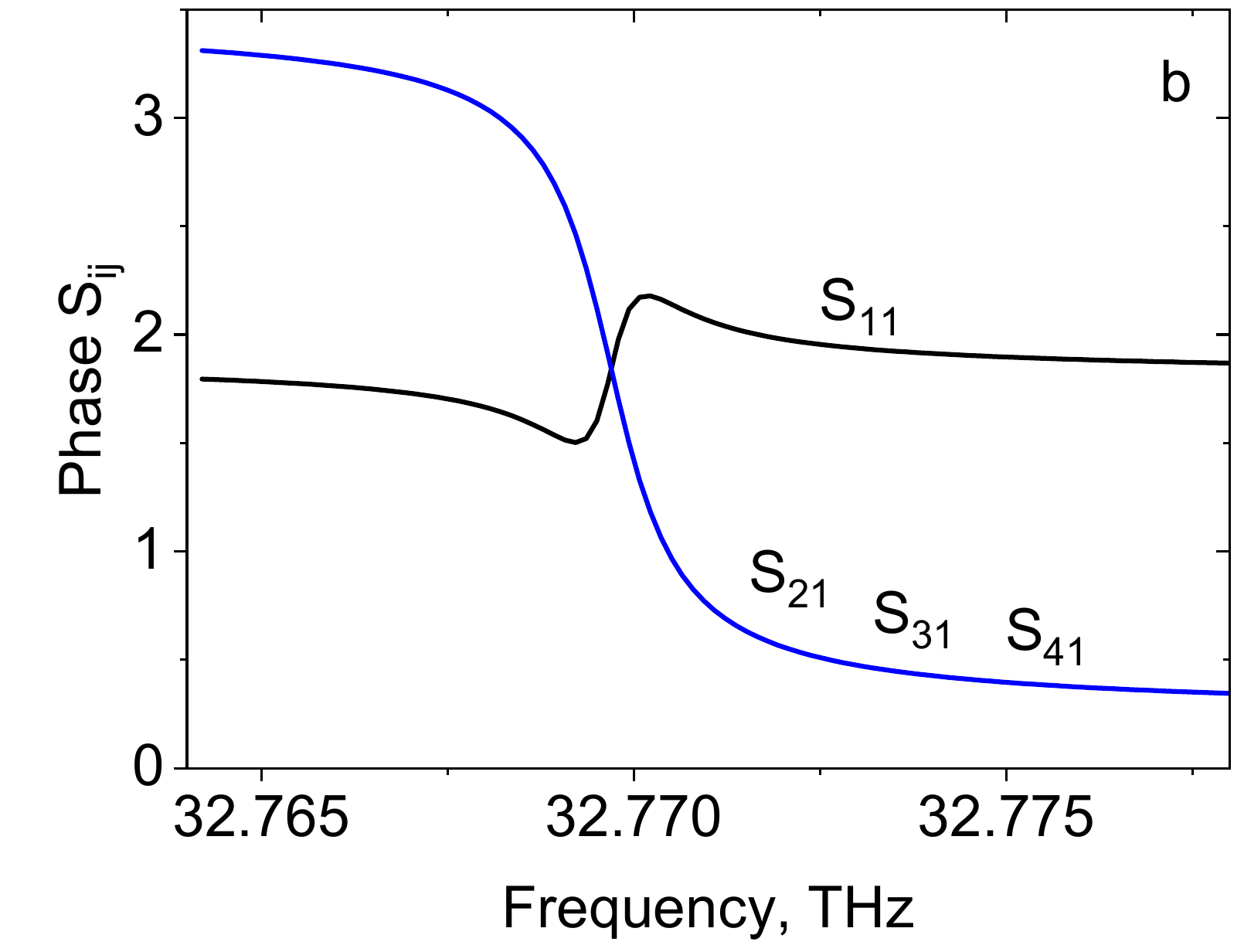} 
	
	\vspace{0.2cm} % Place vertical space *outside* the tabular
	
	\begin{tabular}{cc}
		\includegraphics[width=0.6\columnwidth]{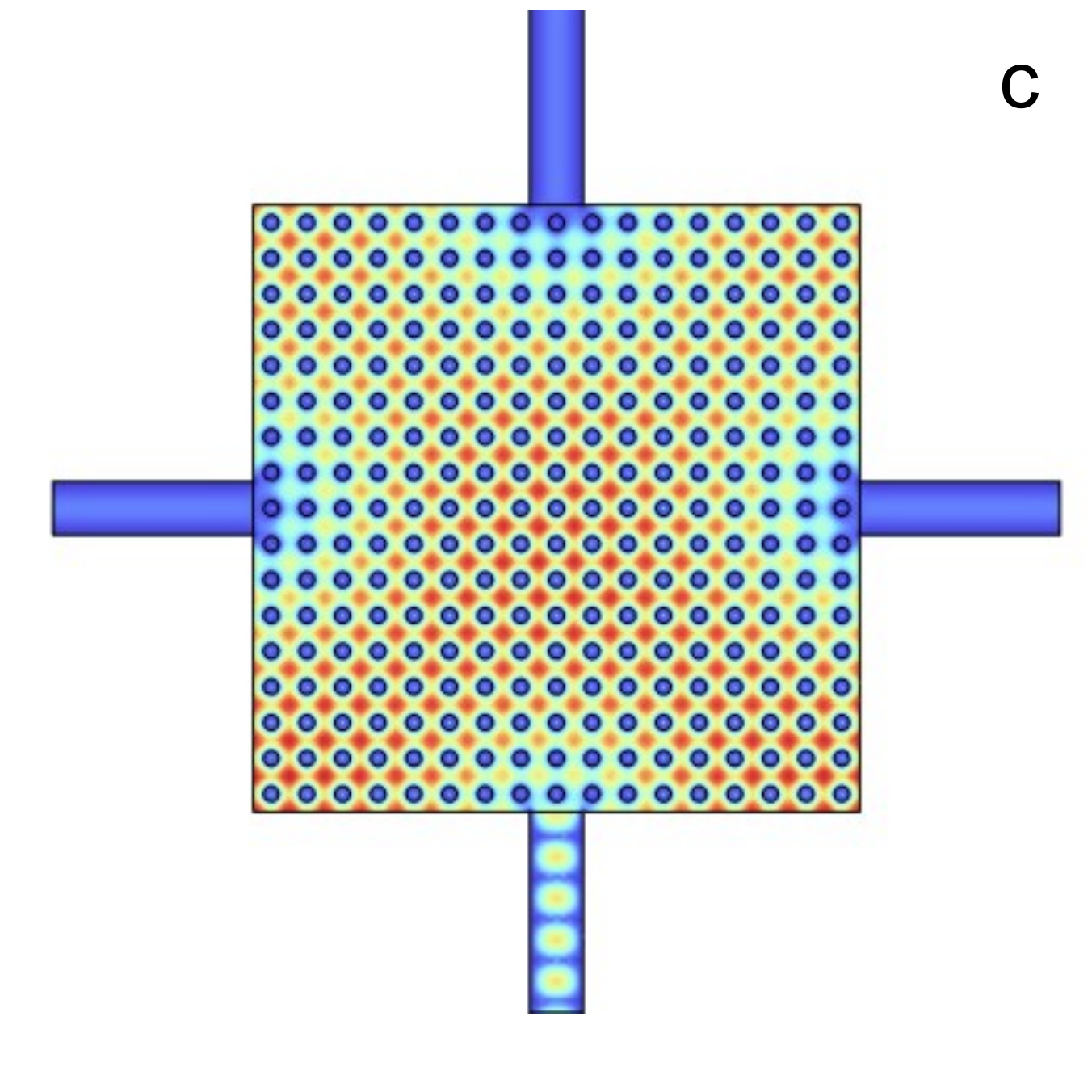} \\
	\end{tabular}
	\caption{The amplitude (a) and phase (b) frequency dependencies for the 2D PC photonic crystal with $17 \times 17$ unit cells. The magnitude and phase of S-parameters at input ($S_{11}$, bottom waveguide) and output ports (left, upper, and right waveguides: $S_{21}, S_{31}, S_{41}$). The electric field distributions are shown at the resonant frequency $f_{17\times 17}=32.77$~THz (this value is for $N=17$, see Figure ~\ref{fig:3} below) for the PMC border structure (c).}
	\label{fig:2}
\end{figure}

The finite PC structure, shown in Figure~\ref{fig:2} were designed to excite the electromagnetic oscillations by an input waveguide (the bottom waveguide) and to receive response by three output waveguides (left, upper, and right ones). Such a structure  has resonant NZERI properties at frequency $ f_{N,N}$ as demonstrated by the equality  of electromagnetic wave amplitude (Figure~\ref{fig:2}a) and  phase (Figure~\ref{fig:2}b) on the PC border at input waveguide channel and 3 different output ones (modules and phases of the reflection and transmission coefficients, respectively, $S_{11}=S_{21}=S_{31}=S_{41}$) at $f_{17,17}=32.77$~THz structure.
In the resonance area module and phase of the transmission coefficient values ($S_{21}=S_{31}=S_{41}$) coincide as well. 
Therefore, the amplitude and phase profiles at the interface between the PC boundaries and the output waveguide channels exhibit perfect coincidence at resonant frequency $f_{N,N}$. 
This demonstrates both amplitude and phase matching, providing clear evidence of the NZERI regime.

Figure~\ref{fig:2}c shows the resonant electric field distribution in the finite PC with PMC boundary conditions, respectively. As expected, the electric field distribution strength values are almost identical in a whole PC area excluding places at the PMC boundaries near the waveguides due to electric field inhomogeneous there.

%F:\MSCA4UA\Publications Eremenko\2024\paper1
%PC 17 17 Amplitude and phase.opju

%2D graphene plane with Si cirles PC 17 17 PEC freq dom  !!!!!!.mph
%D:\COMSOL PROJECTS\2D 3D graphene plane with Si cirles\all symmetric\work 2D

%2D graphene plane with Si cirles own arrays 4  PC 17 17 symm -2 ++ with PMC only PC - !!!.mph
%D:\COMSOL PROJECTS\2D 3D graphene plane with Si cirles\all symmetric

\begin{figure}[!tbh]
	\centering
	{\includegraphics[width=0.9\linewidth]{"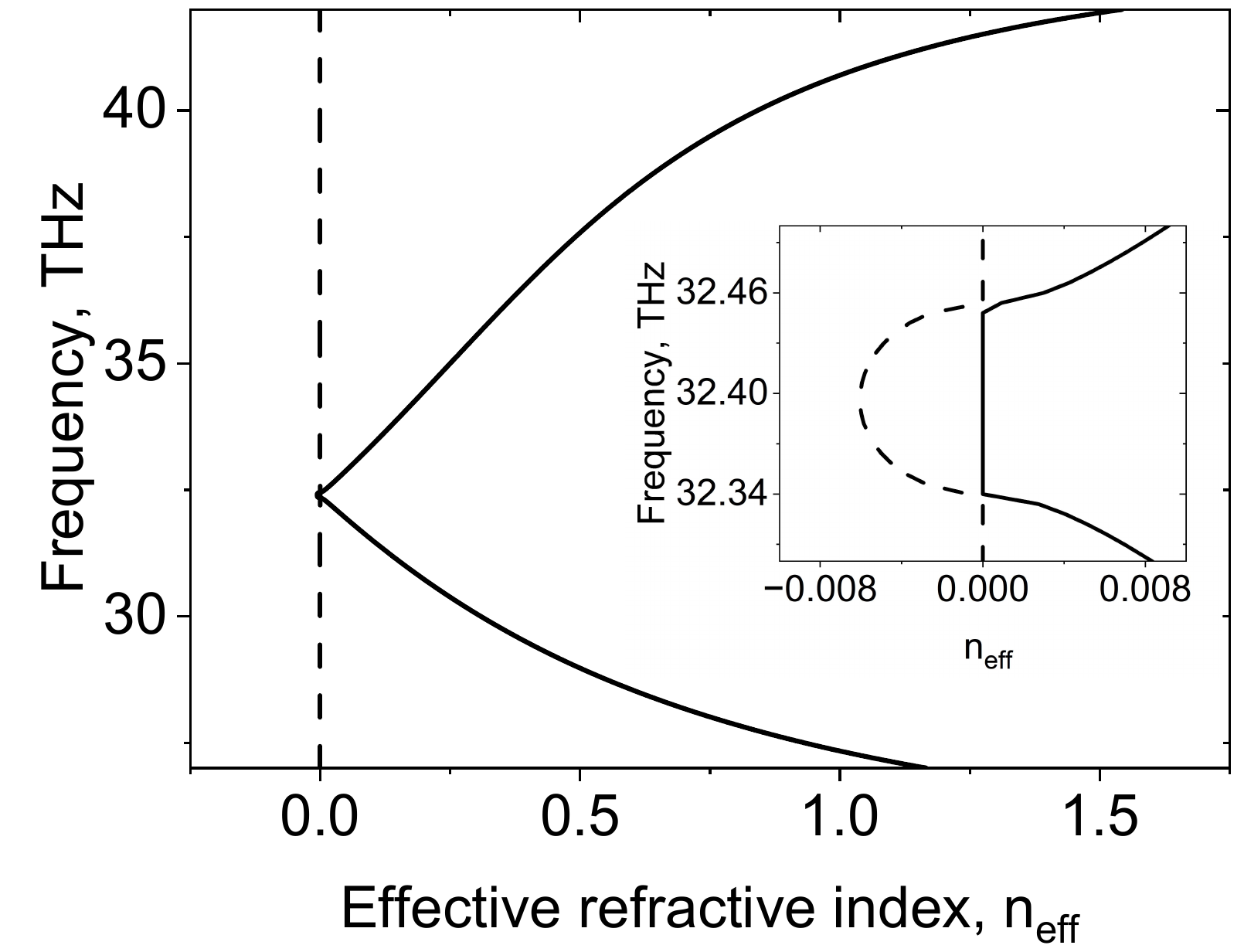"}}
	\caption{The frequency dependence of the complex effective refractive index $n_{eff}$ for a 2D PC with the unit cell shown in Figure~\ref{fig:1} for NZERI frequency area. The real (solid line) and imaginary (dash line) parts of $n_{eff}$. The insert shows a closer view of the area, where real part of $n_{eff}$ is near zero i.e., near $f_D$ frequency.}
	\label{fig:fig4}
\end{figure}

To obtain the effective refractive index $n_{eff}$ for the metasurface in the  NZERI regime frequency area, the effective permittivity $\epsilon_{eff}$ and permeability $\mu_{eff}$ were calculated at Dirac point frequency area using effective medium theory presented in Ref.~\cite{wu2006} with $r_0=2.87r_s$ as a fitting parameter in this theory, where $r_0$ and  $r_s=R$ are outer and inner radii of the unit cell structure elements, respectively.
The results for the 2D metasurface structure are shown in Figure~\ref{fig:fig4} for effective refractive index at NZERI regime frequency band $n_{eff} = \sqrt{\epsilon_{eff} \mu_{eff}}$. We determined that there exists a small frequency area near $f_D$ where the real part of $n_{eff}$ is almost zero and its imaginary part has rather high resonant negative values of order of $-6\times 10^{-3}$ in comparison with the neighboring regions, where its imaginary part much smaller (of order of $-10^{-7}$). 
The real part of $n_{eff}$ reaches air refractive index value equal 1 at $f_b=27.33$~THz and $f_t=40.72$~THz Thus, the NZERI frequency area is between $f_b$ and $f_t$ that can be used for the studied metasurface structure in our case.

% F:\MSCA4UA\Publications Eremenko\2024\paper1
% Ref.24 n=====.opju

\begin{figure}[!tbh]
	\centering
	\includegraphics[width=0.9\columnwidth]{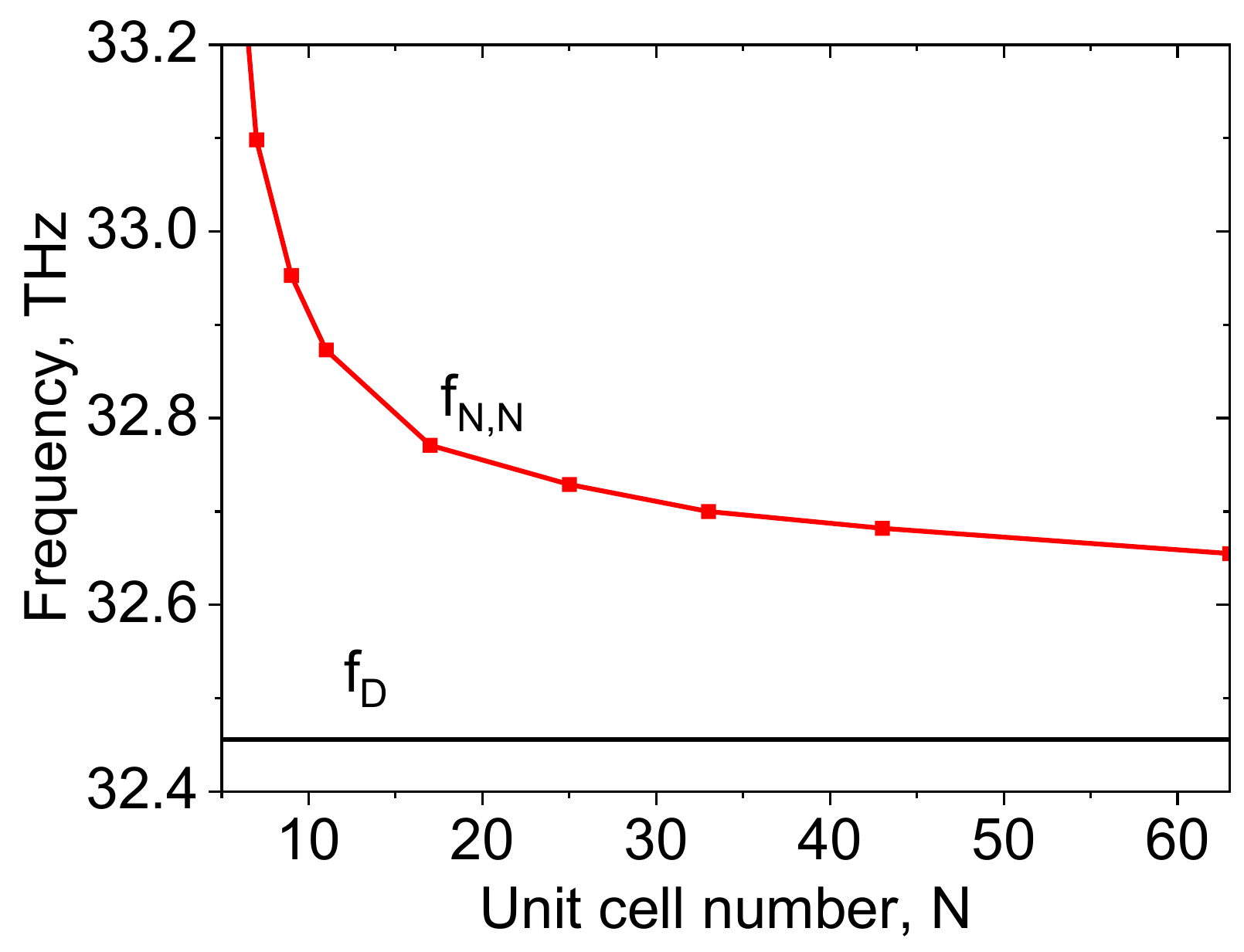}
	\caption{Dependence of the phase-matching frequency $f_{N,N}$ (red line) on the number $N\times N$ of unit cells in the square finite-size PC shown in Figure~\ref{fig:2},  the black line indicates the corresponding frequency $f_D $ obtained from the band structure (Figure~\ref{fig:1}) for the periodic PC.}
	\label{fig:3}
\end{figure}

Figure~\ref{fig:3} shows the dependence of the NZERI resonance frequency $ f_{N,N}$ on the size of the finite 2D PC ($N\times N$ unit cells). As $N$ tends to infinity the resonance frequency of the finite structure approaches $f_D$ of the infinite PC determined above.

We conclude that  2D PC effective refractive index $n_{eff}$ close to zero at the frequency $ f_D $ area. The numerical simulations below demonstrate that a 3D PC composed of dielectric rods arranged in a periodic lattice, exhibits behavior consistent with that of a 2D metasurface configuration, provided that the rod height exceeds six wavelengths (see  Section 3.2). The effective refractive index of a 2D metasurface unit cell model is practically identical to that of a 3D model with sufficiently long rods.
Thus, we examined the PC phase-matching frequency range where $n_{eff} < 1$   (the NZERI regime), which should extend SPP propagation length on graphene layer with such a PC substrate (see Section 3.3).

%2D 3D PC on unit cell numbers.opju
%F:\MSCA4UA\Publications Eremenko\2024\paper1
%3D metasur with rod fresnel_equations  rod height=1lam-eigenfreq alpha - varTM 11.04.24.mph
%D:\COMSOL PROJECTS\3D matesurface from Fresnel equation project\11.04.24

\subsection{3D metasurface structure in NZERI regime}
To determine the frequency dependence of the  effective refractive index 
$n_{eff}$ in the NZERI regime for 3D metasurface, we employed an eigenfrequency solver (to find first triple-degenerate eigenfrequencies in the frequency band) and frequency-domain solver (to calculate S-parameters near resonance conditions) simulations for 3D metasurface unit cells containing silicon rods or spheres. To obtain $n_{eff}$ in the NZERI regime for 3D metasurface we used the independent theoretical frameworks presented in Refs.~\cite{wu2006,martin2019}.
The difference in these two independent  approaches is as follows. 
In Ref. \cite{wu2006} authors presented an effective medium theory for  the electrodynamic properties study in a periodic composite for a specific configuration (a rod or sphere -based unit cell) with complicated Mie resonances. 
In Ref. \cite{martin2019} authors introduced a method for retrieving surface-wave dispersion and a synthesis approach suitable for bianisotropic metasurfaces placed in both symmetric and asymmetric surrounding media. For details see Appendix 2.

The authors in Ref.~\cite{wu2006} used Mie scattering theory and the authors in Ref.~\cite{martin2019} used S-parameters values near resonance conditions for the metasurface unit cell to determine $n_{eff}$.
As known, S-parameters (scattering parameters) are a set of measurable quantities used to characterize the behavior of linear electrical networks, especially at high frequencies (e.g., microwave circuits). They describe how signals propagate and interact within a multi-port network by relating incident and reflected waves. 

\begin{figure*}[!tbh]
	\centering
	\vspace{0.2cm} % Small vertical space between images
	\begin{tabular}{ccc}
		\includegraphics[width=0.28\columnwidth]{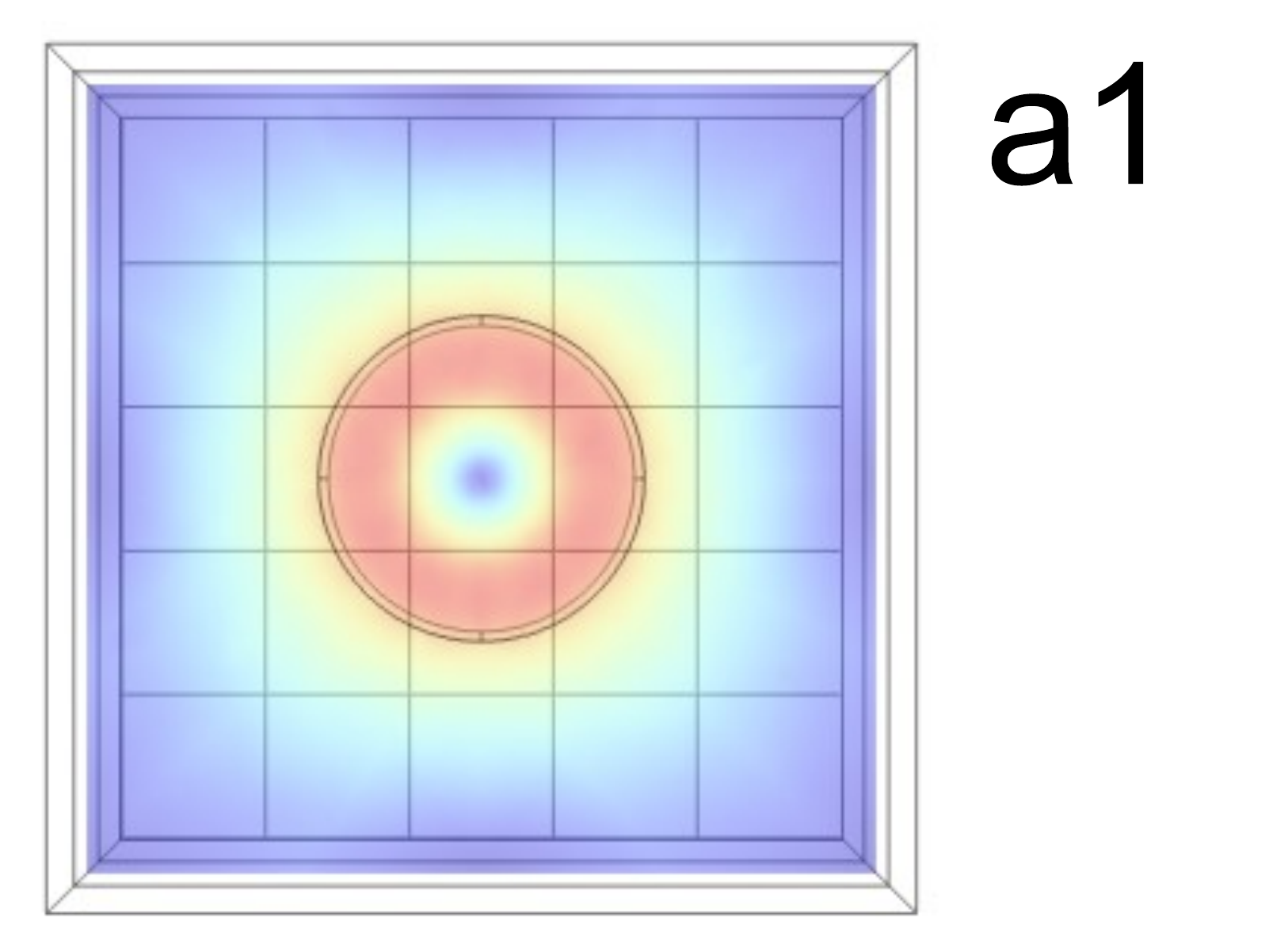} &
		\includegraphics[width=0.28\columnwidth]{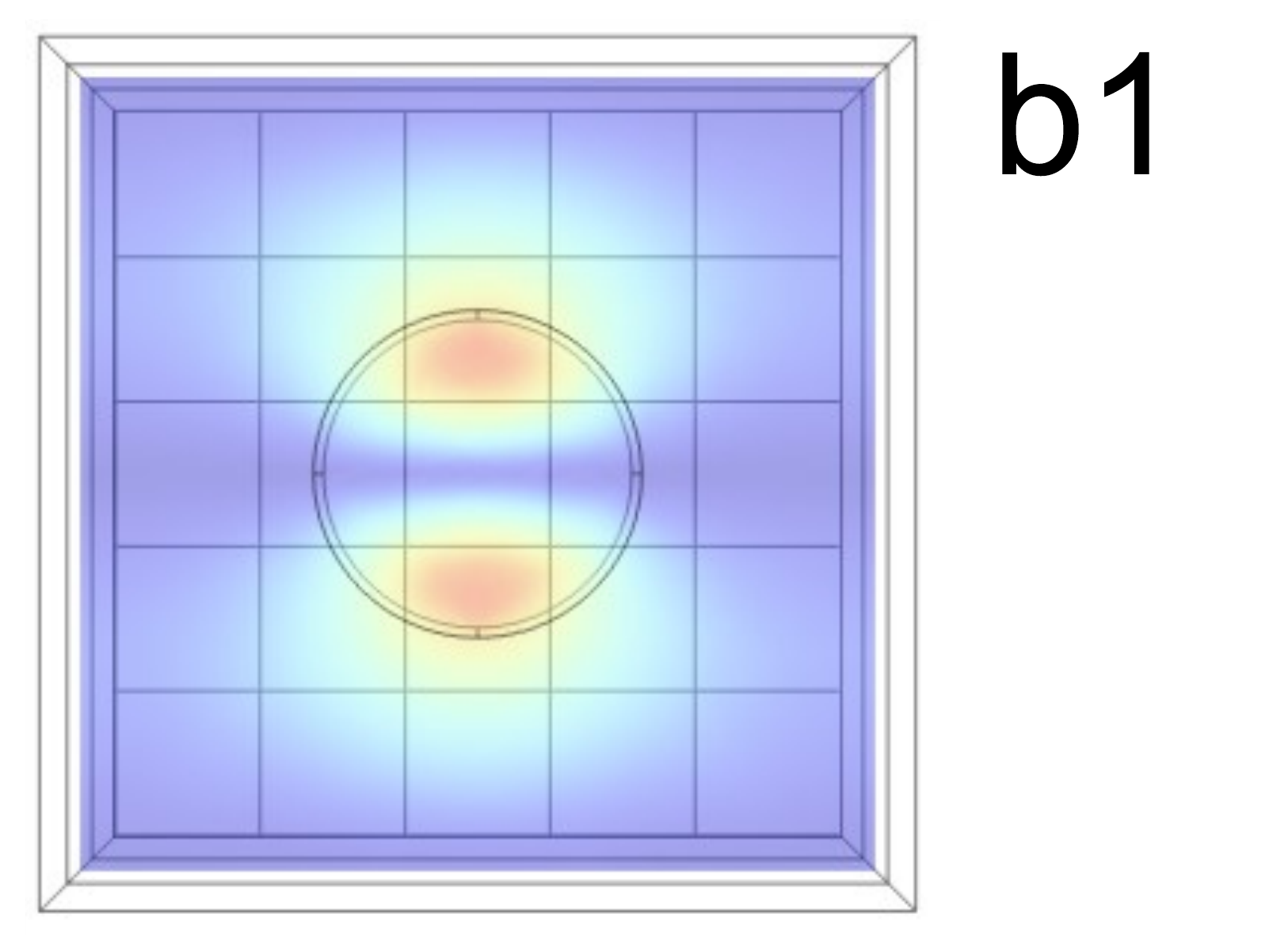} & 
		\includegraphics[width=0.28\columnwidth]{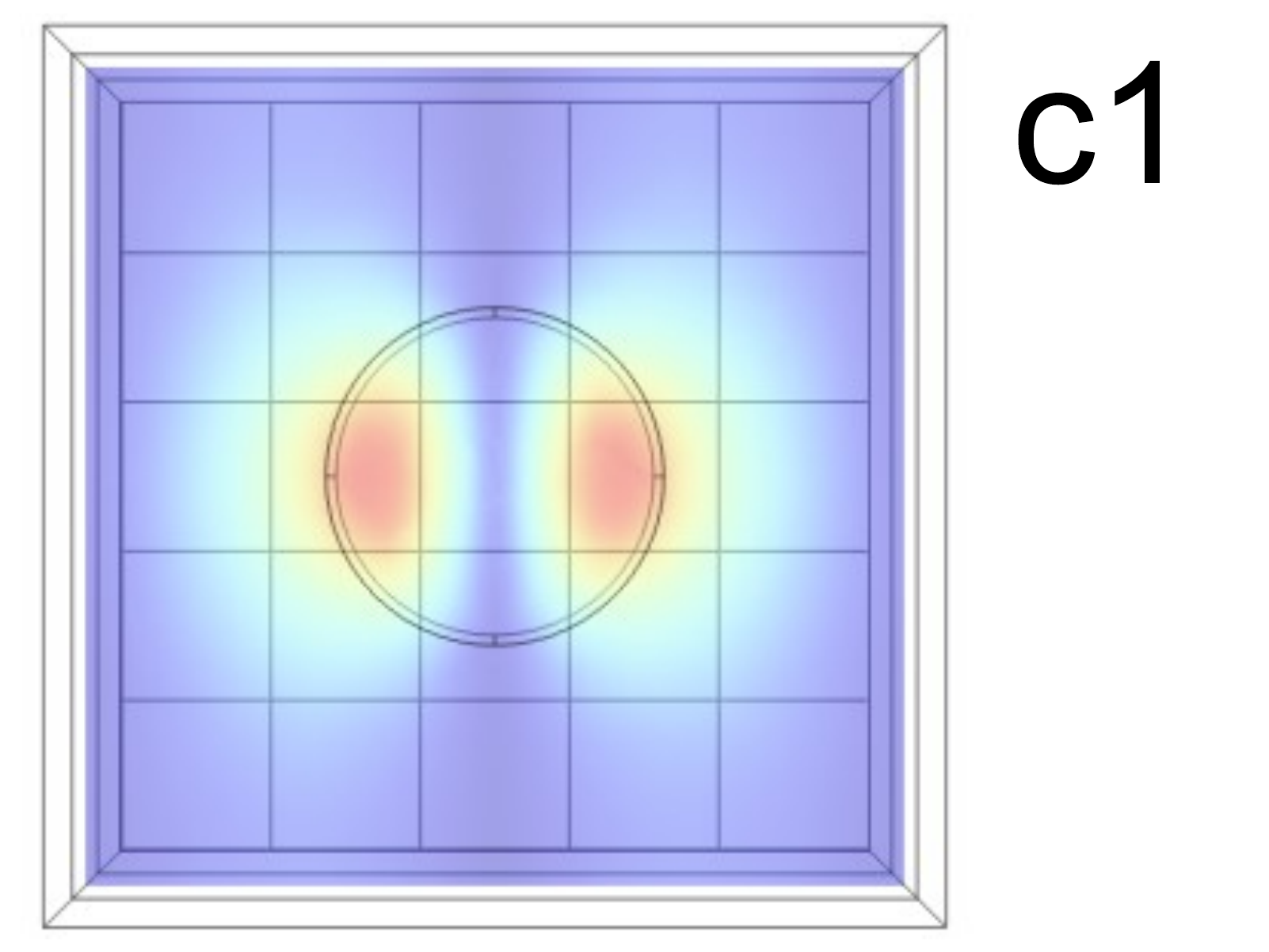} \\
		%(a) First plot & (b) Second plot \\
	\end{tabular}
	\begin{tabular}{ccc}
	\includegraphics[width=0.28\columnwidth]{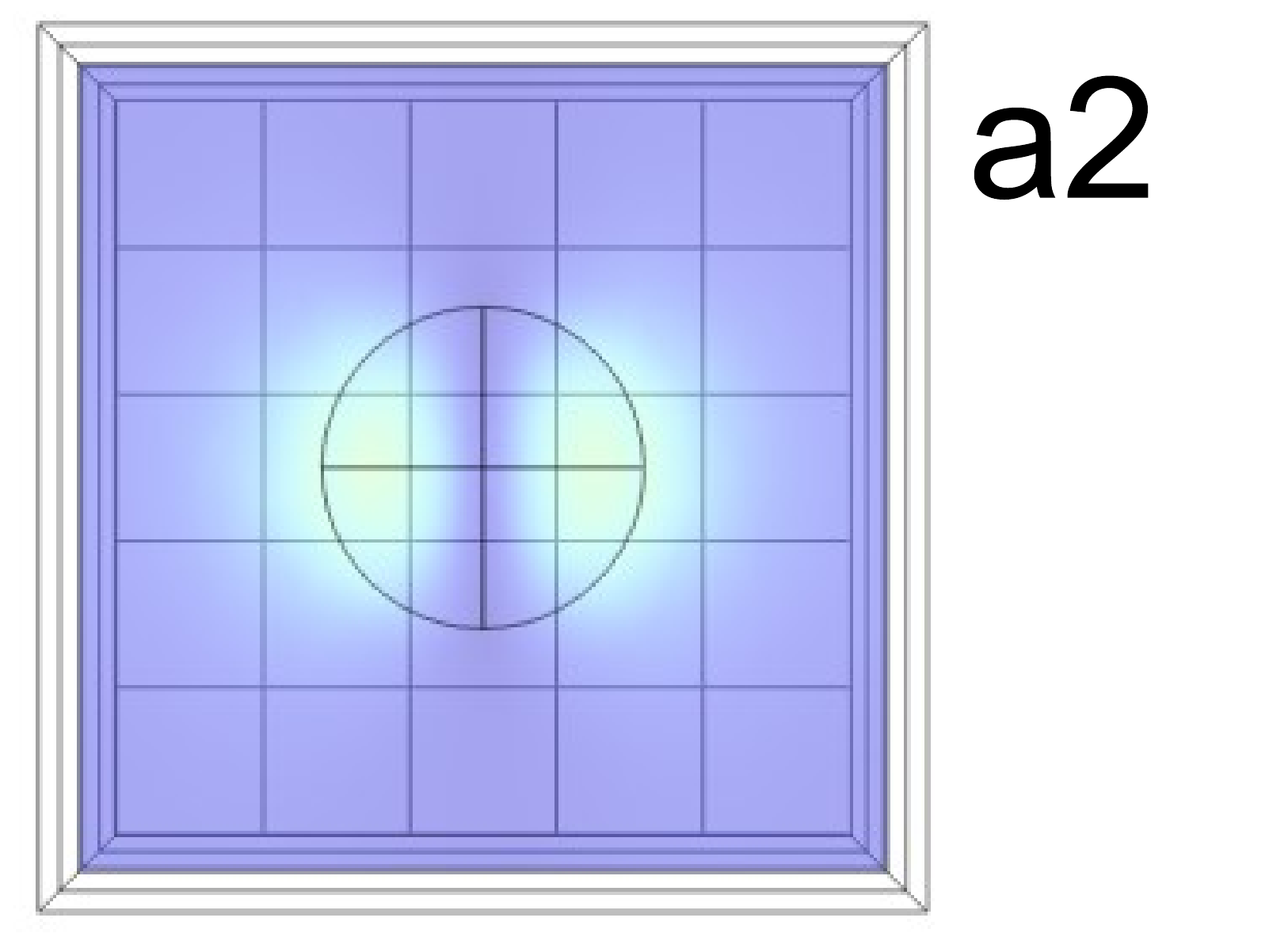} &
	\includegraphics[width=0.28\columnwidth]{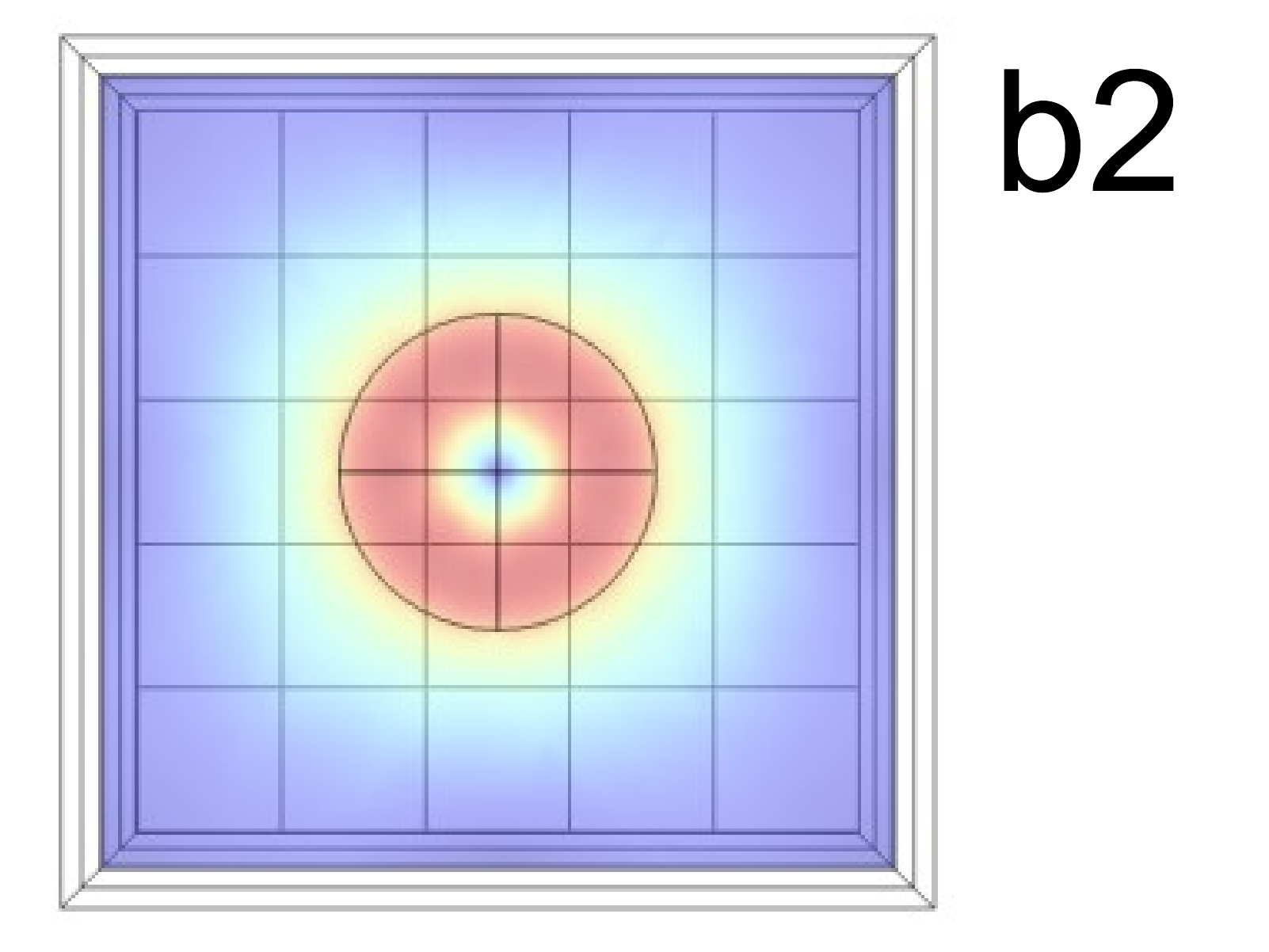} & 
	\includegraphics[width=0.28\columnwidth]{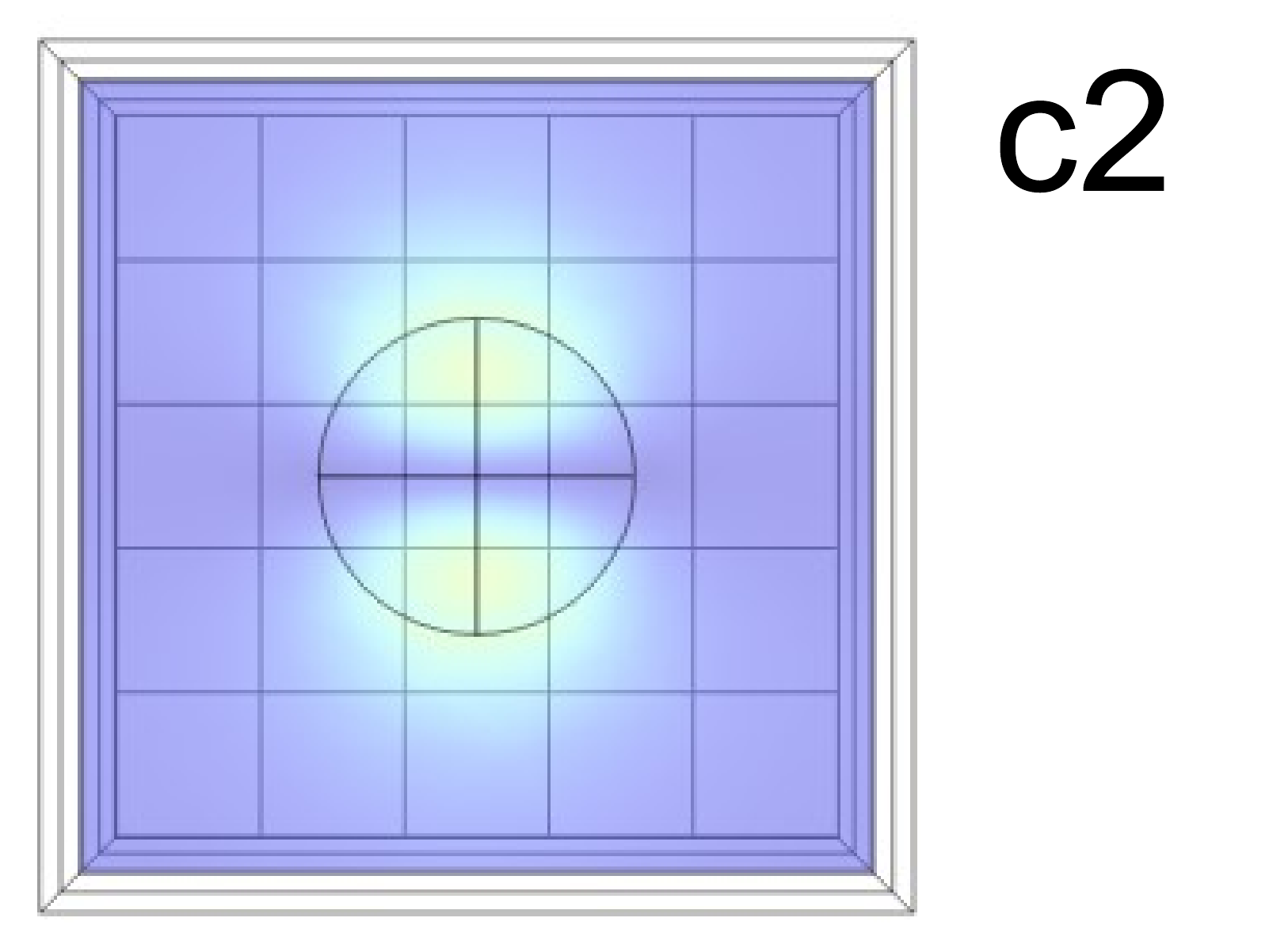} \\
	%(a) First plot & (b) Second plot \\
\end{tabular}
	\begin{tabular}{ccc}
	\includegraphics[width=0.28\columnwidth]{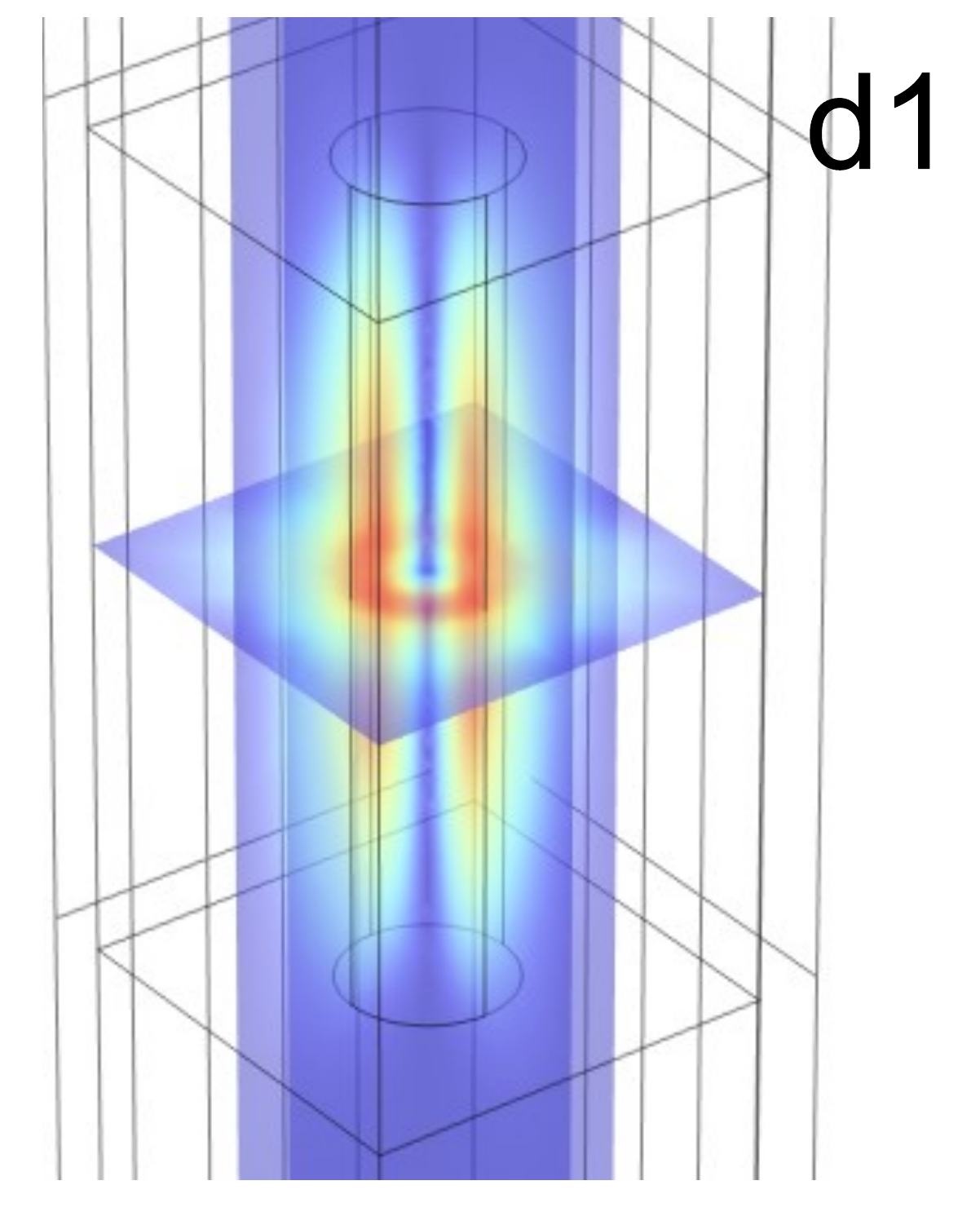} &
	\includegraphics[width=0.28\columnwidth]{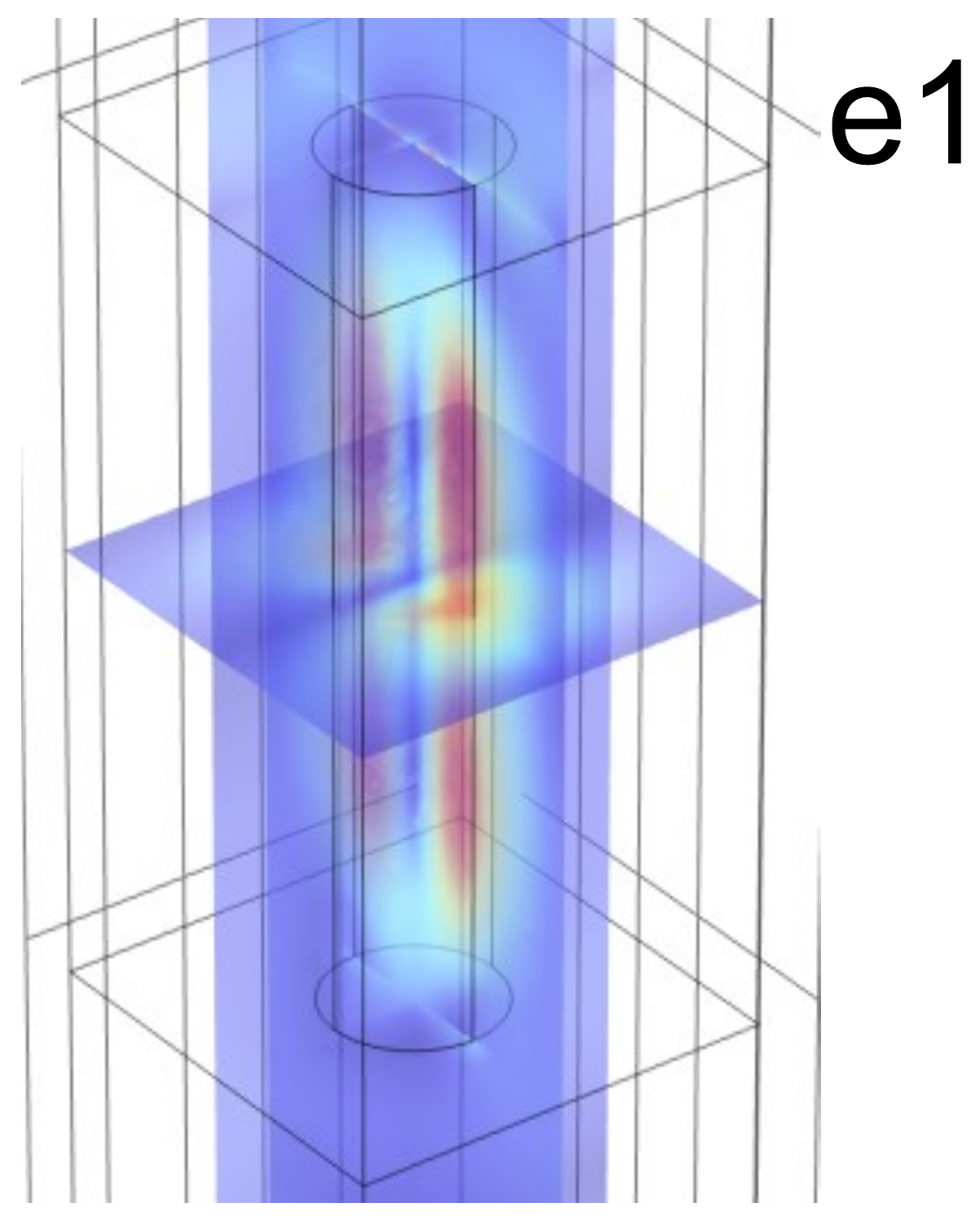} & 
	\includegraphics[width=0.28\columnwidth]{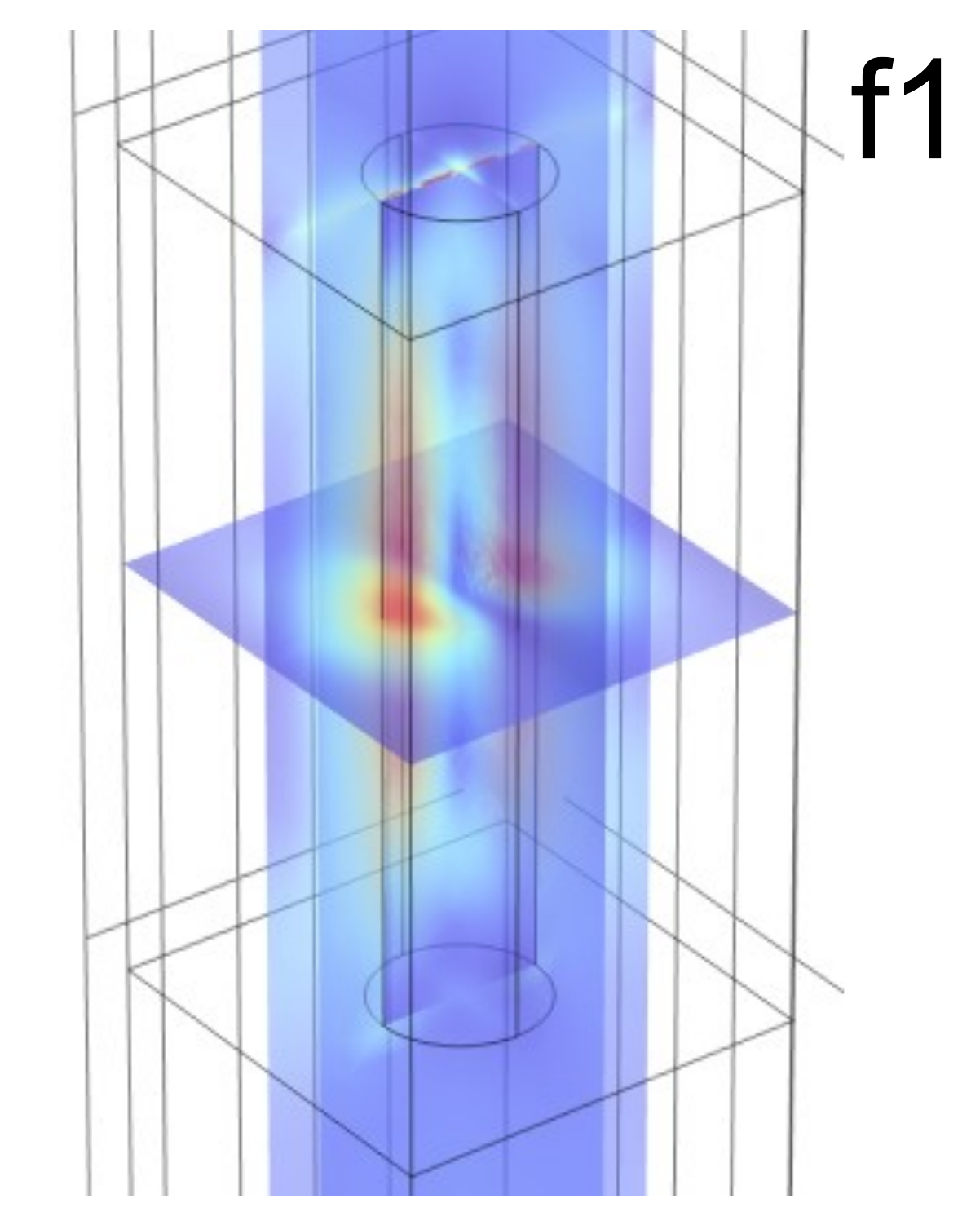} \\
		%(a) First plot & (b) Second plot \\
	\end{tabular}
	\begin{tabular}{ccc}
\includegraphics[width=0.28\columnwidth]{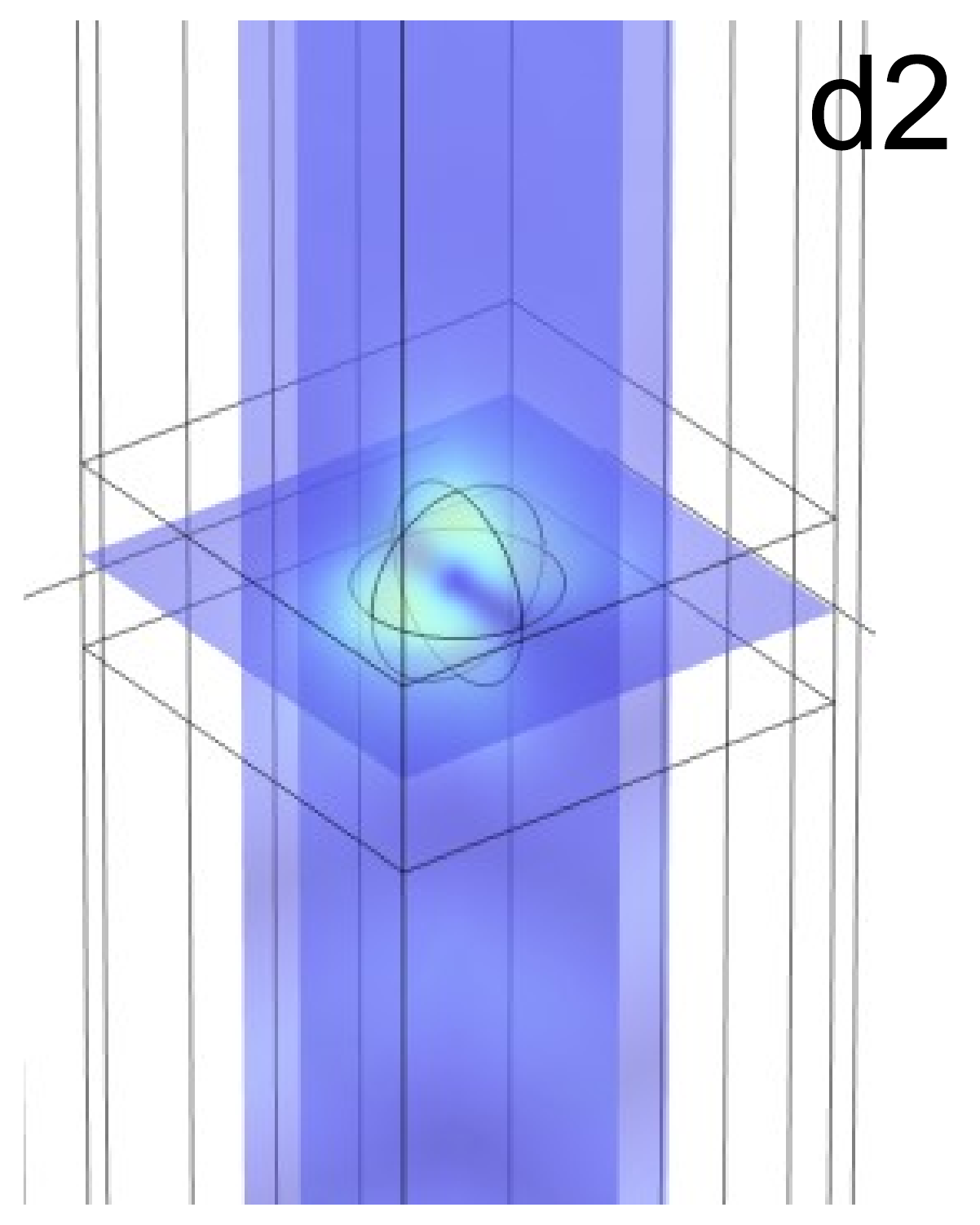} &
\includegraphics[width=0.28\columnwidth]{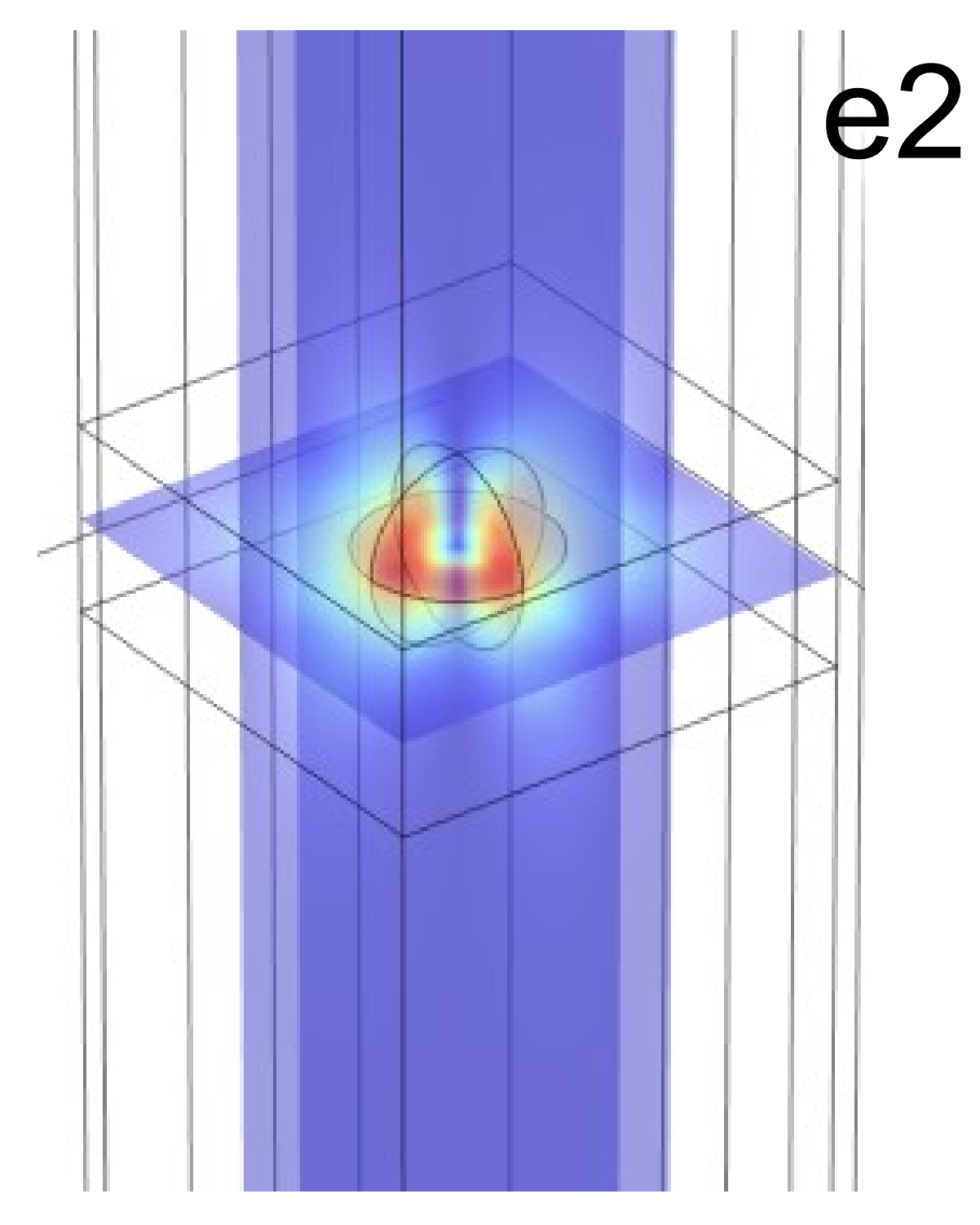} & 
\includegraphics[width=0.28\columnwidth]{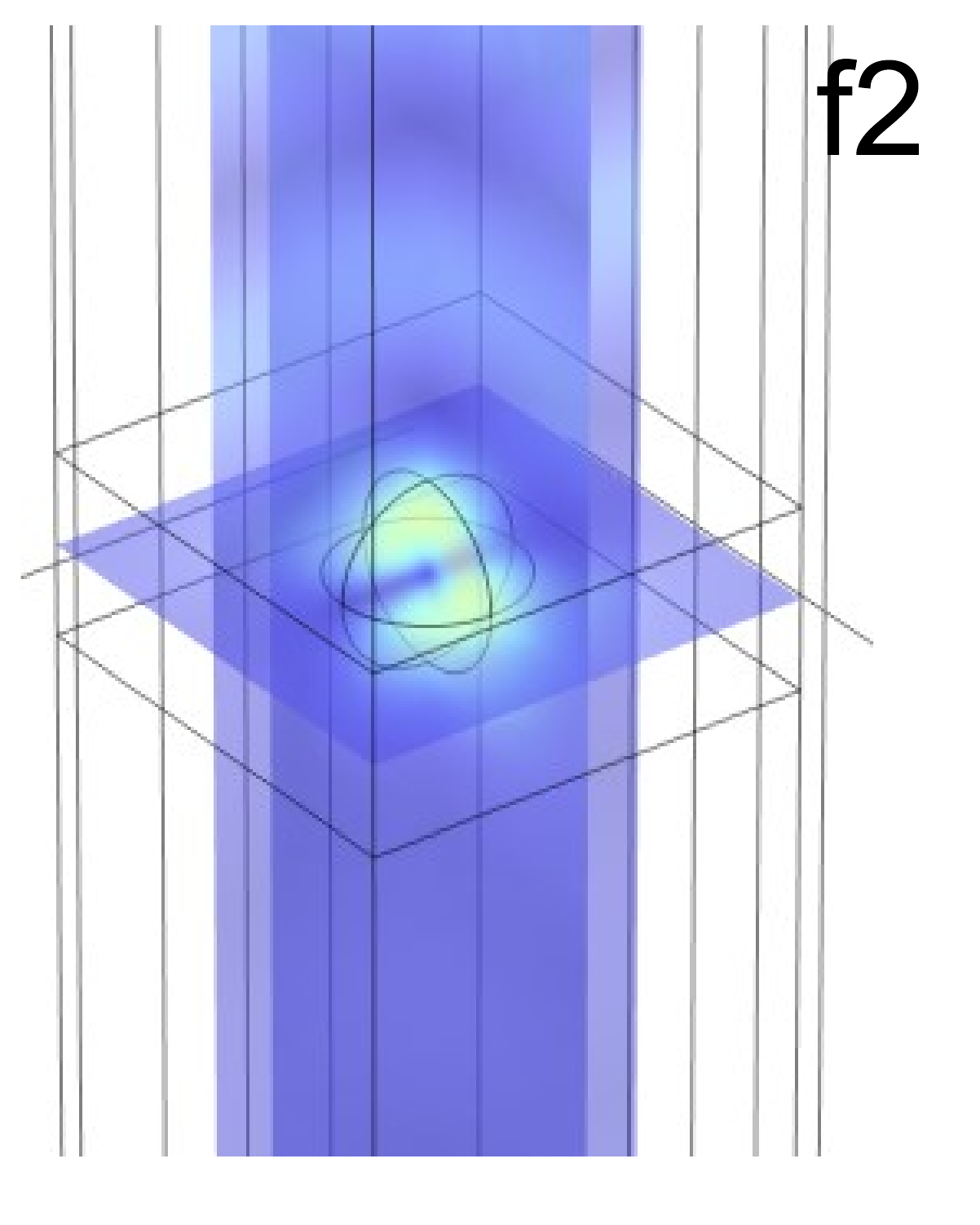} \\
		%(a) First plot & (b) Second plot \\
	\end{tabular}
\caption{The electric field distribution for
	three first resonant eigenfrequencies of the all-dielectric metasurface unit cell with the silicon rod of height of $1\times\lambda$, where $\lambda = 10 \mu m$ and with the sphere of radius $R$ and the unit cell size $a$. The rod unit cell for the monopole (32.916 THz) and the dipole modes (33.048 THz, 33.205 THz) are in $xy$-plane (a1 - c1) and in 3D view (d1 - f1), respectively. The color scale is linear from $4.6\times 10^4$ V/m (blue) to $2.7\times 10^7$ V/m (red) for a1 - f1. The sphere unit cell for the monopole (40.91 THz) and the dipole modes (40.389 Thz, 41.934 Thz) are in $xy$-plane (a2 - c2) and in 3D view (d2 - f2), respectively. The color scale is linear from $2.3 \times 10^3$ V/m (blue) to $7.3 \times 10^6$ V/m (red) for a2 - f2. }
\label{fig:3D_PC_field_rod_sphere}
\end{figure*}

\begin{figure}[!tbh]
	\centering
	\includegraphics[width=0.9\columnwidth]{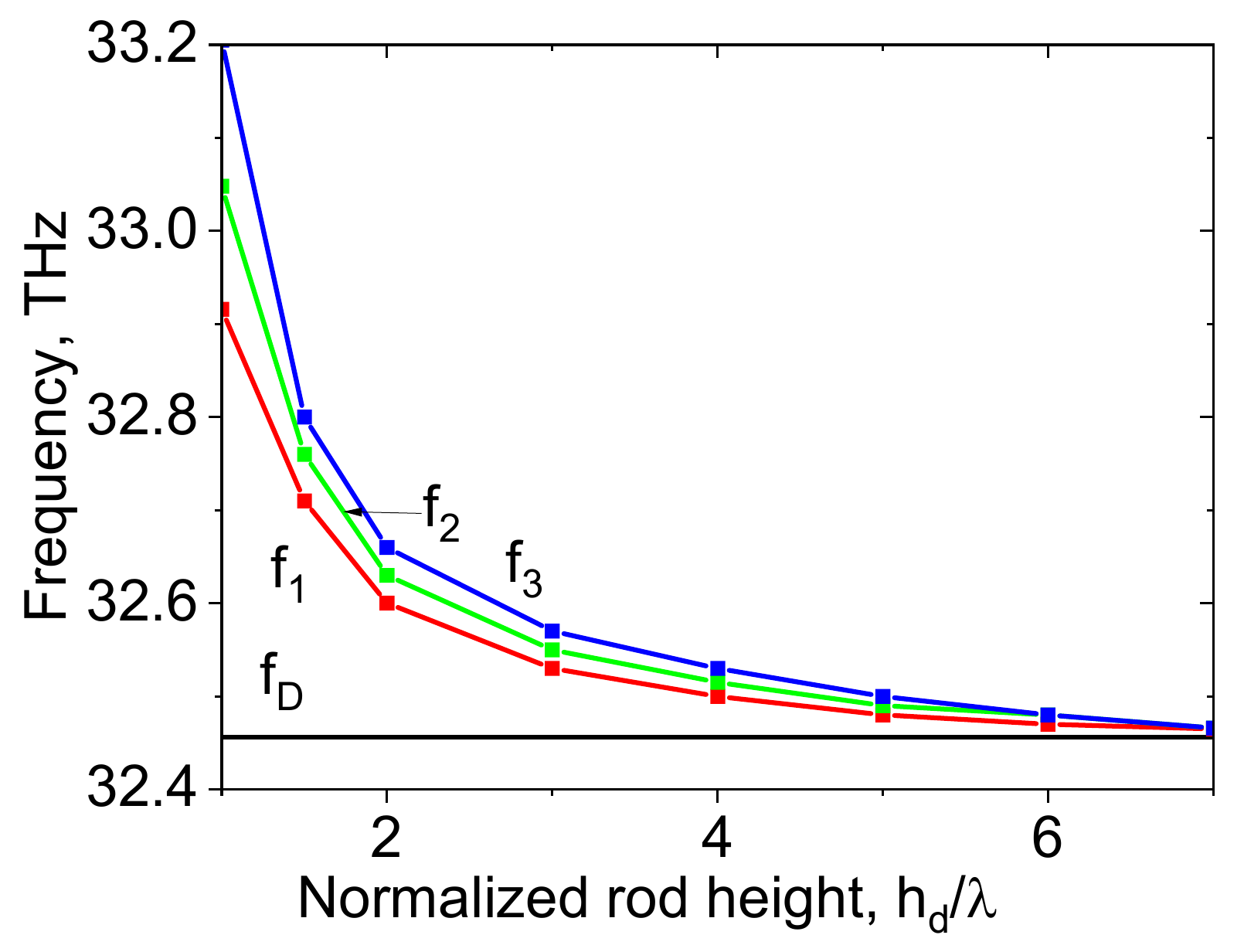} 
	\vspace{0.2cm} % Small vertical space between images
	\caption{
		The dependence of first resonant eigenfrequencies ($f_1, f_2, f_3$) (see Figure~\ref{fig:3D_PC_field_rod_sphere}) of 3D metasurface unit cell with the silicon rod inside on the normalized rod height $h_d/\lambda$,
		$f_D$ indicates the corresponding frequency obtained from the band structure in Figure~\ref{fig:1} for the periodic 2D PC. }
	\label{fig:3D_PC_on_rod_height}
\end{figure}

Figure~\ref{fig:3D_PC_field_rod_sphere} presents the electric field distributions for first almost triple-degenerate eigenfrequencies for the 3D metasurface unit cell with the rod and sphere, respectively. 
Both for the rod and sphere in the metasurface unit cell  the electric field distributions are similar to 2D PC in Figure~\ref{fig:1}). The differences arise from the dipole orientations of the electric field and the frequency response of monopole/dipole modes in 2D/3D rod or sphere unit cells. However, these variations do not alter the fundamental physical interpretation of the studied phenomenon.
It should be noted that the eigenfrequencies of the 3D unit cell with the rod and 2D unit cell coincide when the rod height is sufficiently large (exceeding 6$\lambda$) (see Figure~\ref{fig:3D_PC_on_rod_height}), approaching convergence as the rod height tends to infinity. However, the eigenfrequencies of 3D unit cell with the sphere exhibit slightly different values due to variations in dielectric filling within the unit cell.

%2D 3D PC on unit cell numbers.opju
%F:\MSCA4UA\Publications Eremenko\2024\paper1

%rod
%3D metasur with rod fresnel_equations  rod height=1lam-eigenfreq alpha - varTM 11.04.24 !!!!!.mph
%D:\COMSOL PROJECTS\3D matesurface from Fresnel equation project\11.04.24

%sphere
%D:\COMSOL PROJECTS\3D matesurface from Fresnel equation project\Spheres Array
%3D fresnel_equations  TM No gr spheres array eigenfreq.mph

%2D 3D PC on unit cell numbers.opju
%F:\MSCA4UA\Publications Eremenko\2024\paper1
%sphere eigenfrequencies

%2D 3D PC on unit cell numbers.opju
%F:\MSCA4UA\Publications Eremenko\2024\paper1
%3D metasur with rod fresnel_equations  rod height=1lam-eigenfreq alpha - varTM 11.04.24.mph
%D:\COMSOL PROJECTS\3D matesurface from Fresnel equation project\11.04.24

\begin{figure*}[!htb]
	\centering
	\includegraphics[width=0.9\columnwidth]{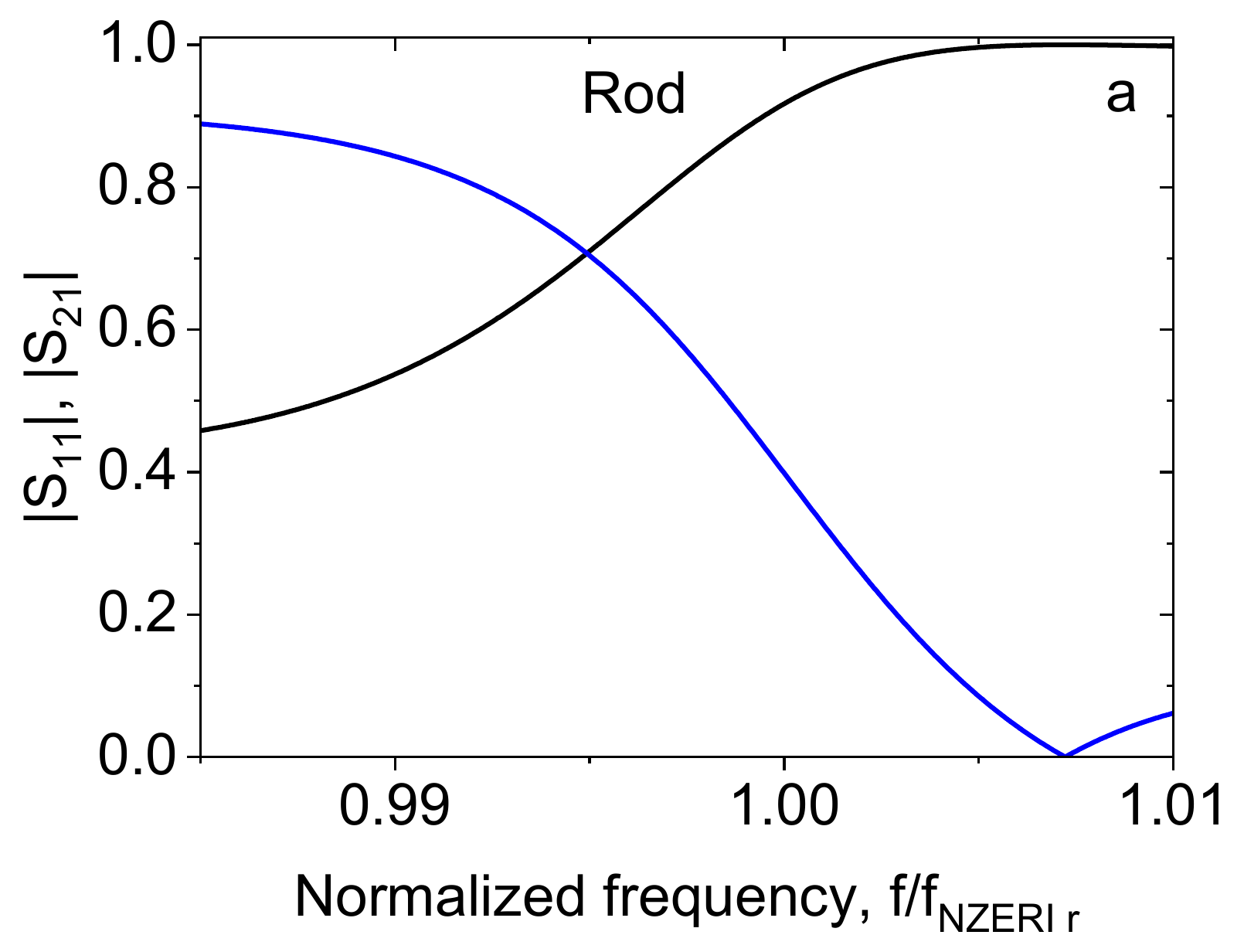}
	\includegraphics[width=0.9\columnwidth]{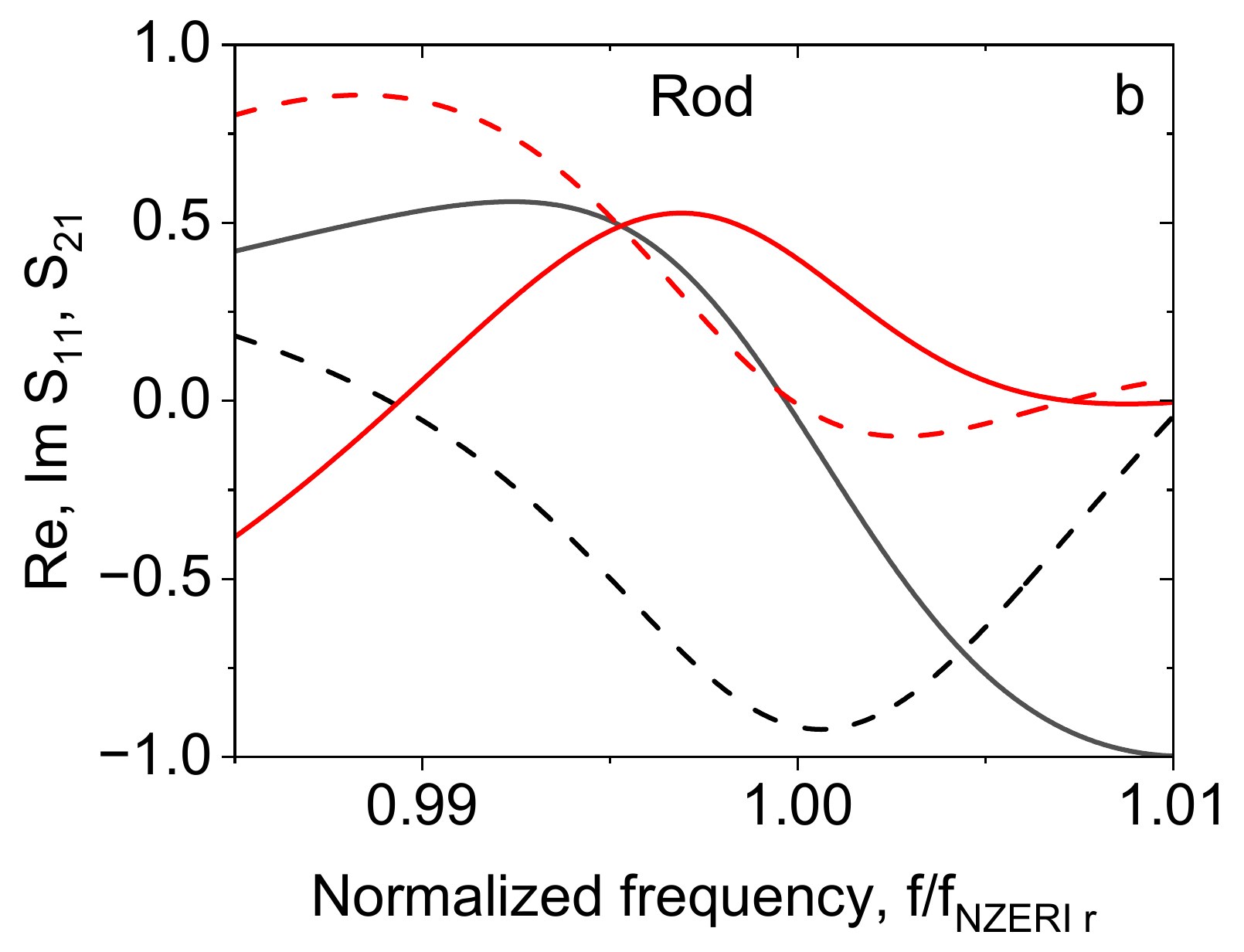}
	\includegraphics[width=0.9\columnwidth]{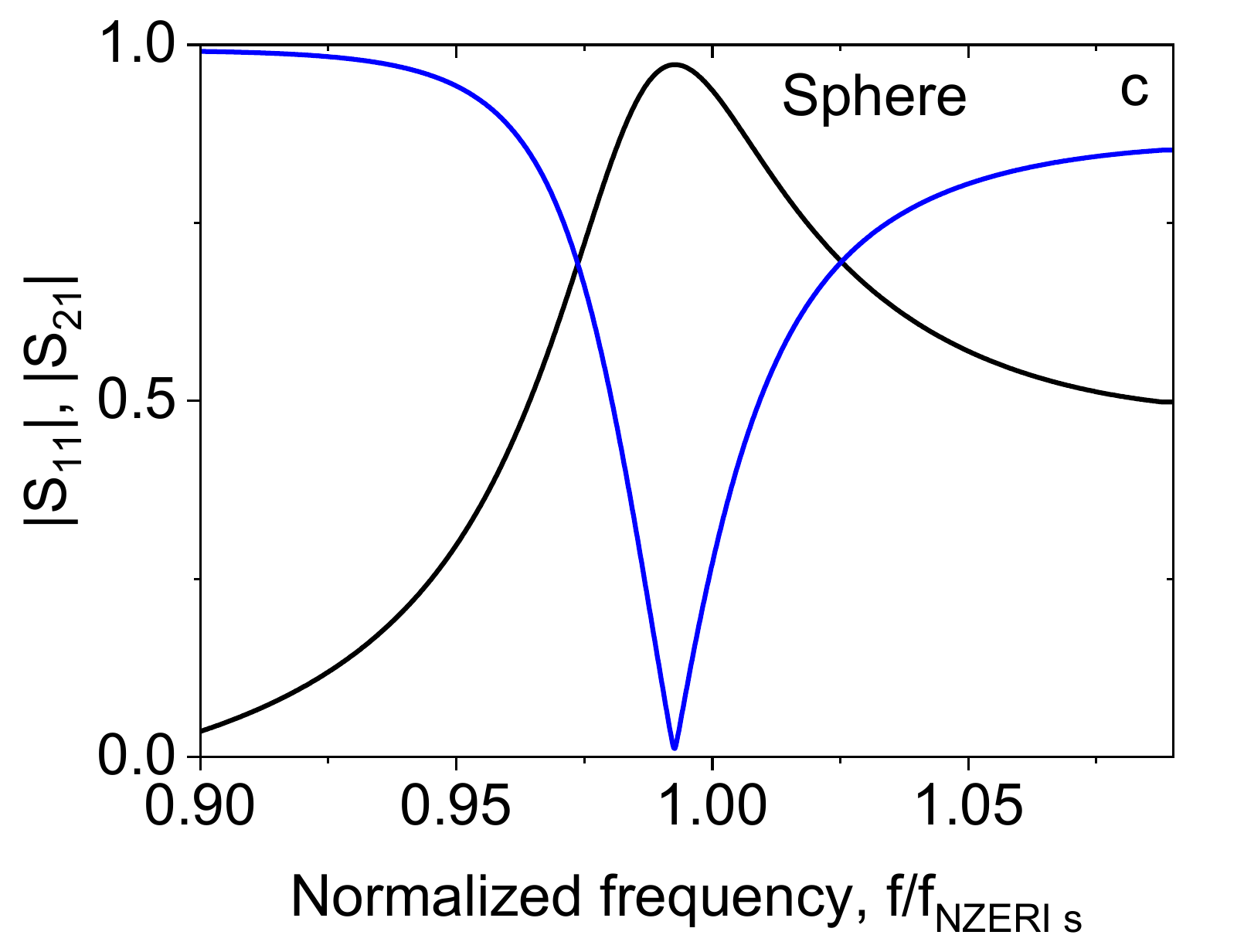}
	\includegraphics[width=0.9\columnwidth]{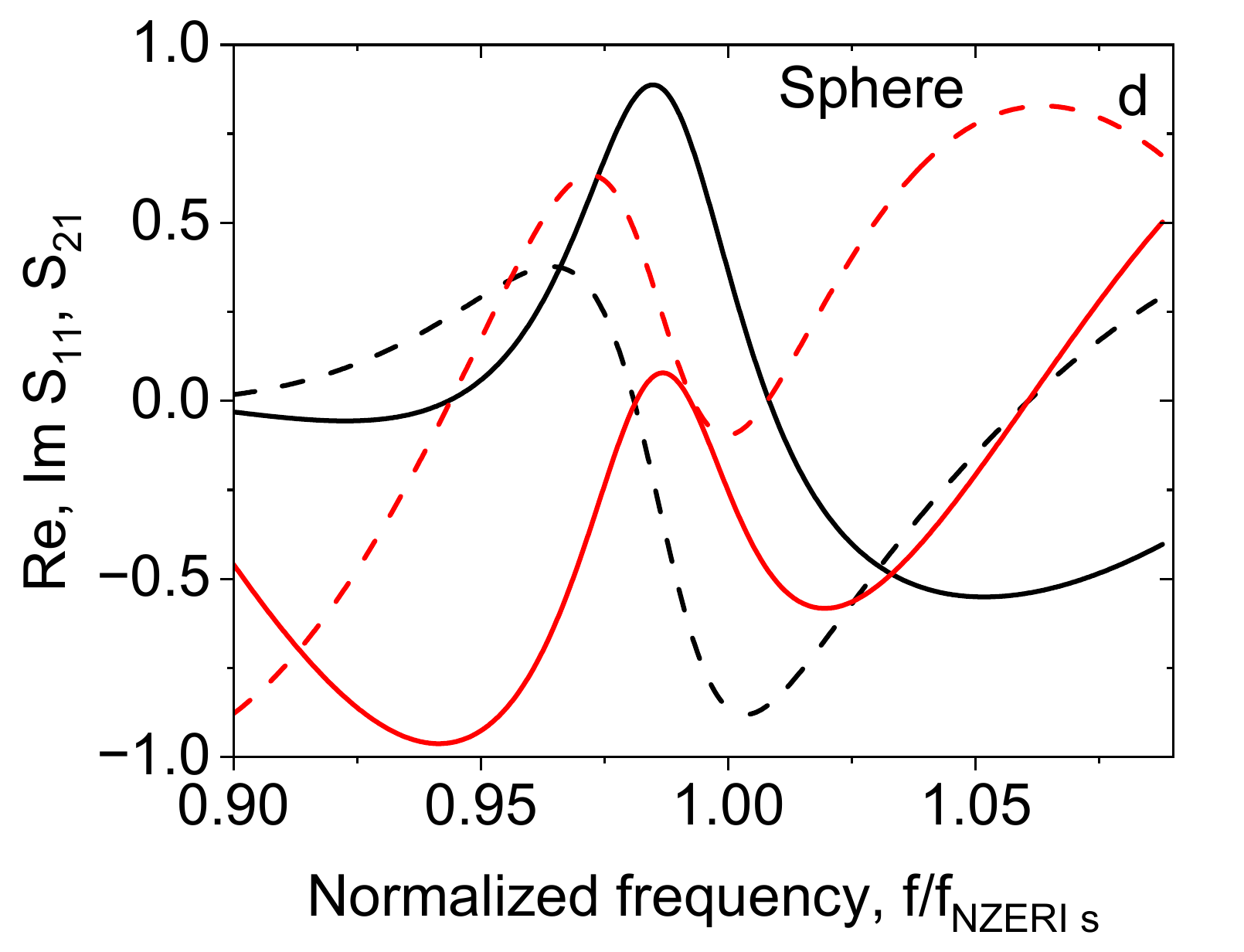} 
\caption{The frequency resonant dependence of $|S_{11}|$ (in black), $|S_{21}|$ (in blue) for the rod (a) and sphere (c) unit cells, respectively. 
Their real (solid lines) and imaginary (dash lines) parts for the rod (b) and sphere (d) unit cell, the real and imaginary parts of $S_{11}$  and $S_{21}$ are in black and red, respectively. The frequency scale normalized on  $f_{NZERI r} = 32.992$~THz for the rod unit cell and $f_{NZERI s} = 41.836$~THz the sphere unit cell.}
	\label{fig:fig_S11_S21}
\end{figure*}

\begin{figure}[!htb]
	\centering
	\centerline
	{\includegraphics[width=0.9\linewidth]{"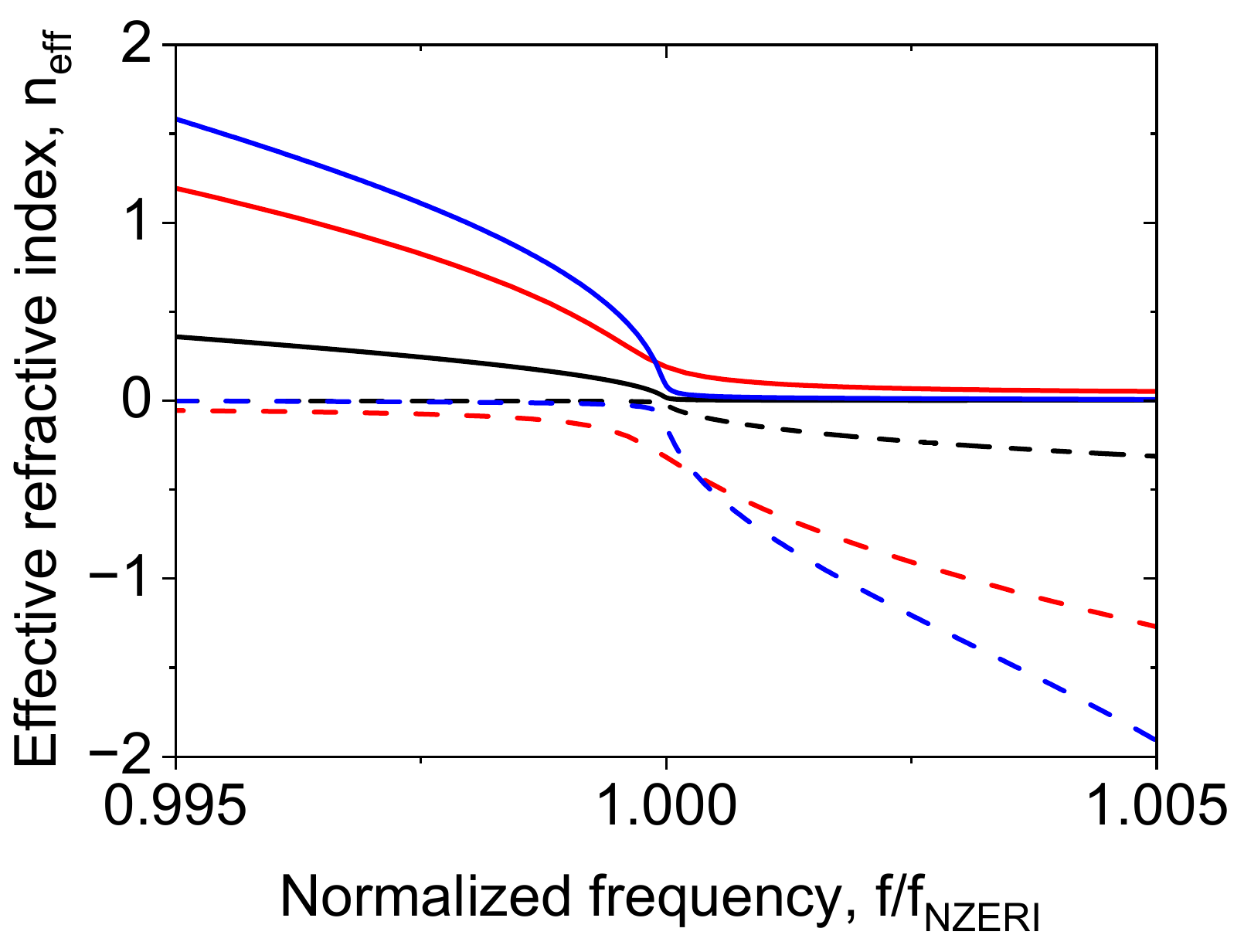"}}
	\caption{The frequency dependence of the real (solid lines) and imaginary (dashed lines) part of the refractive index $n_{eff}$  in the NZERI regime. The black lines  are for 3D metasurface  with sphere in the unit cell using the theory in Ref.~\cite{wu2006}, the red ones  are for the same metasurface structure using the theory in Ref.~\cite{martin2019} ($f_{NZERI s}$) and the blue lines are for 3D metasurface with rod in the unit cell ($f_{NZERI r}$) using the theory in Ref.~\cite{martin2019}. The solid lines are for real parts of $n_{eff}$, the dash lines are for imaginary parts. }
	\label{fig:fig_comp_Ref.24_Martin}
\end{figure}

Figure~\ref{fig:fig_S11_S21} presents the S-parameter analysis for the rod and sphere unit cells revealing characteristic values within the resonant frequency band that correspond to NZERI regime. 
Frequencies  $f_{NZERI r}$ corresponds to the intersections of the real $S_{11}$ and real and imaginary $S_{21}$ parts;  $f_{NZERI s}$ corresponds to the intersections of the real and imaginary $S_{21}$ parts as well and they present the similar frequency for the different metasurface structure configurations in NZERI regime (see Figure~\ref{fig:fig_comp_Ref.24_Martin}).

These asymmetric resonances in all-dieletric metasurfaces, can be referred as Fano resonances~\cite{Limonov2017}. Fano resonances have high Q-factor modes due to low optical losses in dielectrics, strong near-field enhancement with minimal absorption, simultaneous excitation of electric and magnetic dipoles, enables tailored light-matter interaction for SPP coupling. Dielectric metasurfaces with Fano resonances concentrate light at subwavelength scales, enhancing graphene SPP excitation efficiency\cite{Xia2017,Guan2022}.
   
By applying the theoretical frameworks from Refs.~\cite{wu2006,martin2019} to analyze 3D metasurfaces containing both spherical and rod-shaped unit cell elements, we calculated the effective refractive index $n_{eff}$  for the studied 2D and 3D structures and compared the results.
As shown in Figure~\ref{fig:fig_comp_Ref.24_Martin}, both the real and imaginary components of $n_{eff}$ exhibit remarkably similar behavior within the NZERI frequency range. Notably, the lines for rod- and sphere-based unit cells demonstrate close agreement, which we attribute to their common theoretical foundation in Ref.~\cite{martin2019}.

%3D Ref,24 and Martin comparison.opju
%rod
%3D metasur with rod fresnel_equations  rod height=1lam-FREQ DOM alpha - varTM 11.04.24 32-33.5 THz !!!!.mph
%D:\COMSOL PROJECTS\3D matesurface from Fresnel equation project\11.04.24
% rod

%3D Ref,24 and Martin comparison.opju

\subsection{Graphene layer with substrates in NZERI regime}

\begin{figure}[!tbh]
	\centering
		\includegraphics[width=0.9\columnwidth]{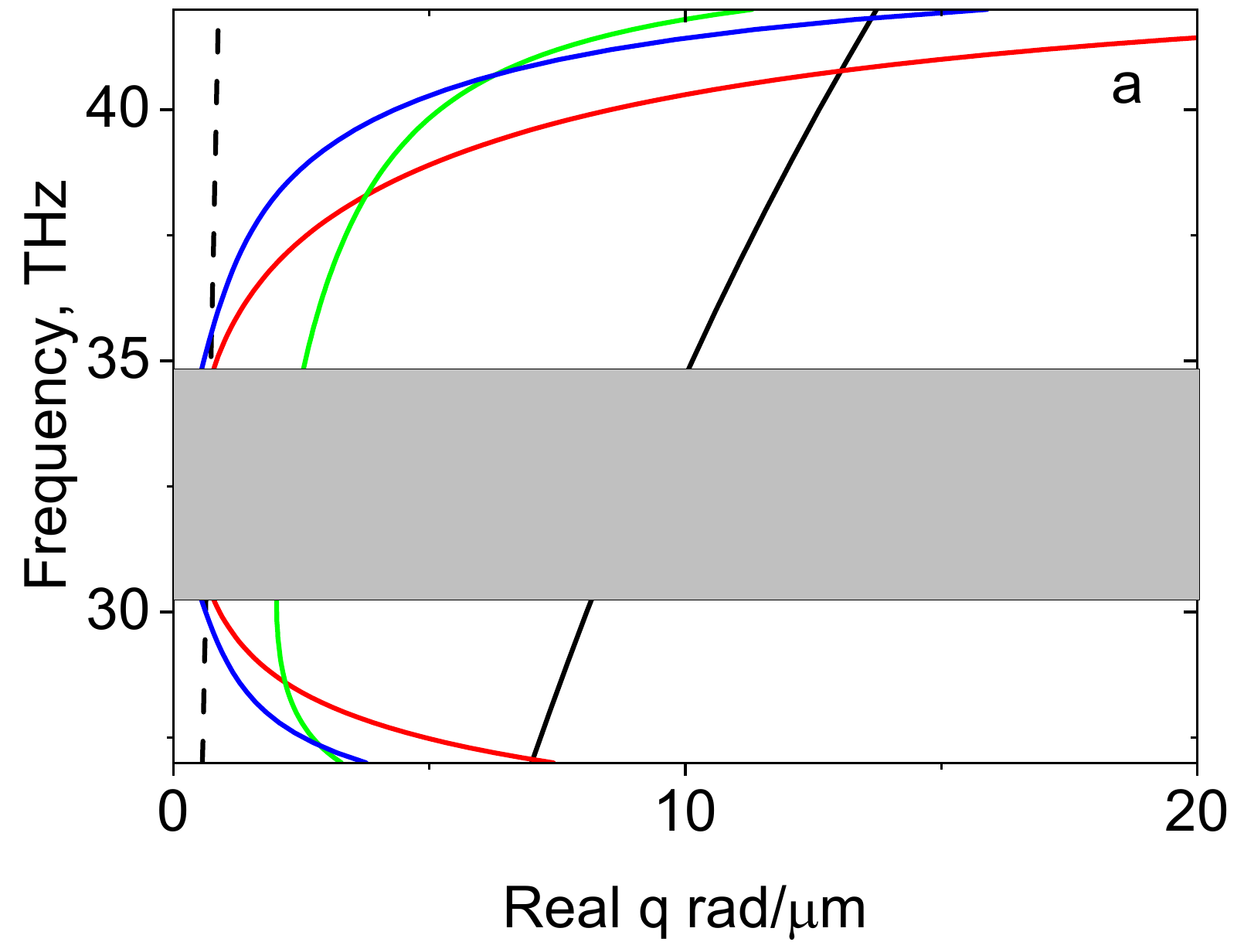}
		\includegraphics[width=0.9\columnwidth]{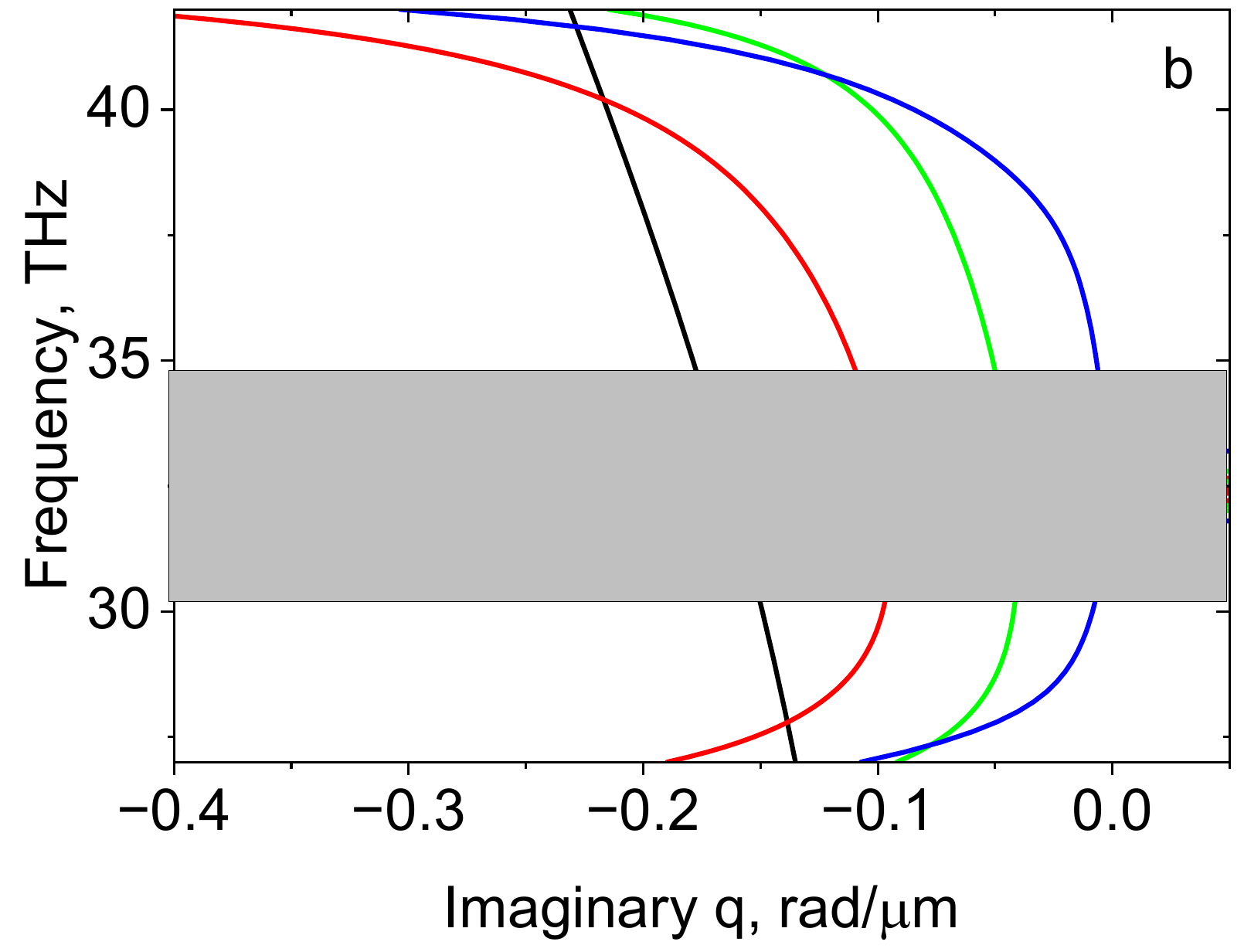} 
\caption{The frequency resonant dependence of real (a) and imaginary (b)  SPP propagation for the graphene layer. The black dashed line is the light line.
	Black solid lines are for the structure (see Figure~\ref{fig:fig7}): 
	PECBC/SBC - air - graphene - air - PECBC/SBC; 
	red: SBC - air - graphene - MS - PECBC; 
	green: PECBC - air - graphene - MS - PECBC; 
	blue: PECBC - MS - graphene - MS - PECBC;
%	magenta: SBC - air - graphene - MS - SBC. 
The red lines calculated by Eqs.~\ref{eq:grdy2} and coinside with magenta ones in (b) and in Figure~\ref{fig:fig9}, accordingly.
Grey shaded areas indicate the regime where computational results are unreliable due to machine precision. }
\label{fig:fig8}
\end{figure}

Notably, certain difficulties exist when performing computer simulation in COMSOL in the frequency range where the medium's NZERI properties are manifested in the structure under consideration. The grey area (Figure~\ref{fig:fig8}) has $n_{eff}$ values of order  $10^{-7}$ and where COMSOL solutions become unreliable due to machine precision, and 
points some numerical instability in the region where the metasurface's effective refractive index is small. These computational difficulties can be explained as follows.
Maxwell's equations in the frequency representation for the vector potential of the electromagnetic field $\vect{A}$ inside the metamaterial can be written in the following form Ref.~\cite{monk2003}
\begin{equation}
	\label{MaxwellA}
	\nabla\times\left(\mu_{\mathrm{eff}}^{-1}\nabla\times\vect{A}\right)-\varepsilon_{\mathrm{eff}}k_0^2\vect{A}=\vect{j}_i,
\end{equation}
where $\vect{j}_i$ is the impressed current density.

Generally, in dielectric metasurface, the impressed current density is relatively small, and both terms in the left-hand side of the Eq.~(\ref{MaxwellA}) are of the same order.
However, in the NZERI region, when $\varepsilon_{\mathrm{eff}}\to 0$ or $\mu_{\mathrm{eff}}\to 0$ (or both $\varepsilon_{\mathrm{eff}}$ and $\mu_{\mathrm{eff}}$ are small)
the first term in left-hand side of the Eq.~(\ref{MaxwellA}) dominates. It leads to the appearance of numerical instability in the COMSOL simulation algorithm. 
While sweeping the frequency to get the dispersion curve, some irregular spikes may appear at the frequency in the NZERI region. 
When $n_{eff} > 0.01$ simulation becomes regular.

%F:\MSCA4UA\Publications Eremenko\2024\paper1\tau=2!e-13
% tau 2E-13 beta Lan SPP, L SPP 2mikm.opju

%graphene spp air-air Ef=0.8 eV gr layer -0.345nm air-air  27-42 Thz!!!!! L3mikm.mph black
%D:\COMSOL PROJECTS\2D graphene substrate\Ef´=0.8eV

%2D graphene spp TWO substrate - neff  -2D Ref.24 ONE sub ONE PEC.mph red
%2D graphene spp TWO substrate - neff  -2D Ref.24 TWO sub No PEC.mph magenta
%2D graphene spp TWO substrate - neff  -2D Ref.24 ONE sub TWO PEC.mph green
%2D graphene spp TWO substrate - neff  -2D Ref.24 TWO sub TWO PEC.mph blue
%D:\COMSOL PROJECTS\2D graphene substrate\Ef=0.8eV with substrate - neff\Comparison trans bound conditions

\begin{figure}[!tbh]
	\centering
	{\includegraphics[width=0.9\linewidth]{"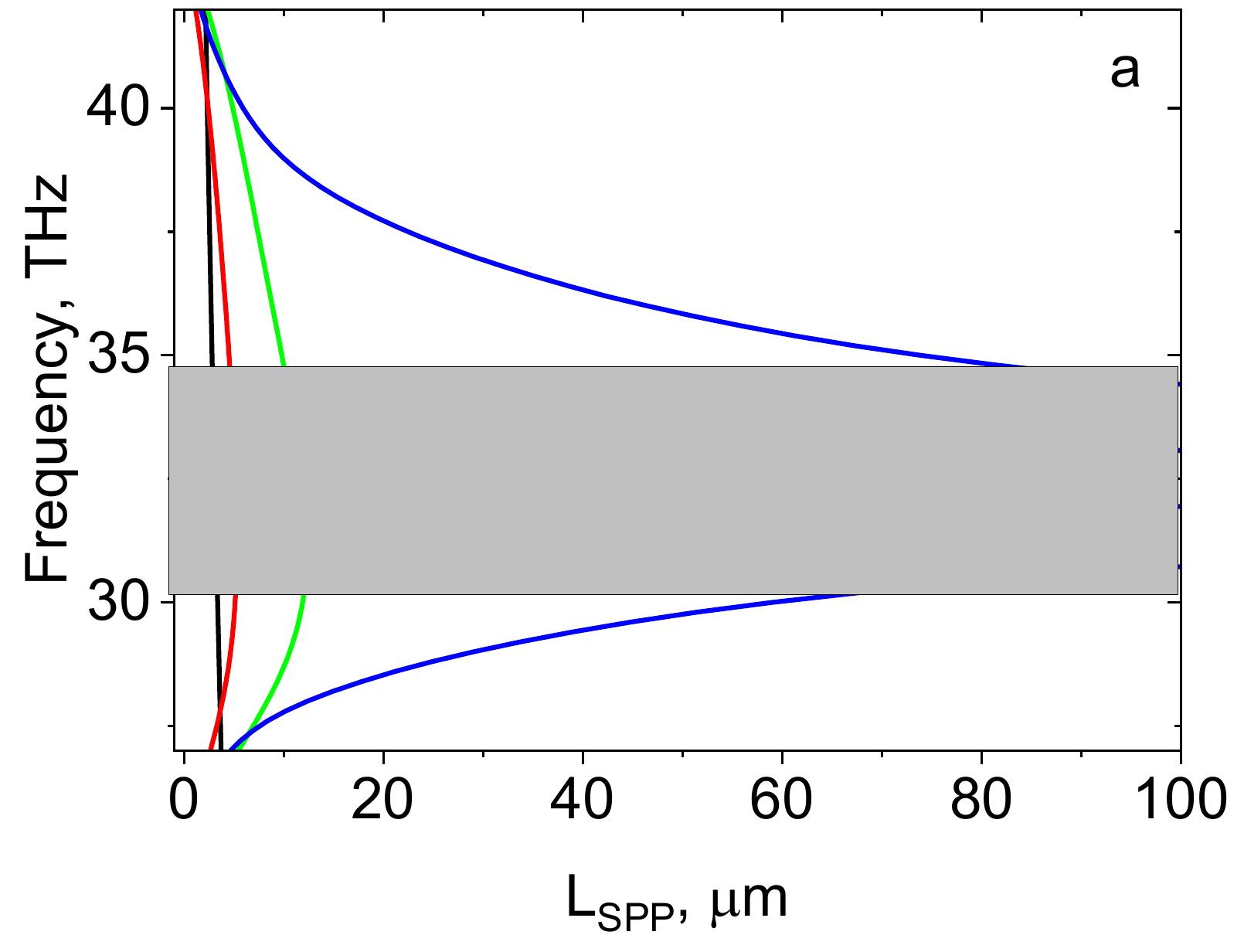"}} 
	{\includegraphics[width=0.9\linewidth]{"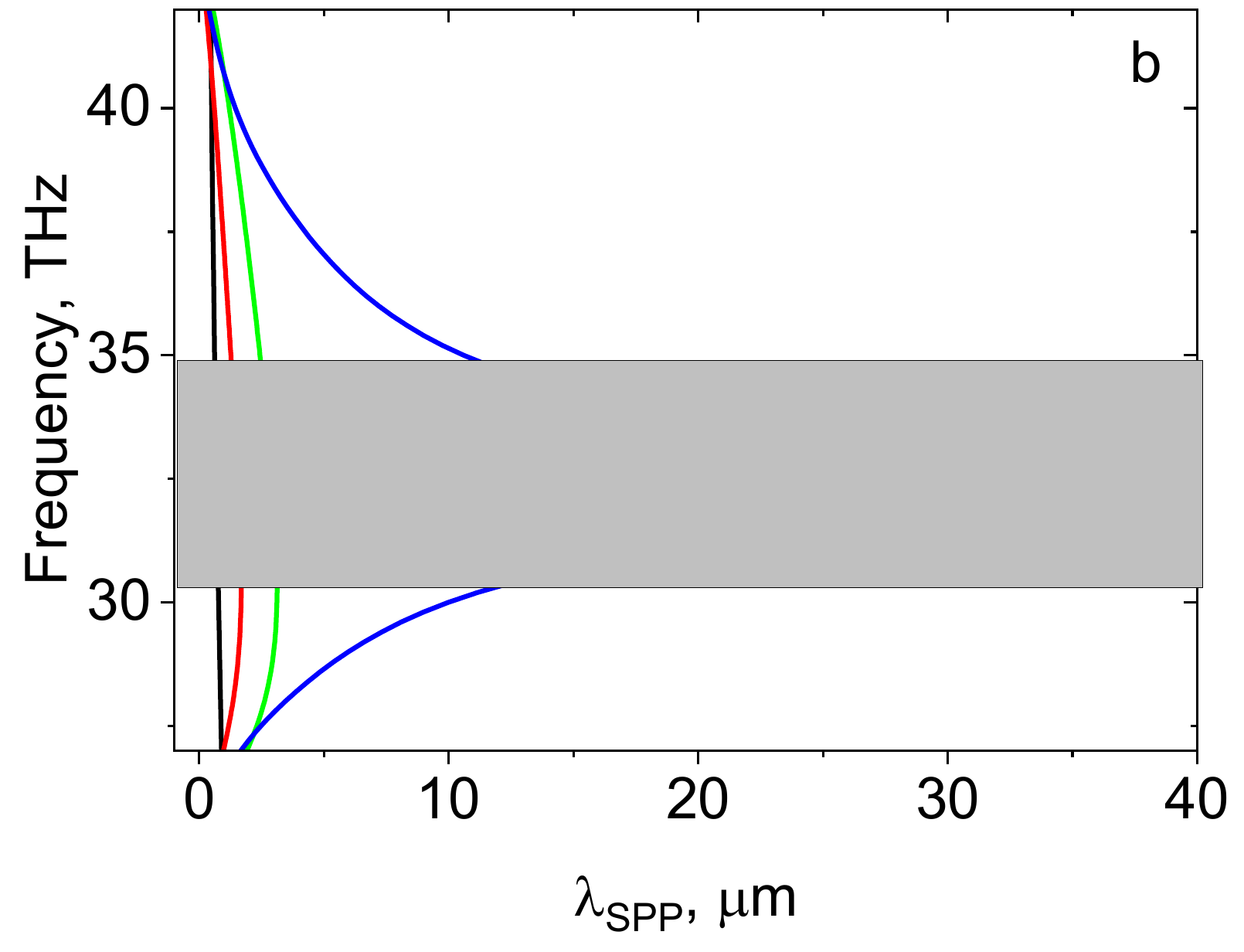"}}
\caption{The  frequency dependence of the SPP propagation length $L_{SPP}$ (a) and SPP wavelength $\lambda_{SPP}$ (b). 
The color lines denote the same conditions as in Figure~\ref{fig:fig8}.}
	\label{fig:fig9}
\end{figure}

Figure~\ref{fig:fig8} shows the dispersion dependences calculated by Comsol for several graphene structure configurations with PECBC or SBC and metasurfaces in NZERI regime as a substrate (MS). 
The real and imaginary parts of the dispersion relations (Figure~\ref{fig:fig8}) intersects with the black line when the effective refractive index reaches  $n_{eff}=1$ at frequencies $f_b$ and $f_t$ bands for all structure cases. Wave vector $q$ values approaches the light line ($k_0$) when  $n_{eff}<1$ for the cases correspond to red and blue lines.  Such fast waves are commonly referred to leaky modes because they can easily decouple and be radiated from the graphene surface into free space as shown in  Ref.~\cite{Low2018}.
However, as will be described below (see Figure~\ref{fig:fig9}c), at NZERI regime such fast waves save surface character and can be considered as SPPs.
Importantly, the surface behavior of SPP on graphene  at NZERI substrates sharply changed in the NZERI region that is demonstrated in Figure~\ref{fig:fig9} and Figure~\ref{fig:fig_2D_fields}. 

Figure~\ref{fig:fig9} presents the frequency dependence  of the SPP propagation length ($L_{SPP}=1/(2 q^{\prime\prime)})$) on graphene, the SPP wavelength ($\lambda_{SPP}=2 \pi/q^{\prime}$) and their ratio in the NZERI regime for graphene layer with different substrates. Notably, the  SPP propagation length for the case of graphene sandwiched between two identical NZERI regime metasurfaces (Figure~\ref{fig:fig9}, blue line) are the greatest and can reach values up to approximately 70~$\mu m$. SPP wavelength also sharply increases (can reach up to 15 $\mu m$) in this region, but not so greatly as SPP propagation length. As the results, in NZERI area their ratio is from 4 to 7. For comparison, as previously mentioned, the SPP propagation length on the suspended graphene is approximately $3\mu m$ at 30 THz under the considered parameters~\cite{Hu2022}. 
We obtained similar results using corresponding physical parameters of the graphene layer, see Figure~\ref{fig:fig9}a, black line.
For the rest of structure cases studied here SPP propagation length also greater than for the suspended graphene (see Figure~\ref{fig:fig9}).   

Notably,  SPP propagation length increases when $real(q)\rightarrow k_0$  owing to the SPP electric field volume part becomes smaller in the graphene and bigger in outer space, thus, the SPP attenuation decreases.

\begin{figure}[!tbh]
	\centering
	{\includegraphics[width=0.9\linewidth]{"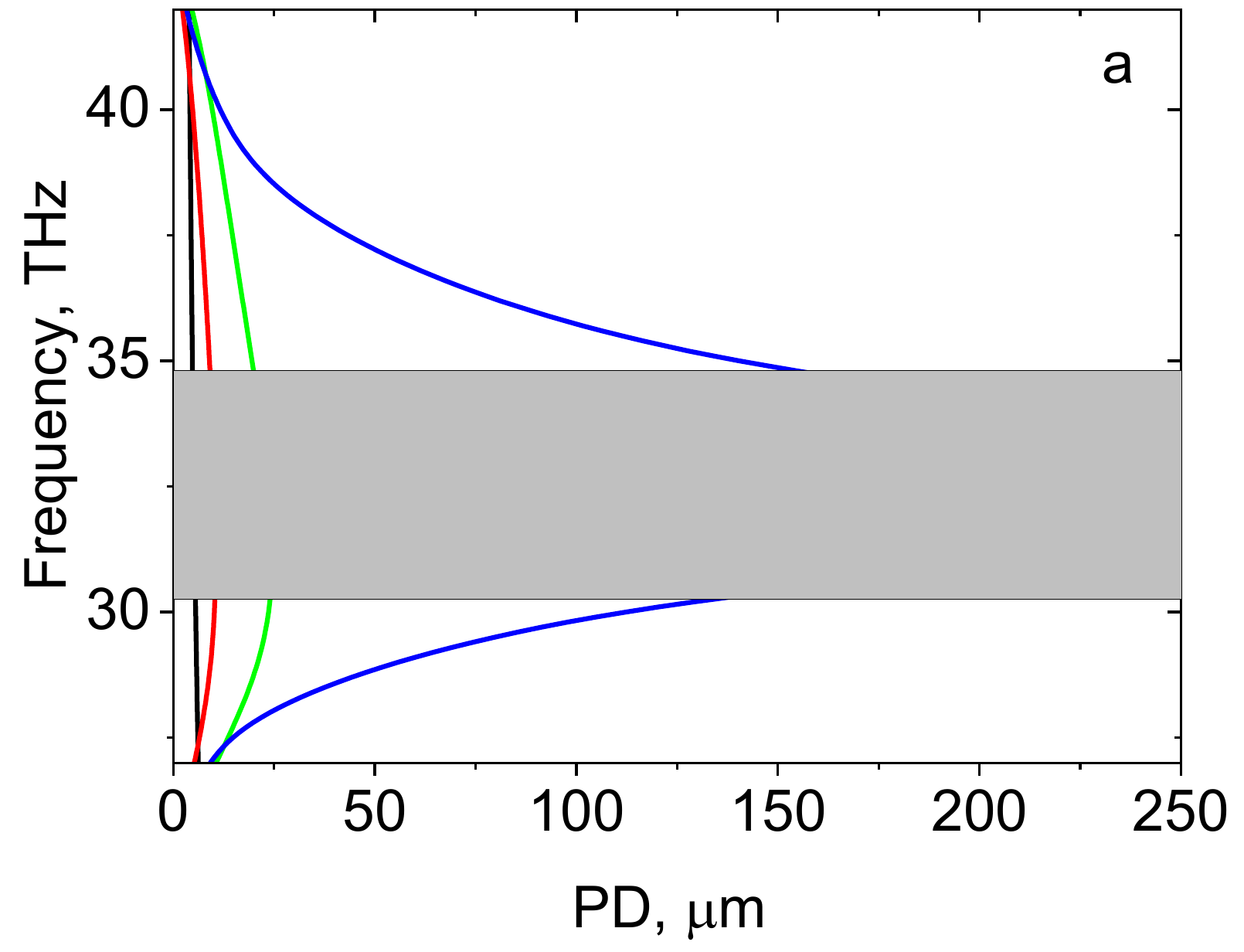"}} 
%	{\includegraphics[width=0.9\linewidth]{"FiguresForPaper1/fig_PD_SPP_L_SPP.pdf"}}
	{\includegraphics[width=0.9\linewidth]{"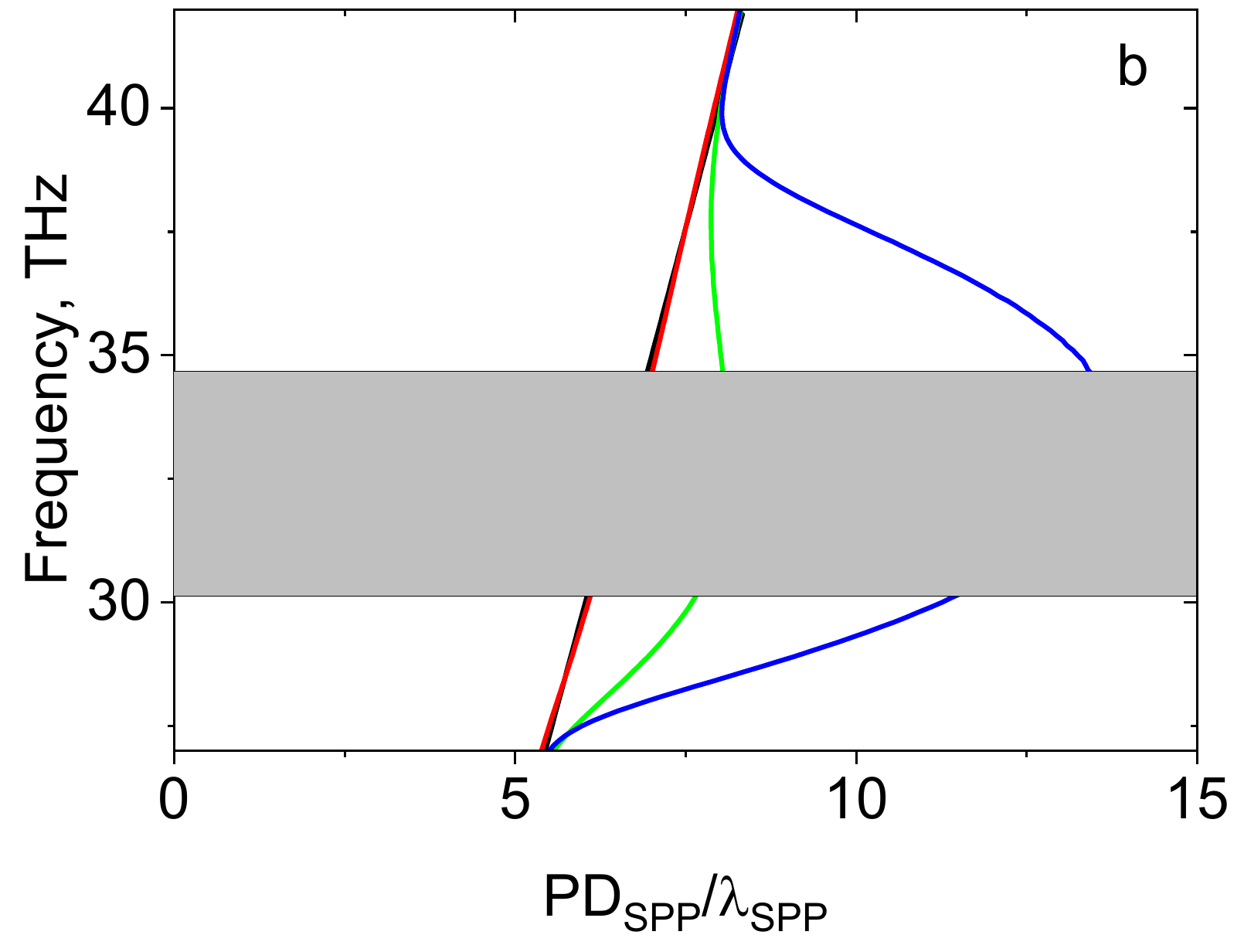"}}
	\caption{The frequency dependence of the wave penetration depth $PD_{SPP}$ (a) 
%the ratio $PD_{SPP}/L_{SPP}$ (b) 
and the ratio of $PD_{SPP}/\lambda_{SPP}$  (b) dependences in the NZERI frequency area. The color lines denote the same conditions as in Figure~\ref{fig:fig8}.}
	\label{fig:figPD}
\end{figure}

\begin{figure*}[!tbh]
	\centering
		{\includegraphics[width=0.22\linewidth]{"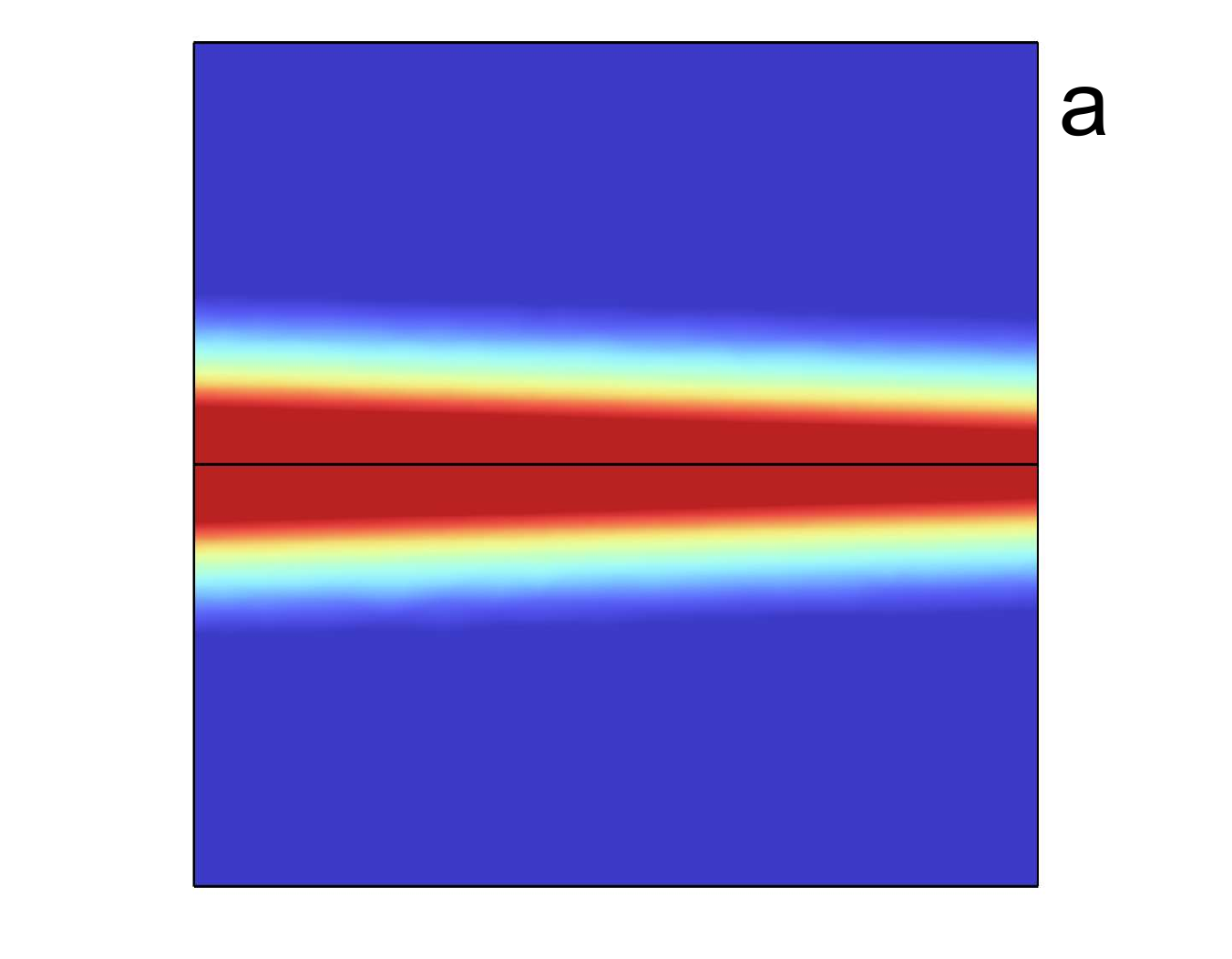"}} 
		{\includegraphics[width=0.22\linewidth]{"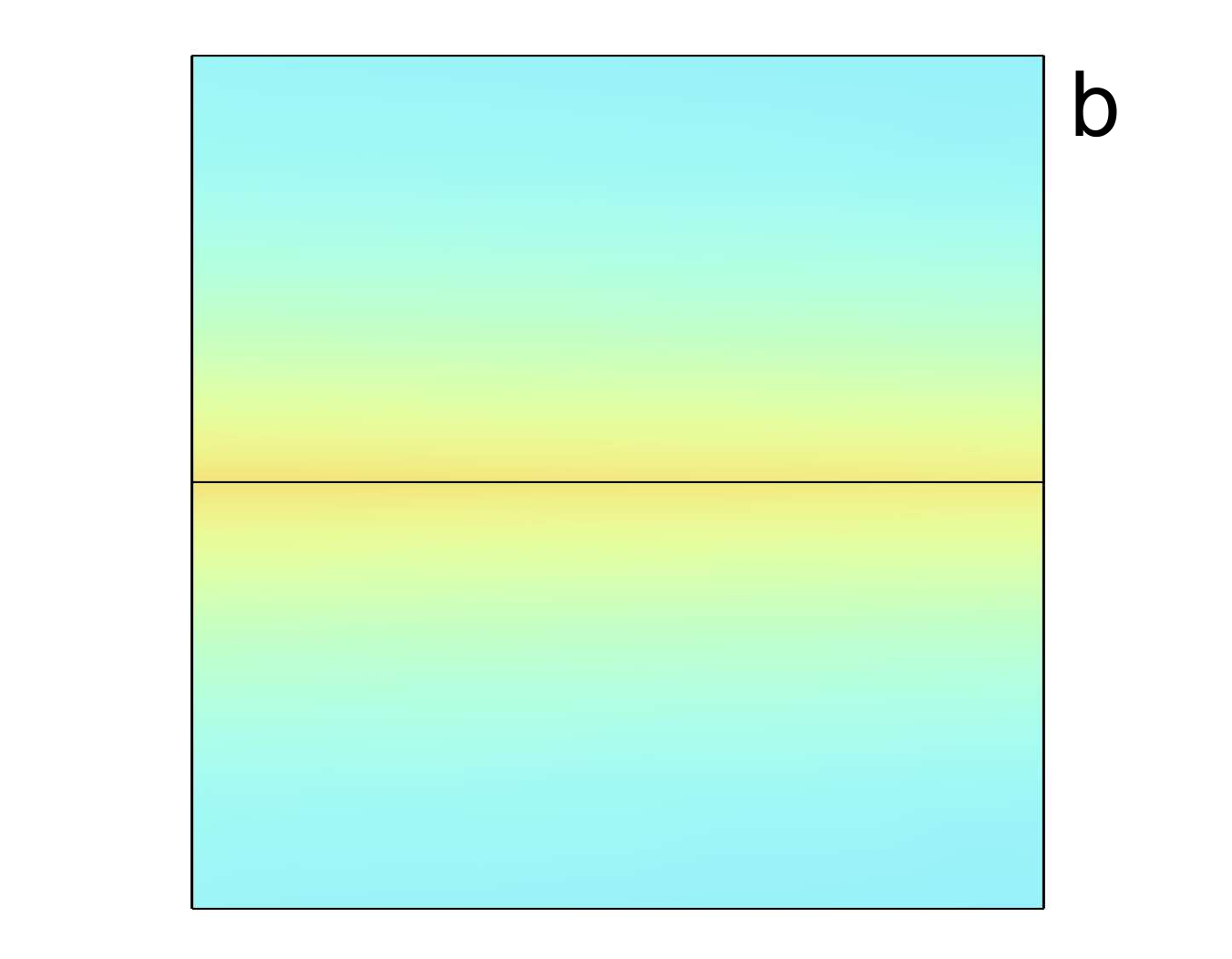"}} 
		{\includegraphics[width=0.22\linewidth]{"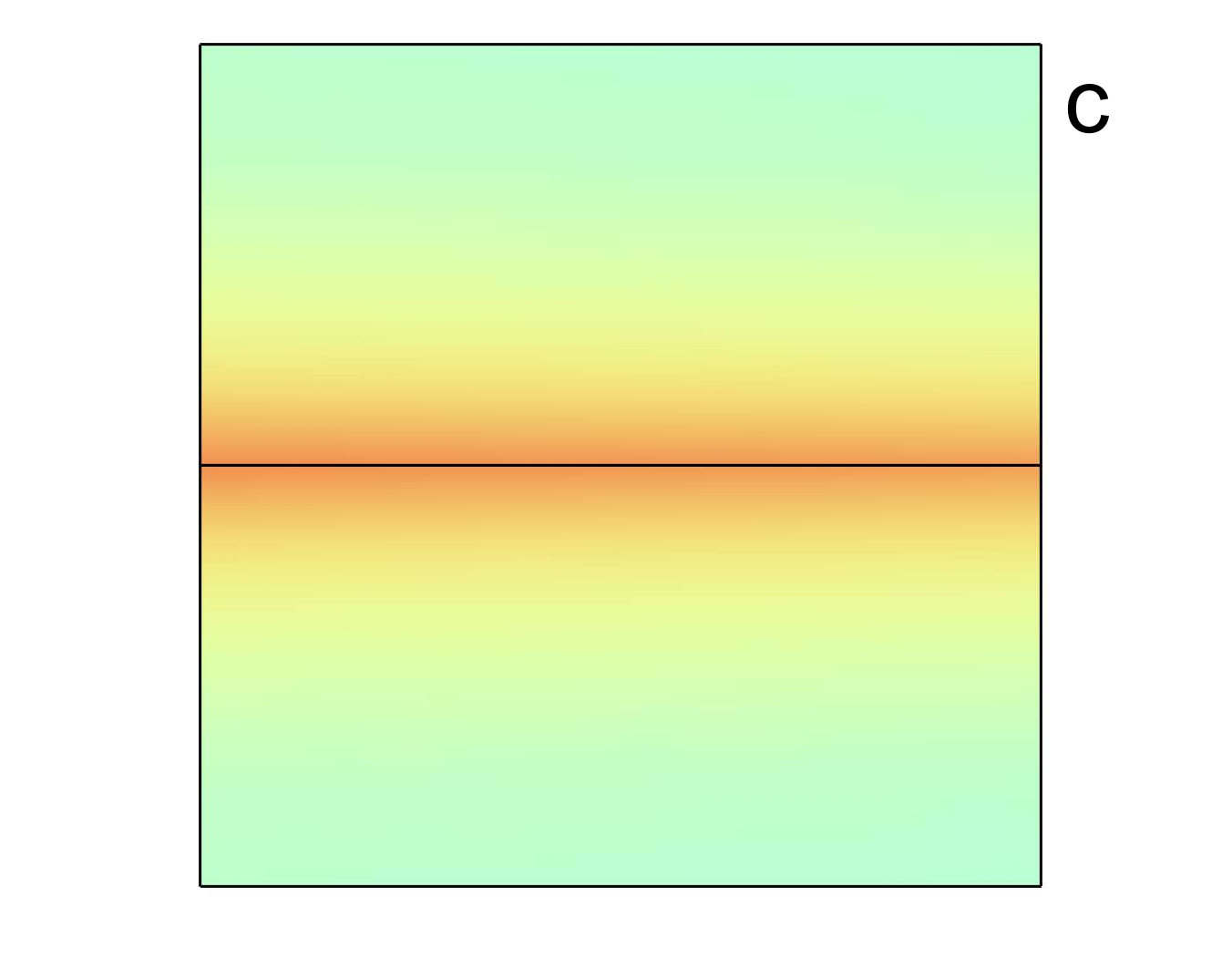"}} 
		{\includegraphics[width=0.22\linewidth]{"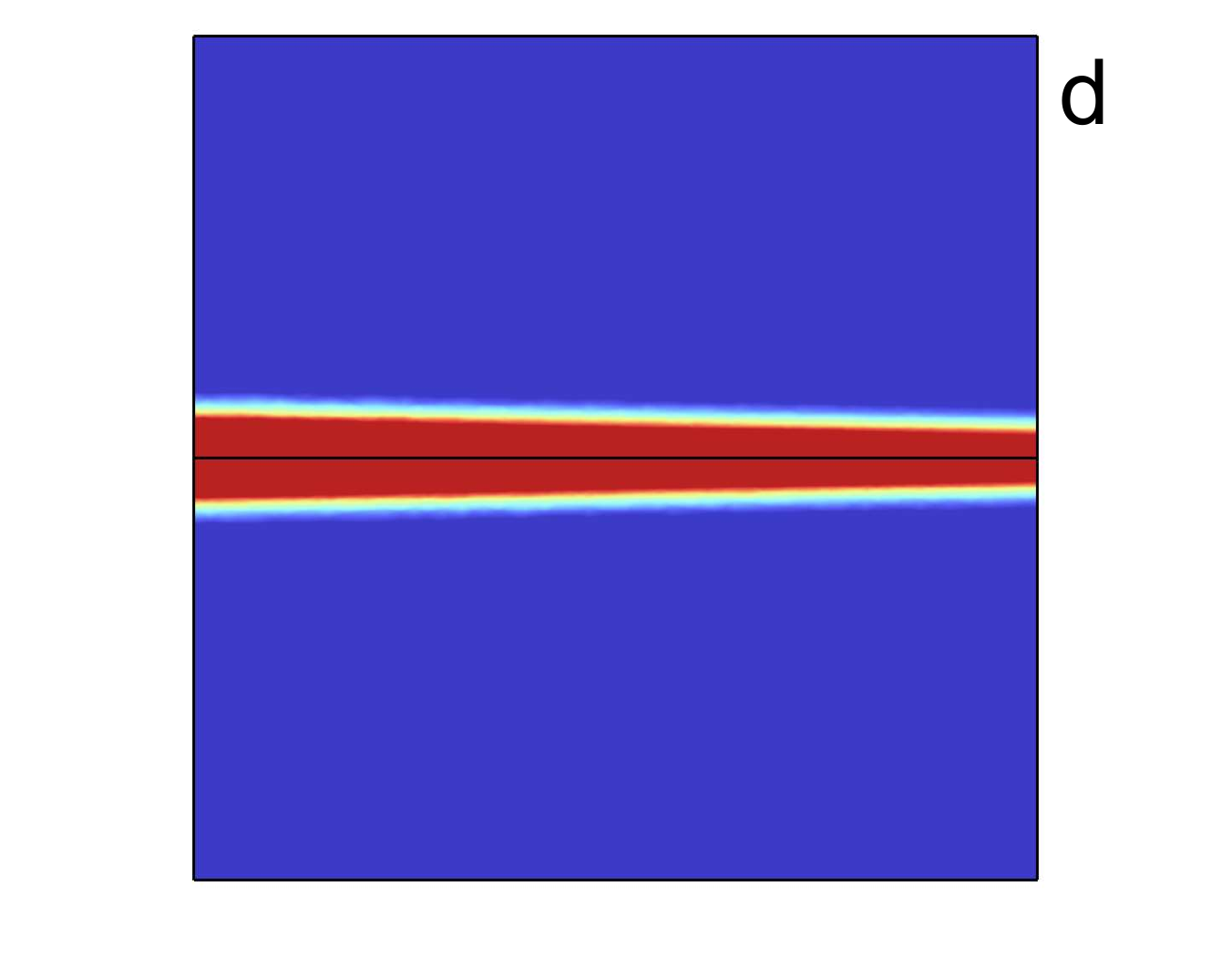"}}  
	\caption{The electric field  module ($|E|$)  distribution of SPPs on graphene layer sandwiched between two identical NZERI regime metasurfaces for 2D model structure. The plots (a) and (d) are at $f_b$ and $f_t$ ($n_{eff}=1$), the plots (b) and (c) are at $f_{m1}=30$~THz and $f_{m2}=35$~THz ($n_{eff}<~1$), respectively. The electric field strength minimum (in blue) corresponds to $4\times 10^4$ V/m, the maximum one (in red) is $\times 10^5$ V/m. The SPP propagation calculated distance is $2\mu m$. }
	\label{fig:fig_2D_fields}
\end{figure*}

Figure~\ref{fig:figPD} presents SPP penetration depth into outer substrate medium 
$PD_{SPP}=1/k_i^{"}$  (see Eqs.~(\ref{eq:grdy1a})) and its ratio on $\lambda_{SPP}$. 
We found that the values of 
the ratio of $ PD_{SPP}/\lambda_{SPP}$ for the structure with two NZERI regime substrates are approximately bigger in two times than for the suspended graphene case and, thus, the surface character of such SPPs is less than for the suspended graphene. Our analysis reveals that the ratio of the SPPs penetration depth to the SPP wavelength, $PD_{SPP}/\lambda_{SPP}$, for structures with dual NZERI regime substrates is approximately twice as large as that of suspended graphene. Consequently, SPPs in the dual-NZERI configuration exhibit:
\begin{itemize}
	\item Reduced surface confinement compared to suspended graphene.
	\item Greater field penetration into the surrounding media.
	\item Modified dispersion properties due to substrate interactions.
\end{itemize}

%F:\MSCA4UA\Publications Eremenko\2024\paper1\tau=2!e-13
% tau 2E-13 beta Lan SPP, L SPP 2mikm.opju
%tau 2E-13 penetration depth 2mikm.opju

%graphene spp air-air Ef=0.8 eV gr layer -0.345nm air-air  27-42 Thz.mph
%D:\COMSOL PROJECTS\2D graphene substrate\Ef´=0.8eV

%2D graphene spp TWO substrate - neff  -2D Ref.24 ONE sub ONE PEC.mph
%2D graphene spp TWO substrate - neff  -2D Ref.24 TWO sub No PEC.mph
%2D graphene spp TWO substrate - neff  -2D Ref.24 ONE sub TWO PEC.mph
%2D graphene spp TWO substrate - neff  -2D Ref.24 TWO sub TWO PEC.mph
%D:\COMSOL PROJECTS\2D graphene substrate\Ef=0.8eV with substrate - neff\Comparison trans bound conditions

%F:\MSCA4UA\Publications Eremenko\2024\paper1\tau=2!e-13
% tau 2E-13 beta Lan SPP, L SPP 2mikm.opju
%tau 2E-13 penetration depth 2mikm.opju

%graphene spp air-air Ef=0.8 eV gr layer -0.345nm air-air  27-42 Thz.mph
%D:\COMSOL PROJECTS\2D graphene substrate\Ef´=0.8eV

%2D graphene spp TWO substrate - neff  -2D Ref.24 ONE sub ONE PEC.mph
%2D graphene spp TWO substrate - neff  -2D Ref.24 TWO sub No PEC.mph
%2D graphene spp TWO substrate - neff  -2D Ref.24 ONE sub TWO PEC.mph
%2D graphene spp TWO substrate - neff  -2D Ref.24 TWO sub TWO PEC.mph
%D:\COMSOL PROJECTS\2D graphene substrate\Ef=0.8eV with substrate - neff\Comparison trans bound conditions

Figure~\ref{fig:fig_2D_fields} presents the SPP propagation view for the electric field strength module along the graphene layer sandwiched between two identical NZERI regime metasurfaces for 2D model structure (see Figure~\ref{fig:fig8}, the blue line) at pointed frequencies for two cases: $n_{eff}= 1$ ($f_b$, $f_t$) and  $n_{eff}<1$ ( $f_{m1}$, $f_{m2}$ ). This demonstrates the variation in SPP propagation characteristics within the NZERI frequency regime: the electric field is more strongly confined near the graphene layer at frequencies  $f_b$ and $f_t$ frequencies than at $f_{m1}$, $f_{m2}$ ones due to SPP wave propagation number $q$ dependence (Figure~\ref{fig:fig8}).

%2D graphene spp TWO substrate - neff  -2D Ref.24 TWO sub TWO PEC.mph
%D:\COMSOL PROJECTS\2D graphene substrate\Ef=0.8eV with substrate - neff\Comparison trans bound conditions

%E field  2D neff.opju
%F:\MSCA4UA\Publications Eremenko\2024\paper1

\begin{figure*}[!tbh]
	\centering
		{\includegraphics[width=0.2\linewidth]{"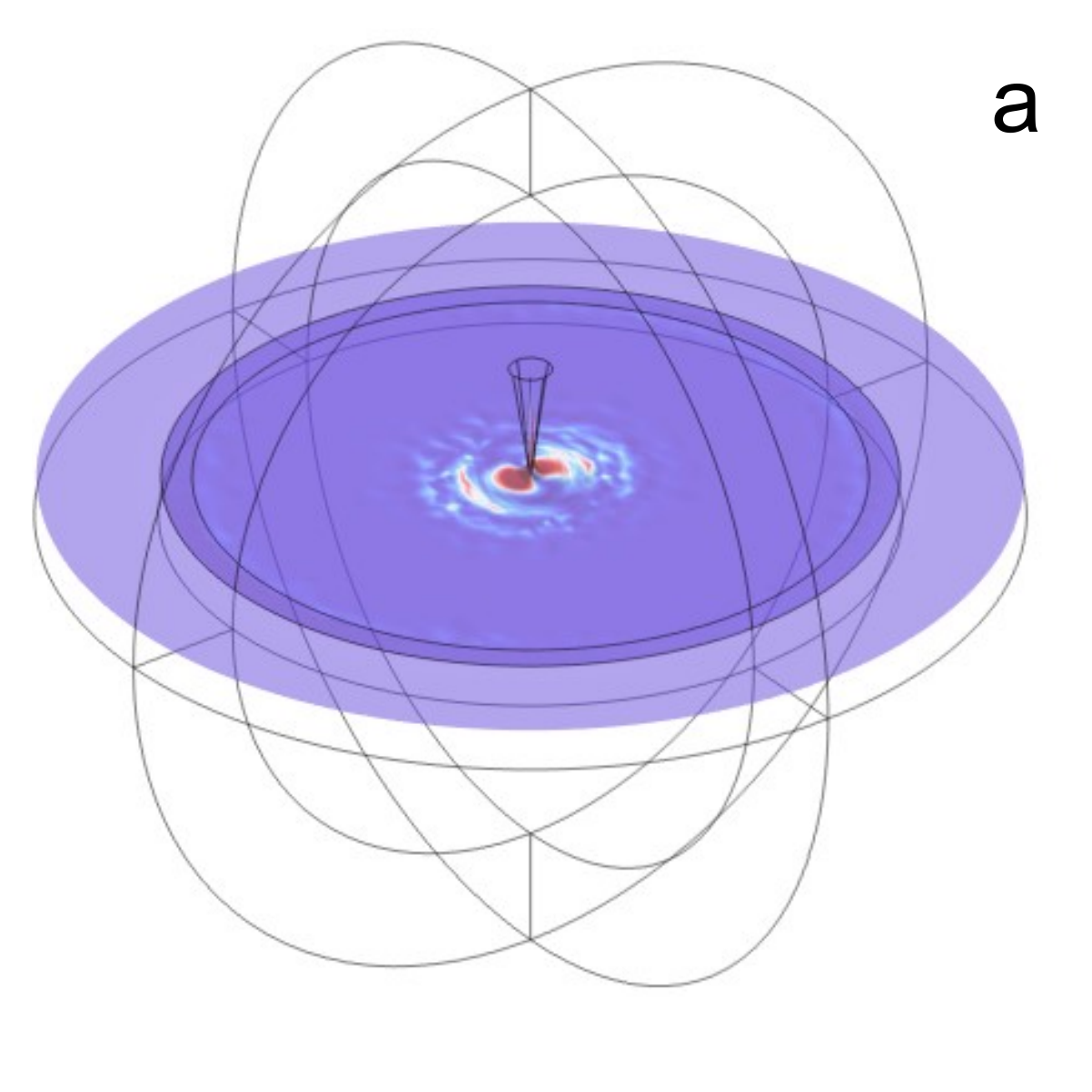"}}  
		{\includegraphics[width=0.2\linewidth]{"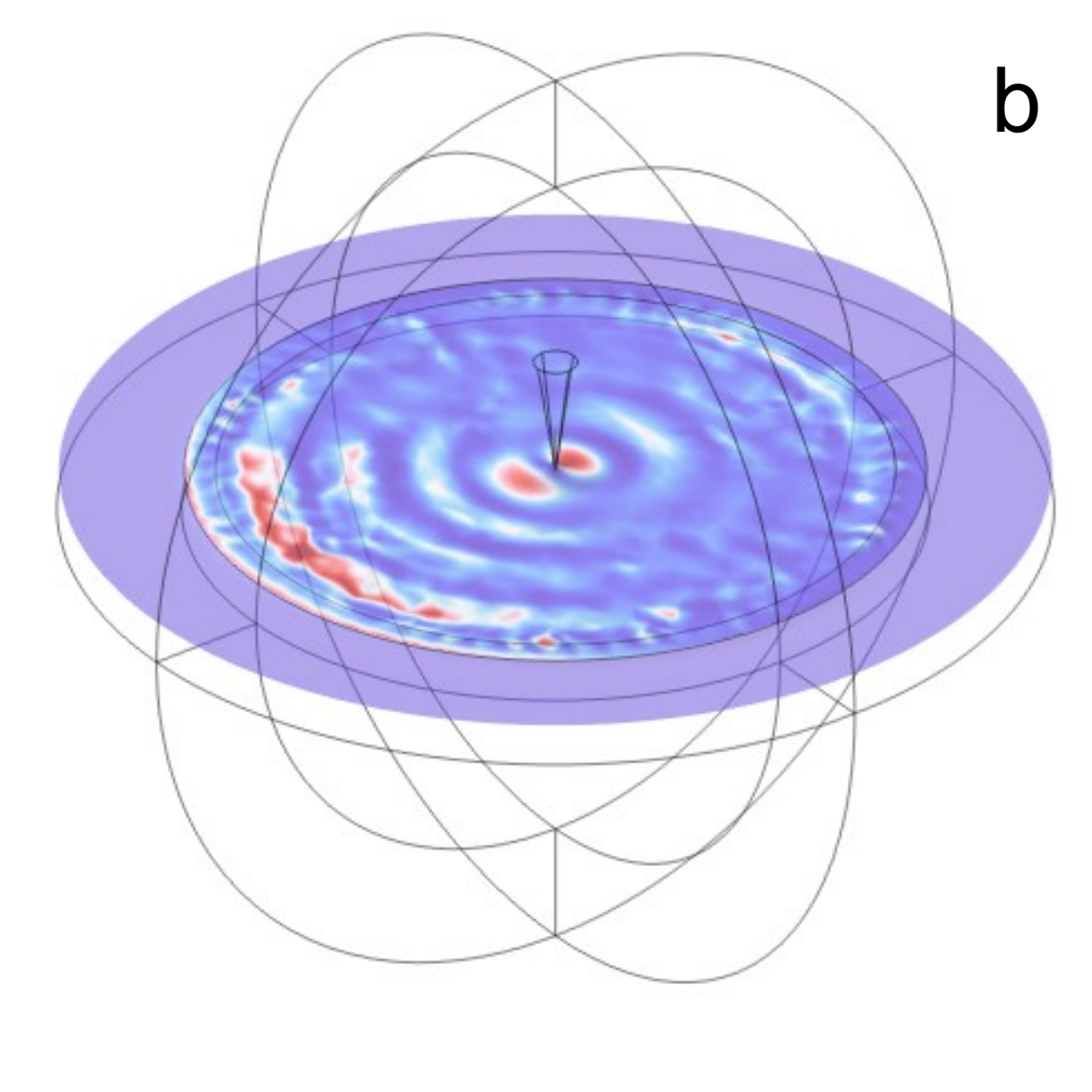"}} 
		{\includegraphics[width=0.2\linewidth]{"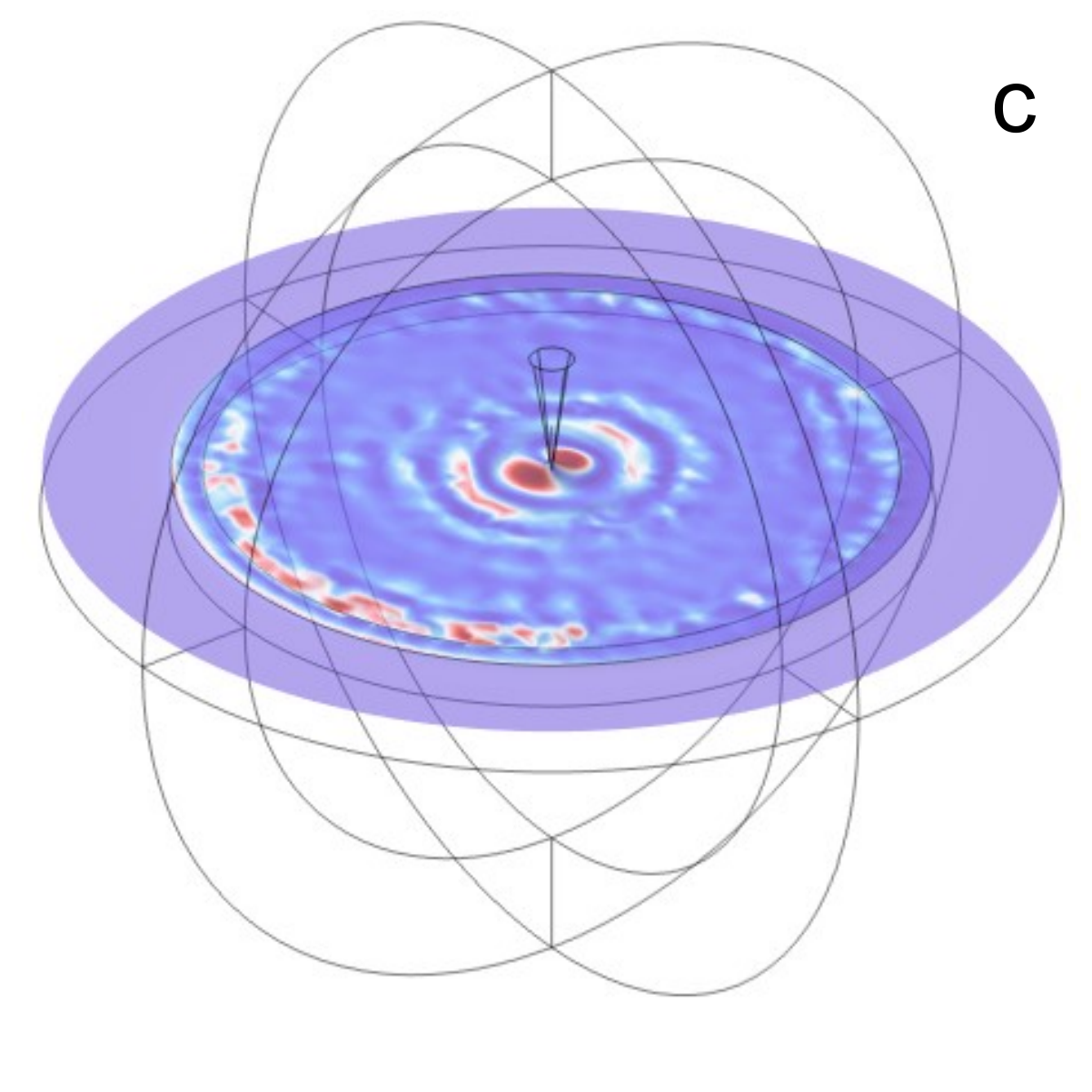"}} 
		{\includegraphics[width=0.2\linewidth]{"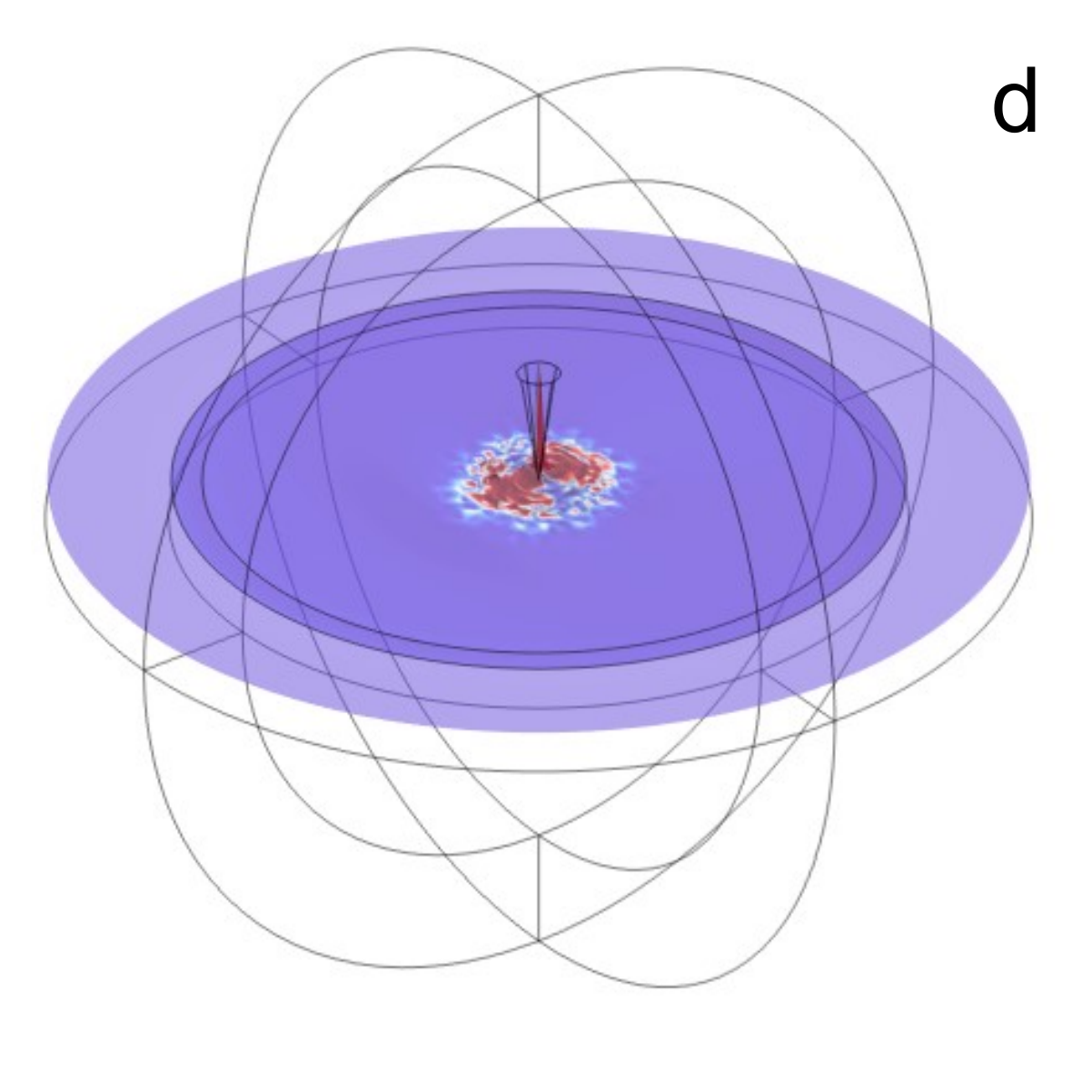"}}
	\caption{3D view of SPP electric field distribution $|E_{x}|-|E_{bgx}|$, where $E_{x}$ and $E_{bgx}$ are the electric field component and the background one along $x$-axis, respectively) in the round graphene layer ($6.8\mu m$ radius) on the substrate as metasurface in NZERI regime with SBC at frequencies $f_b$ (a) ($n_{eff}= 1$), $f_{m1}$ (b) ($n_{eff}= 0.31$),  $f_{m2}$ (c) ($n_{eff}= 0.246$) and $f_t$ (d) ($n_{eff}= 1$). SPPs were excited by the gold tip spherical edge ($0.44\mu m$ radius).  The electric field strength minimum (in blue) corresponds to $0$ V/m, the maximum one (in red) is $2.5$ V/m. The spherical contour lines around the graphene are perfect matched layer (PML) in Comsol model. }
	\label{fig:fig10}
\end{figure*}

Figure~\ref{fig:fig10} displays the 3D electric field distribution, demonstrating the propagation of SPPs along the graphene layer atop the metasurface substrate at the frequency in NZERI region.
For this model structure (with the tip excitation the surface waves) the exciting electromagnetic wave angle is of 80 degrees from the normal to the graphene layer~\cite{McLeod2014},
the wave vector is directed along the $x$-axis, the plane wave falls on a golden tip and excites SPPs on the graphene. The gap between a tip round edge (of the radius  $10\mu m$) and the round graphene layer is of $10\mu m$ radius. 

As we can see, at the reference points ($f_b$ and $f_t$) when $n_{eff}=1$ the SPPs propagate in a small graphene area around the tip increases together with the SPP wavelength ($\lambda_{SPP}$) as we calculated for 2D model (see Figure~\ref{fig:fig9} the frequency intersection of the red and green lines with the blue one). Thus, these 3D results confirm the increasing of the SPP propagation length on graphene with the substrate as metasurface in NZERI regime. 

%3D graphene with neff substrate.opju 
%F:\MSCA4UA\Publications Eremenko\2024\paper1
% 3D with ordinary graphene layer neff in substrate.mph
%D:\COMSOL PROJECTS\2D graphene substrate\Ef=0.8eV with substrate - neff\3D neff -substrate

\subsection{Physics of NZERI regime}
The NZERI regime in all-dielectric metasurfaces emerges from the unique dispersion characteristics near the G-point of the photonic band structure. At the G-point, the electromagnetic modes exhibit zero phase difference across unit cells, corresponding to spatially uniform phase distributions within the metasurface with special geometric and material parameters. This leads to a vanishingly small propagation constant and an effectively infinite phase velocity, resulting in an effective refractive index approaching zero.

Physically, the NZERI regime arises due to strong resonant interactions within the periodic dielectric structure, often dominated by electric and magnetic Mie-type resonances in high-index dielectric elements. These resonances couple coherently across the lattice, producing collective electromagnetic modes with nearly flat dispersion around the G-point. The flatness of the dispersion curve (i.e., small slope, see Figure~\ref{fig:1}) indicates a low group velocity, while the near-zero slope of the frequency–wavevector relation implies an extremely long effective wavelength within the material.

Unlike metallic plasmonic systems where NZERI behavior is associated with negative permittivity and high losses, the all-dielectric realization enables low-loss operation and better compatibility with photonic integration.

Therefore, the NZERI regime at the G-point in all-dielectric metasurfaces represents a purely geometric and modal effect that enables efficient manipulation of light, enhanced nonlinear optical responses, and collective emitter behavior—all of which are critical for next-generation photonic and quantum devices.

\section{Computational Methods}
The graphene substrate in the NZERI regime is modeled using the finite element method (FEM) implemented in the Wave Optics Module of the COMSOL Multiphysics 6.2.
To determine the intrinsic resonant modes of the 2D/3D metasurface unit cells (see 
Figure~\ref{fig:1},
Figure~\ref{fig:3D_PC_field_rod_sphere},
Figure~\ref{fig:3D_PC_on_rod_height}), 
we employed the Eigenfrequency Solver, which solves Maxwell’s equations in the absence of external excitation to compute the system’s natural eigenmodes and corresponding complex eigenfrequencies.
Key setup details include the following.
For the  metasurface unit cell with dielectric rod (sphere) we use periodic Floquet boundary conditions applied along the in-plane directions ($x-y$ plane).
Material properties are realistic dispersionless material parameters were used to isolate structural resonances from material resonances.
The eigenfrequency search range was specified to focus the computation around expected resonance bands.
For 2D structure we use scattering boundary conditions with perfectly matched layer (PML) or PEC ones. PML is needed to satisfy the Sommerfeld boundary conditions.
The Sommerfeld boundary conditions (also known as Sommerfeld radiation conditions) are used in scattering theory and wave propagation problems to ensure that solutions represent outgoing waves at infinity, avoiding unphysical incoming waves from infinity. These conditions are crucial in electromagnetics, acoustics, and quantum mechanics.
For 3D structure for that purposes we use PML boundary conditions that are employed to terminate the computational domain along the $z$-axis. These layers are designed to match the optical index at the interface, ensuring minimal reflection. As waves enter PML, they are exponentially attenuated, effectively absorbing outgoing radiation and preventing unphysical back-reflections from the outer boundaries.

To obtain the mode electromagnetic field profiles at both ports  (see  Figure~\ref{fig:fig8}, Figure~\ref{fig:fig9}, Figure~\ref{fig:fig_2D_fields}) two Boundary Mode Analysis study steps are used one for each port.
Boundary Mode Analysis as a Comsol technique is used to find intrinsic modes that can naturally exist and propagate along a boundary or interface without requiring external excitation. Additionally, a Frequency Domain study step is included to simulate SPP propagation and to extract the dispersion relation, i.e., the dependence of SPP frequency on the wavevector. The SPPs arise from the coupling between surface plasmon and incident electromagnetic wave. In this model, we focus on the p-polarized case ($TM$ mode).

The resonant response of the 3D metasurface unit cells (see 
Figure~\ref{fig:fig_S11_S21}) was numerically studied using COMSOL Multiphysics' Frequency Domain Solver. The simulation setup is as follows.
An incident electromagnetic wave was introduced via an input port at normal along $z$-axis,
the excitation wavelength was set to $\lambda$,
the input and output port were positioned to measure S-parameters, including:
reflection coefficient ($S_{11}$),
transmission coefficient ($S_{21}$).
For the  metasurface unit cell we use periodic Floquet boundary conditions. A separation of 
$1\lambda$,
was maintained between the 3D metasurface unit cell and input/output ports to minimize undesired boundary reflections and ensure accurate field propagation.
The electromagnetic properties of the materials and geometrical features of the unit cell were accurately defined based on the intended physical structure.
This frequency-domain approach allowed us to capture the steady-state electromagnetic field distributions and extract the transmission and reflection spectra across a range of frequencies.

In 3D structure for the SPP excitation in graphene layer with substrate in  NZERI region we use a  gold tip (see Figure~\ref{fig:fig10}) to have electric field concentration on the graphene surface.  There is a PML sphere outer layer with a scattering boundary conditions on an inner sphere and the sphere space with the studied structure inside. The incident electromagnetic wave falls on the structure under certain angle (80 degree from normal in $z$-axis) to excite SPP on greaphene by the tip . 

To excite SPPs in the graphene-substrate system within NZERI regime using a gold tip (Figure~\ref{fig:fig10}), we implement the following computational framework such as
a spherical PML layer serves as the outer boundary to absorb scattered radiation.
An inner PML sphere with scattering boundary conditions encloses the graphene-substrate structure to minimize numerical reflections.

Notably, we model the all-dielectric metamaterial operating in the NZERI regime as an effective graphene substrate with the effective refractive index $ n_{eff} $ as a representative physical parameter of the metamaterial, to support SPP propagation, without explicitly simulating a finite number of unit cell elements. This approach avoids excessive computational time and memory usage.

\section{Conclusions}

In this work, we demonstrate that the effective refractive index of an all-dielectric metasurface in the NZERI regime can be used to significantly extend the propagation length of graphene  SPPs. Using several independent theoretical methods, we characterized the effective refractive index of the metasurface in the NZERI frequency range for both 2D and 3D graphene-layer metasurface structures.

Our numerical simulations further confirm that stacking graphene on/or between all-dielectric metasurfaces operating in the NZERI regime results in a significant increase in the SPP propagation length, by at least an order of magnitude. In particular, the SPP propagation length, which is approximately $3 \mu m$ for graphene with air substrates (30 THz, $E_F~ =~0.8$ eV, $\tau~ =~2\cdot 10^{-13}$ s, $d_{eff}=0.345$ nm). 
For a structure with an air superstrate and a metasurface substrate in the NZERI regime placed on a PEC boundary, the SPP propagation length can reach approximately $6{\mu m}$.  
For a similar configuration with PEC boundaries on both the top and bottom, the propagation length increases to about $10{\mu m}$. 
Finally, when the graphene layer is sandwiched between two NZERI substrates and enclosed by PEC boundaries on both sides, the propagation length can reach up to $70{\mu m}$.

Furthermore, although the SPP wavelength also increases in this regime (up to $15\mu m$), the propagation length enhancement is much more pronounced, with the SPP wavelength to propagation length ratio ranging from 4 to 7.

We present the first demonstration of a NZERI metasurface employed as a substrate for graphene to enhance the propagation of SPP.  A quantitative framework is developed to evaluate the trade-off between SPP propagation length and field confinement. Additionally, we propose practical design guidelines that are compatible with current nanofabrication technologies.

Our analysis reveals that the surface behavior of SPP on graphene undergoes a distinct transformation in the NZERI region. SPP propagation length increases when SPP propagation number tends to light line owing to the SPP electric field volume part becomes smaller in the graphene and bigger in outer space, thus, the SPP attenuation decreases.This critical effect highlights the unique optical behavior of graphene SPPs in the presence of NZERI substrates, opening new avenues for the manipulation and control of plasmonic wave propagation in graphene-based photonic devices.

%%%%%%%%%%%%%%%%%%%%%%%%%%%%%%%%%%%%%%%%%%%%%%%%%%%%%%%%%%%%%%%%%%%%%
%% The same is true for Supporting Information, which should use the
%% suppinfo environment.
%%%%%%%%%%%%%%%%%%%%%%%%%%%%%%%%%%%%%%%%%%%%%%%%%%%%%%%%%%%%%%%%%%%%%
%\section*{Associated content}
%\subsection*{Supporting Information}

%This will usually read something like: ``Experimental procedures and
%characterization data for all new compounds. The class will
%automatically add a sentence pointing to the information on-line:
%The Supporting Information is available free of charge at \url{https://pubs.acs.org/doi/...}.

%\subsection*{ORCID}
%Zoya E. Eremenko: https://orcid.org/0000-0002-9635-2611\\
%Igor N. Volivichev: https://orcid.org/0000-0003-1136-1920 \\
%Aliaksei Charnukha:

%\subsection*{Notes}

%The authors declare no competing financial interest.

\begin{acknowledgement}
Z.E.  acknowledges the funding from the European Union under the Marie Skłodowska-Curie grant agreement no. MSCA4Ukraine project number 1.4 - UKR - 1232611 - MSCA4Ukraine (IFW Dresden) and Scholarship grant (MPI PKS, Dresden). Views and opinions expressed are however those of the author(s) only and do not necessarily reflect those of the European Union, the European Research Executive Agency or the MSCA4Ukraine Consortium. Neither the European Union nor the European Research Executive Agency, nor the MSCA4Ukraine Consortium as a whole nor any individual member institutions of the MSCA4Ukraine Consortium can be held responsible for them.
 We sincerely thank Dr. Aliaksei Charnukha, Prof. Vladimir R. Tuz, Prof. Alexander I. Nosich, and Dr. Yuri M. Savin for the fruitful discussions that contributed to this work.
\end{acknowledgement}

\subsection*{Appendix 1}
\setcounter{equation}{0}
\renewcommand{\theequation}{A\arabic{equation}}
\setcounter{table}{0}
\renewcommand{\thetable}{A\arabic{table}}
We conducted our eigensolver analysis in the Irreducible Brillouin Zone for a 2D square periodic structure using a parametric sweep with a parameter, \( \mathbf{k} = (k_x, k_y) \), ranging from 0 to 3. This range is divided into three segments~\cite{joannopoulos2011photonic,Sakoda2005}. 
\begin{itemize}
	\item $ 0 \leq k \leq 1 $ corresponds to the wave number along  $\Gamma$-$X$,
	\item $ 1 \leq k \leq 2 $ corresponds to the wave number along  $X$-$M$, and
	\item $ \leq k \leq 3 $ corresponds to the wave number along  $M$-$\Gamma$.
\end{itemize}

For each value of $ \mathbf{k} $
% (Eqs.~(\ref{eq:kx1}, ~(\ref{eq:ky1}), 
we computed the lowest natural frequencies and plotted the wave propagation frequencies. Any band gap is identified as a frequency range where no wave propagation branches exist.

%\begin{equation} \label{eq:kx1}
%	k_x = 
%	\begin{cases}
%		(\pi/a )\cdot (0 - k), & \text{if } k < 0, \\
%		(\pi/a )\cdot (k - 0), & \text{if } 0 \leq k < 1, \\
%		(\pi/a ), & \text{otherwise.}
%	\end{cases}
%\end{equation} 

%\begin{equation} \label{eq:ky1}
%	k_y = 
%	\begin{cases}
%		(\pi/a ) \cdot (0 - k), & \text{if } k < 0, \\
%		0, & \text{if } 0 \leq k < 1, \\
%		(\pi/a ) \cdot (k - 1), & \text{otherwise.}
%	\end{cases}
%\end{equation}

\subsection*{Appendix 2}
Now several methods have been developed for determining the effective parameters of metamaterials (and, in particular, metasurfaces),
describing their electrodynamic properties~\cite{alu2011,martin2019,wu2006}. 
These methods give equivalent results in the limit of an infinitely large wavelength. 
When the wavelength decreases, the results obtained by these methods begin to differ~\cite{gozh2013}. This discrepancy's origin is related to how the field average is defined in Ref.~\cite{alu2011h,gozh2013}.
The choice of a suitable method becomes most relevant for resonant structures when the wavelength in the metamaterial is comparable to the dimensions of the structural elements of the unit cell.
Elucidating whether effective parameters can be used beyond the long-wave approximation is important for two reasons. First, describing them in terms of effective susceptibilities provides a clearer physical interpretation. Second, in computer simulations of structures containing metamaterials, replacing the unit cell array with a single medium with given effective parameters drastically reduces the required computational resources.

Most of the closed-form methods are restricted by $\lambda/a\gtrsim 3$~\cite{alu2011h,martin2019}. For specific unit cell geometry, some methods extend this range, e.g. method in Ref.~\cite{wu2006} remains valid up to $\lambda/a>0.8$.

In this study, we compare the results of two methods that give similar results in the frequency range under consideration, validating
their applicability. Both of these methods demonstrate the existence of the NZERI region in the considered all-dielectric metasurface.

One of the methods aimed at studying the electrodynamic properties of 2D objects and metasurfaces is reported in Ref.~\cite{martin2019}. It is
based on generalized sheet transition conditions (GSTCs)~\cite{idemen2011}. GSTCs equations relate the fields on the opposite sides of the metasurface under study:
\begin{equation}\label{lb1}
	\boldsymbol{\hat{z}}\times \Delta\boldsymbol{H} = j\omega\varepsilon_0\,\dbar{\chi}_{\mathrm{ee}}\cdot\vect{E}^\mathrm{av}+
	jk\,\dbar{\chi}_{\mathrm{em}}\cdot\vect{H}^\mathrm{av},
\end{equation}
\begin{equation}\label{lb2}
	\boldsymbol{\hat{z}}\times \Delta\boldsymbol{E} =
	-j\omega\mu_0\,\dbar{\chi}_{\mathrm{mm}}\cdot\vect{H}^\mathrm{av}-
	jk\,\dbar{\chi}_{\mathrm{me}}\cdot\vect{E}^\mathrm{av},
\end{equation}
where $\omega$ is the frequency; $\varepsilon_0$ and $\mu_0$ are the vacuum permittivity and permeability, respectively; $k$ is the wave vector;  
$\Delta\boldsymbol{E}$ and $\Delta\boldsymbol{H}$ are the differences of the electric and
magnetic fields on top and bottom of the metasurface, respectively; $\dbar{\chi}_{\mathrm{ee}}$, $\dbar{\chi}_{\mathrm{em}}$,
$\dbar{\chi}_{\mathrm{me}}$ and $\dbar{\chi}_{\mathrm{mm}}$ are  the metasurface bianisotropic surface susceptibilities;
$\vect{E}^\mathrm{av}$ and $\vect{H}^\mathrm{av}$ are the arithmetic averages of the electric
and magnetic fields across the metasurface, respectively:
\begin{align}
	&\Delta \vect{E} = \vect{E}^{+}-\vect{E}^{-},\\
	&\Delta \vect{H} = \vect{H}^{+}-\vect{H}^{-},\\
	&\vect{E}^\mathrm{av} = \frac{1}{2}\left(\vect{E}^{+}+\vect{E}^{-}\right),\label{Eav}\\
	&\vect{H}^\mathrm{av} = \frac{1}{2}\left(\vect{H}^{+}+\vect{H}^{-}\right).\label{Hav}
\end{align}
Superscripts ``$+$'' and ``$-$'' denote fields on the top ($z=+0$) and the bottom ($z=-0$) sides of the metasurface.

Maxwell's equations and the GSTC equations form a boundary value problem that can be solved by standard methods.
Following the results in Ref.~\cite{martin2019} let us express the fields on the top ($z = +0$) and the bottom ($z = -0$) sides of the metasurface as follows:
\begin{align}
	\vect{E}^{+} =& \left(
	\boldsymbol{\hat{x}}\frac{k_{z}}{k_0}\eta_0 A^{+}_\mathrm{TM}+\boldsymbol{\hat{y}} A^{+}_\mathrm{TE}
	\right)\mathrm{e}^{-jk_xx},\\
	\vect{H}^{+} =& \left(
	-\boldsymbol{\hat{x}}\frac{k_{z}}{k_0\eta_0} A^{+}_\mathrm{TE}+\boldsymbol{\hat{y}} A^{+}_\mathrm{TM}
	\right)\mathrm{e}^{-jk_xx},\\
	\vect{E}^{-} =& \left(
	-\boldsymbol{\hat{x}}\frac{k_{z}}{k_0}\eta_0 A^{-}_\mathrm{TM}+\boldsymbol{\hat{y}} A^{-}_\mathrm{TE}
	\right)\mathrm{e}^{-jk_xx},\\
	\vect{H}^{-} =& \left(
	\boldsymbol{\hat{x}}\frac{k_{z}}{k_0\eta_0} A^{-}_\mathrm{TE}+\boldsymbol{\hat{y}} A^{-}_\mathrm{TM}
	\right)\mathrm{e}^{-jk_xx},
\end{align}
where $k_0^2=k_x^2+k_z^2$ is the wave vector in the vacuum, $k_0=\omega/c$; $c$ is the velocity of light; $\eta_0$ is the vacuum impedance;
$A^{-}_\mathrm{TE}$, $A^{-}_\mathrm{TM}$, $A^{+}_\mathrm{TE}$ and $A^{+}_\mathrm{TM}$ are the
complex amplitudes that can be written in matrix form:
\begin{equation}
	\vect{x}^\mathrm{T}=\left(A^{-}_\mathrm{TE},A^{-}_\mathrm{TM},A^{+}_\mathrm{TE},A^{+}_\mathrm{TM}\right),
\end{equation}

Then, Eqs.~(\ref{lb1})-(\ref{lb2}) can be represented as the following matrix equation for a eigenvalue problem:
\begin{equation}\label{egprob}
	\dbar{A}\cdot\vect{x}=k_z\, \dbar{B}\cdot\vect{x}, 
\end{equation}

Birefringent metasurfaces have only four nonzero susceptibilities components 
$\chi_{\mathrm{ee}}^{xx}$, $\chi_{\mathrm{ee}}^{yy}$, $\chi_{\mathrm{mm}}^{xx}$, $\chi_{\mathrm{mm}}^{yy}$.
In this case,
%\begin{strip}
\begin{equation*}
	\dbar{A}=\left(
	\begin{smallmatrix}
		0 &
		-2 &
		0 &
		2 \\
		j\omega\varepsilon\chi_{\mathrm{ee}}^{yy}&
		0 &
		j\omega\varepsilon\chi_{\mathrm{ee}}^{yy} &
		0 \\
		2 &
		0 &
		-2 &
		0 \\
		0 &
		j\omega\mu\chi_{\mathrm{mm}}^{yy} &
		0 &
		j\omega\mu\chi_{\mathrm{mm}}^{yy}
	\end{smallmatrix}\right),
\end{equation*}
\begin{equation*}
	\dbar{B}=\frac{1}{k\eta}\left(
	\begin{smallmatrix}
		0 &
		j\omega\varepsilon\eta^2\chi_{\mathrm{ee}}^{xx} &
		0 &
		-j\omega\varepsilon\eta^2\chi_{\mathrm{ee}}^{xx} \\
		-2 &
		0 &
		-2 &
		0 \\
		-j\omega\mu\chi_{\mathrm{mm}}^{xx}&
		0 &
		j\omega\mu\chi_{\mathrm{mm}}^{xx} &
		0 \\
		0 &
		-2\eta^2 &
		0 &
		-2\eta^2
	\end{smallmatrix}\right).
\end{equation*}
%\end{strip}

Solution to the eigenvalue problem in Eq.~(\ref{egprob}) is presented in the Table~\ref{tbl:eg1}.

\begin{table}
	\caption{Eigenvalue problem solution for birefringent metasurfaces}
	{
		\label{tbl:eg1}
		\centering
		\begin{tabular}{|l|c|c|}
			\hline
			\multicolumn{1}{|c|}{Mode} & $\vect{x}^\mathrm{T}$ & $k_z$\rule[-2ex]{0pt}{5ex} \\
			\hline
			TE, symmetric& (-1, 0, 1, 0)  &  $2 j/\chi_{\mathrm{mm}}^{xx}$\rule[-1.5ex]{0pt}{4ex}\\
			\hline
			TM, symmetric& (0, -1, 0, 1)  &  $2 j/\chi_{\mathrm{ee}}^{xx}$\rule[-1.5ex]{0pt}{4ex}\\
			\hline
			TE, asymmetric& (1,0,1,0) &  $-jk_0^2\chi_{\mathrm{ee}}^{yy}/2$\rule[-1.5ex]{0pt}{4ex}\\
			\hline
			TM, asymmetric& (0,1,0,1) &  $-jk_0^2\chi_{\mathrm{mm}}^{yy}/2$\rule[-1.5ex]{0pt}{4ex}\\
			\hline
		\end{tabular}
	}
\end{table}

In the case of an isotropic metasurface, $\chi_{\mathrm{ee}}^{xx}=\chi_{\mathrm{ee}}^{yy}=\chi_{\mathrm{ee}}^{\phantom{xx}}$ and
$\chi_{\mathrm{mm}}^{xx}=\chi_{\mathrm{mm}}^{yy}=\chi_{\mathrm{mm}}^{\phantom{yy}}$. The surface susceptibilities can be found from
the S-parameters (which in this case satisfy the following relations: $S_{11}=S_{22}$, $S_{21}=S_{12}$) as follows~\cite{martin2019}:
\begin{align}
	&\chi_{\mathrm{ee}}^{\phantom{xx}}=\frac{2j}{k_0}\ \frac{S_{11}+S_{12}-1}{S_{11}+S_{12}+1}, \\
	&\chi_{\mathrm{mm}}^{\phantom{xx}}=\frac{2j}{k_0}\ \frac{S_{11}-S_{12}+1}{S_{11}-S_{12}-1}.
\end{align}

For the isotropic case, the wave eigennumbers $k_z$ are presented in the Table~\ref{tbl:eg2}.

\begin{table}
	\caption{Eigenvalue problem solution for isotropic metasurfaces}
	\label{tbl:eg2}
	\centering
	\begin{tabular}{|l|c|}
		\hline
		\multicolumn{1}{|c|}{Mode} & $k_z$ \rule[-2ex]{0pt}{5ex}\\
		\hline
		TE, symmetric&  $\dfrac{S_{11}-S_{12}-1}{S_{11}-S_{12}+1}k$\rule[-3ex]{0pt}{7ex}\\
		\hline
		TM, symmetric&  $\dfrac{S_{11}+S_{12}+1}{S_{11}+S_{12}-1}k$\rule[-3ex]{0pt}{7ex}\\
		\hline
		TE, asymmetric&  $\dfrac{S_{11}+S_{12}-1}{S_{11}+S_{12}+1}k$\rule[-3ex]{0pt}{7ex}\\
		\hline
		TM, asymmetric&  $\dfrac{S_{11}-S_{12}+1}{S_{11}-S_{12}-1}k$\rule[-3ex]{0pt}{7ex}\\
		\hline
	\end{tabular}
\end{table}

The effective refractive index $n_{eff}$ of the investigated metasurface can be expressed in terms of its electric susceptibility $\chi_{\mathrm{ee}}$ and magnetic susceptibility $\chi_{\mathrm{mm}}$ as:

\begin{equation}
    	n_{eff} = \sqrt{\epsilon_{eff}\mu_{eff}},
\end{equation}

where $\epsilon_{eff} = 1 + \chi_\mathrm{ee}$ and $\mu_{eff} = 1 + \chi_\mathrm{ee}$ represent the effective permittivity and permeability, respectively. This description accounts for both the electromagnetic response and subwavelength unit cell interactions of the metasurface.
 
The primary challenge in applying the GSTCs to resonant metasurfaces lies in their key assumption that the average local field within the metasurface equals the arithmetic mean of the fields on either side, as expressed in Eqs.~(\ref{Eav})-(\ref{Hav}). Under resonant conditions, the validity of this approximation becomes questionable, since the electromagnetic field exhibits strong spatial variation across the metasurface thickness.
 
\begin{figure}[!tbh]
	\centering
	\centerline{\includegraphics[width=0.9\linewidth]{"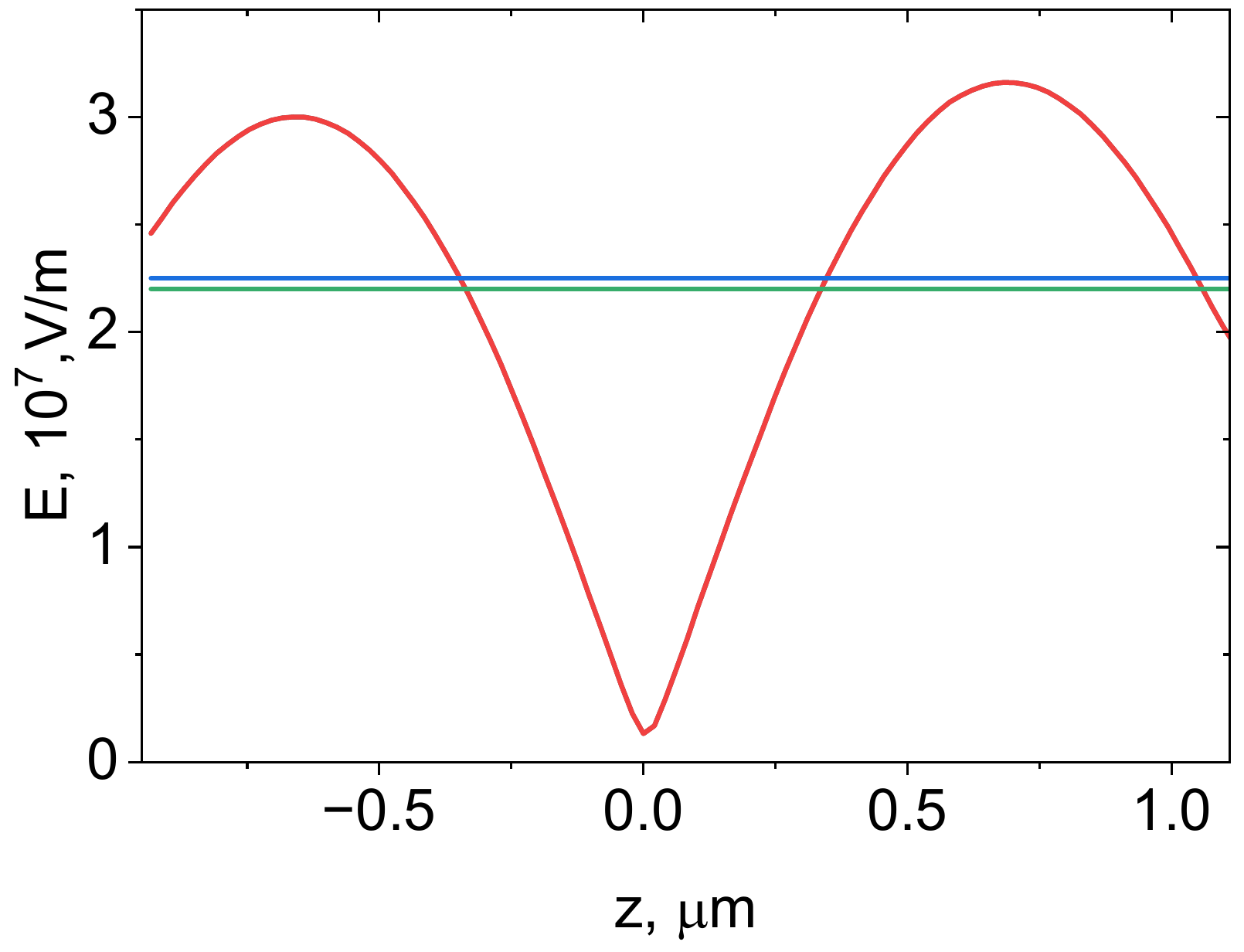"}} 
	\caption{The electric field distribution $E_{norm}~=~(E_x E_x^*+E_y E_y^*+E_z E_z^*)^{1/2}$ (red line), $E^{av}_z$ (blue line) as an average value along $z$-axis inside the metasurface unit cell, and $E^{av}$ according to Eqs.~(\ref{Eav}) (green line) in 3D metasurface unit cell with silicon sphere inside at the resonant frequency $f_{NZERI s}$.}
	\label{fig:fig11}\end{figure}

Figure~\ref{fig:fig11} presents the results of full-wave simulations of the electromagnetic field distribution within the 3D unit cell. These results indicate that, for the unit cell operating in its resonant mode, the electric field averaged across the metasurface thickness closely matches the value predicted by Eqs.~(\ref{Eav})-(\ref{Hav}). This agreement can be attributed to two key factors: (1) the unit cell is symmetric with respect to the plane $z=0$; and (2) the dielectric inclusion has a spherical shape, whose eigen modes are antisymmetric along the z-axis. Together, these properties provide a physical justification for the use of GSTCs in analyzing the metasurface under resonant conditions, supporting the reliability of the method in yielding physically meaningful results.

%F:\MSCA4UA\Publications Eremenko\2024\paper1
% Ex field distribution 3D sphere unit cell.opju

%D:\COMSOL PROJECTS\3D matesurface from Fresnel equation project\Spheres Array
%3D fresnel_equations  TM No gr spheres array freq dom.mph
%3D fresnel_equations  TM gr spheres array freq dom.mph

To verify the validity of using the GSTCs method~\cite{martin2019} to study a resonant metasurface of finite thickness and to verify the obtained results, we used another independent method presented in Ref.~\cite{wu2006}. This method does not rely on GSTCs, and, according to the authors, it applies to cell sizes of the order of a wavelength. However, it was developed only for a unit cell containing a sphere or disk. 

The main concepts of the method described in Ref.~\cite{wu2006} are as follows:
(I) A test object embedded in a medium does not scatter an incident plane wave if its refractive index matches that of the surrounding medium. Based on this, the $n_{eff}$ of a metamaterial can be defined as the refractive index of a test object whose value is chosen to eliminate scattering.
(II) A structural element of the unit cell, when surrounded by a vacuum gap, can be treated as the test object.
(III) If the unit cell consists of a dielectric sphere or rod, its scattering behavior is primarily governed by the first Mie scattering coefficient. Consequently, the $n_{eff}$ of such a structure can be determined as the refractive index of the coated sphere (or rod) that nullifies the first Mie scattering coefficient.

The scattering cross section of the coated spherical particle is given by the following equation:
%\vspace*{2cm}
%\vfill\eject
%\newpage
%\columnbreak
\begin{equation}\label{Csc3D}
	C_{sc}=\frac{2\pi}{k_0^2n_{\mathrm{eff}}^2}\sum\limits_{l=1}^\infty(2l+2)
	\left(|D_{l,\mathrm{tot}}^{(E)}|^2+|D_{l,\mathrm{tot}}^{(H)}|^2\right),
\end{equation}
where $n_{\mathrm{eff}}=\sqrt{\varepsilon_{\mathrm{eff}}\mu_{\mathrm{eff}}}$ is the metasurface effective refractive index; $\varepsilon_{\mathrm{eff}}$ and $\mu_{\mathrm{eff}}$ are the metasurface effective permittivity and permeability, respectively; 
$D_{l,\mathrm{tot}}^{(E)}$ and $D_{l,\mathrm{tot}}^{(H)}$ are the Mie scattering coefficients of the whole coated sphere for TM and TE modes, respectively.

In the sum in Eq.~(\ref{Csc3D}) two terms with $l=1$ dominate. Therefore, the effective medium conditions are $D_{1,\mathrm{tot}}^{(E)}=0$ and $D_{1,\mathrm{tot}}^{(H)}=0$.
After some algebra (for details see Ref.~\cite{wu2006}) we get the following equations for the  metasurface effective permittivity and permeability:
\begin{align}
	&\varepsilon_{\mathrm{eff}}=2\varepsilon_0B_1^{(E)}\left(B_1^{(E)}+k_0r_0B_2^{(E)}\right)^{-1},\\
	&\mu_{\mathrm{eff}}=2\mu_0B_1^{(H)}\left(B_1^{(H)}+k_0r_0B_2^{(H)}\right)^{-1},\\
	%\end{align}
	%\begin{align}
	&B_1^{(E)}=j_1(k_0r_0)+jF^{(E)}y_1(k_0r_0),\\
	&B_1^{(H)}=j_1(k_0r_0)+jF^{(H)}y_1(k_0r_0),\\
	&B_2^{(E)}=j_1^\prime(k_0r_0)+jF^{(E)}y_1^\prime(k_0r_0),\\
	&B_2^{(H)}=j_1^\prime(k_0r_0)+jF^{(H)}y_1^\prime(k_0r_0),\\
	%\end{align}
	%\begin{align}
	&D_1^{(E)}=\frac{-\mu_0D_{\psi\psi}(k_s,k_0)+\mu_sD_{\psi\psi}(k_0,k_s)}
	{\mu_0k_sD_{\psi\xi}-\mu_sk_0D_{\xi\psi}},\\
	&D_1^{(H)}=\frac{-\mu_sD_{\psi\psi}(k_0,k_s)+\mu_0D_{\psi\psi}(k_s,k_0)}
	{\mu_sk_0D_{\psi\xi}-\mu_0k_sD_{\xi\psi}},\\
	%\end{align}
	%\begin{align}
	&D_{\psi\psi}(k_1,k_2)=k_1\psi_1(k_1r_s)\psi_1^\prime(k_2r_s),\\
	&D_{\psi\xi}=\psi_1(k_sr_s)\xi_1^\prime(k_0r_s),\\
	&D_{\xi\psi}=\xi_1(k_0r_s)\psi_1^\prime(k_sr_s),
\end{align}
\noindent where $j_1$ and $y_1$ are the spherical Bessel and Neumann functions, respectively; $D_1^{(E)}$ and $D_1^{(H)}$ are the Mie scattering coefficients of the spherical inclusion (uncoated);
$F^{(E)}=D_1^{(E)}/(1+D_1^{(E)})$, $F^{(H)}=D_1^{(H)}/(1+D_1^{(H)})$,
$\psi_1(x)=xj_1(x)$, $\xi_1(x)=xh_1(x)$, $h_1(x)=j_1(x)+jy_1(x)$.

%%%%%%%%%%%%%%%%%%%%%%%%%%%%%%%%%%%%%%%%%%%%%%%%%%%%%%%%%%%%%%%%%%%%%
%% The appropriate \bibliography command should be placed here.
%% Notice that the class file automatically sets \bibliographystyle
%% and also names the section correctly.
%%%%%%%%%%%%%%%%%%%%%%%%%%%%%%%%%%%%%%%%%%%%%%%%%%%%%%%%%%%%%%%%%%%%%
\bibliography{ACS_SPPgr}
%\end{multicols}
\end{document}